\documentclass[a4paper, 12pt]{article}
\usepackage[dvips]{graphics}

\begin{document}

\title{\huge Spiral Multi-component Structure in Pad\'{e} - Approximant QCD}

\author{ F.  N.  Ndili\\
T.  W.  Bonner Nuclear Laboratory, MS315, Physics Department,\\
Rice University, Houston, TX.77251-1892, USA.}

\date{November, 2000}

\maketitle

\begin{abstract}
We present a  graphical method of analyzing the infra-red fixed point structure of Pad\'{e}
approximant QCD.  The analysis shows a spiral multi-component couplant structure as well as an
infra-red attractor behavior of PQCD couplant for all flavors $0 \le N_{f} \le 16$.  

\end{abstract}

{\it{Keywords: Pad\'{e} QCD couplant multiplicity structure}\/}\\
{\bf PACS: 12.38.-t}\\
E-mail: ndili@hbar.rice.edu

\newpage

\section{INTRODUCTION}
In two recent papers Elias et. al.~\cite{Elias98,Elias99} showed that by analyzing the sequence of 
occurrence of the first zeros of the denominator and numerator of a Pad\'{e} approximant beta function, one 
can separate out two possible behaviors of Pad\'{e} QCD couplant in the infra-red (IR) region.  The couplant
may exhibit an  IR attractor behavior of Kogan-Shifman type~\cite{Kogan95} in which the PQCD couplant
bifurcates at some critical  cut-off momentum $\mu_{c}$ into a second  (upper) branch, in a manner
that leaves the infra-red region $0 \le \mu < \mu_{c}$ totally free of any PQCD color force and its bifurcated
branch. We may call this scenario II.  The alternative scenario I is that the PQCD couplant freezes at 
the critical momentum point  $\mu = \mu_{c}$, being its stable infra-red fixed point, and remains 
so frozen for all $0 \le \mu < \mu_{c}$ in the manner of Mattingly-Stevenson~\cite{Mattingly92,Mattingly94}.

The finding of Elias et. al. by their denominator-numerator zero approach is  
that scenario II holds for all $0 \le N_{f} \le 5$, regardless of which
 Pad\'{e} approximants, $[1|2], [2|1], [1|3], [3|1]$, or $[2|2]$ they used.  On the other hand, 
 they found  that scenario I holds for all  $9 \le N_{f} \le 16$. For the flavor numbers, 
 $N_{f} = 6, 7, 8$, their method led to indeterminate results.  Their positive denominator zeros ceased to 
 exist for $N_{f} \geq 6$, but then their numerator zeros were found to be negative for $N_{f} = 6, 7, 8$. 
 As a result their method did not resolve the above question for these flavor states of  Pad\'{e}  QCD.
 
 We have re-examined the above question, using a graphical method to analyze the Pad\'{e} QCD
 couplant equation of evolution, and  applying this  not only to the NNNLO  $[2|1], [1|2]$
 Pad\'{e} approximants studied by Elias et. al. but also to the optimized  $[1|1]$ Pad\'{e} approximant
 more directly related to the optimized NNLO truncated PQCD beta function studied by Mattingly 
 and Stevenson.

 While confirming some of the findings of Elias et. al., we found substantially new features. In
 particular, we found for all flavors, not just the Elias et. al. bifurcation at momentum point
 $Q = \mu = \mu_{c} = Q_{\mathrm{min}}$, but a second  bifurcation at $Q = \mu = Q_{\mathrm{max}}$,
 where in all cases $Q_{\mathrm{max}} > Q_{\mathrm{min}}$.  More explicitly, we found the Pad\'{e} 
 couplant equation yielding three independent solutions or component couplants $a_{1}, a_{2}, a_{3}$ 
 that are however joined together into a continuous chain-like spiral structure at the two momentum points
 $Q_{\mathrm{min}}$ and $Q_{\mathrm{max}}$ that can be called  couplant bifurcation points.

 The $a_{3}$  solution arising from the second bifurcation, is found to increase towards lower
 energies,  running  past the original PQCD bifurcation point at $Q = \mu_{c}$, into the region 
 $\mu < \mu_{c}$, the implication being that this $a_{3}$ second bifurcation upper branch 
 is either a genuine NPQCD component, or else a persisting Landau pole structure coming from the original
 truncated PQCD beta function on which the Pad\'{e} beta function was based. 

By noting from our graphical structures and plots, that the momentum gap $0 \le Q  <Q_{\mathrm{min}}$
is not zero for all $9 \le N_{f} \le 16$, we found that  scenario II actually holds also 
for the flavor states $9 \le N_{f} \le 16$ as for all $0 \le N_{f} \le 8$, in difference to the findings
of Elias et. al. The only amplification is that one has to go  to sufficiently low
 momentum to see the persisting infra-red attractor points at $\mu = \mu_{c} = Q_{\mathrm{min}}$.
 This apart, the gap $0 \le \mu < \mu_{c}$ remains finite and empty of any PQCD couplant even
 for the highest flavors $N_{f} = 16, 15, 14 ..$.  The result is our new feature that there is
 intrinsically in Pad\'{e} QCD, no couplant freezing  for all flavors $0 \le N_{f} \le 16$.
 
We present the above analysis and findings as follows.  In section 2,  we collect together the
basic QCD couplant equations we need, together with their Pad\'{e} approximants, and the optimized
formulation of the particular NNLO Pad\'{e} beta function.  In section 3 , we present our 
numerical and graphical method of solving  the Pad\'{e} QCD couplant equations of orders
$[1|1], [2|1], [1|2]$ that we considered. Then  in sections 4 and 5, we state the main features
and results we found with these Pad\'{e} QCDs.  In section 6, we use our graphical method to  show
explicitly that infra-red scenario II  holds intrinsically for all flavors $0 \le N_{f} \le 16$.
In section 7, we relate our results and findings to  the IR fixed point results of Banks and
Zaks~\cite{Banks82}, and  Stevenson et. al.~\cite{Mattingly92,Mattingly94,Stevenson94,Kubo84}.   
Our summary and conclusions are stated in section 8.

\section{BASIC QCD COUPLANT EQUATIONS AND THEIR PAD\'{E} APPROXIMANTS}

\subsection{The basic dynamical Equation of evolution of QCD coupling Constant}
The starting point is  the basic RG  dynamical equation of evolution of QCD coupling constant given by: 

\begin{equation}\label{eq: ndili1}  
\mu \frac{\partial \alpha_{s}(\mu)}{\partial \mu}  =  \beta (\alpha_{s}(\mu))         
\end{equation}

where $\mu$ is some arbitrary  energy  renormalization  scale, and $\beta(\alpha_{s}(\mu))$ is the (abstract) QCD 
beta function.  Eqn.~(\ref{eq: ndili1}) as it stands is believed to have a universal validity for all of
QCD, that is to hold over a wide range of values of momentum scale $\mu$, except if  the beta function is 
later specially  parameterized or truncated. In the asymptotically free (AF) region of QCD,  the  beta 
function  is parameterized in the perturbation series form:

\begin{equation}\label{eq: ndili2}
\beta (a) = -ba^2 (1 + ca + c_{2}a^2 + c_{3}a^3 + c_{4}a^4 + \cdots  \rightarrow \infty )       
\end{equation}

where $a$ is the QCD coupling  constant in the form $a = \alpha_{s}/ \pi$,  while the first few expansion 
coefficients $b, c, c_{2}, c_{3}$ have the specific values~\cite{Gross73,Caswell74,Tarasov80}:

\begin{equation}\label{eq: ndili3} 
b = \frac{33 - 2N_{f}}{6}     = 2\beta_{0}                                      
\end{equation}

\begin{equation}\label{eq: ndili4} 
c = \frac{153 - 19 N_{f}}{2(33 - 2 N_{f})}   = \beta_{1}/ \beta_{0}                          
\end{equation}

\begin{equation}\label{eq: ndili5}
c_{2}(\overline{MS}) = \frac{3}{16(33 - 2 N_{f})} \left [ \frac{2857}{2} - \frac{5033}{18} N_{f} + \frac{325}{54} N_{f}^2  \right ] 
\end{equation}

as well as the four-loop term $c_{3}(\overline{MS})$ computed not so long ago by Ritbergen
et. al~\cite{Ritbergen97}, and given by  $c_{3}(\overline{MS}) = \beta_{3}/ \beta_{0}$, where

\begin{equation}\label{eq: ndili5a} 
\beta_{3} = 114.23033 - 27.133944 N_{f}  + 1.5823791 N_{f}^2 + 5.85669582 \times 10^{-3} N_{f}^3                             
\end{equation}

The rest of the coefficients $c_{i}, i = 4, 5, 6, \dots  \infty$ are unknown so that  QCD in making any direct use of
 eqn~(\ref{eq: ndili2}) can at most use only the following truncated beta functions;

\begin{enumerate}

\item The leading order form (LO), where $\beta (\alpha_{s})$ is truncated to:

\begin{equation}\label{eq: ndili6}
\beta (a) \approx  -ba^2 = \beta^{(1)} (a)                                   
\end{equation}

\item The next-to-leading order form (NLO), where $\beta (\alpha_{s})$ is truncated to:

\begin{equation}\label{eq: ndili7}
\beta (a)  \approx  -ba^2(1 + ca ) =  \beta^{(2)}(a)                           
\end{equation}

\item The third order form or the NNLO truncation where we approximate the QCD beta function by :

\begin{equation}\label{eq: ndili8}
\beta (a)  \approx  -ba^2(1 + ca + c_{2}a^2 )  =  \beta^{(3)} (a)                     
\end{equation}

\item  The fourth order form or the NNNLO  truncation  where we exhaust all presently  known terms of eqn.~(\ref{eq: ndili2})
and approximate $\beta (\alpha_{s})$ as:

\begin{equation}\label{eq: ndili9}
\beta (a)  \approx  -ba^2(1 + ca + c_{2}a^2 + c_{3}a^3 )  =   \beta^{(4)} (a)                 
\end{equation}

\end{enumerate}

These truncations now invariably restrict the range of validity of eqn.~(\ref{eq: ndili1}) to only the
purely PQCD region.

The Pad\'{e} approximant for QCD is a means of extending any one of these truncated series into a 
specific infinite series of Pad\'{e}  form, that is hoped to correct to some extent, the effects of truncation
on QCD dynamics given by eqn.~(\ref{eq: ndili1}), compared to any direct use of
the pure truncated  eqns.~(\ref{eq: ndili6}) to~(\ref{eq: ndili9}).  Such Pad\'{e} extensions
are then substituted into the general solution of eqn.~(\ref{eq: ndili1}), given in one form as:

\begin{equation}\label{eq: ndili13} 
\tau = b \ln \frac{Q}{\tilde \Lambda} = b\int_0^a \left [ \frac{1}{\beta (x)} - \frac{1}{\beta^{(2)}(x)} \right ] dx - b\int_a^\infty \frac{dx}{\beta^{(2)}(x)}
\end{equation}

where $\tilde \Lambda$ is related to $\Lambda_{\mathrm{QCD}}$ by

\begin{equation}\label{eq: ndili14}
\ln \frac{\Lambda_{\mathrm{QCD}}}{\tilde \Lambda}  =  \frac{c}{b} \ln \frac{2c}{b}                 
\end{equation}

If we substituted eqns.~(\ref{eq: ndili6}) to~(\ref{eq: ndili8}) in succession into eqn.~(\ref{eq: ndili13}),
we will obtain PQCD couplant equations of the form:

\begin{equation}\label{eq: ndili13a} 
\tau = b \ln \frac{Q}{\tilde \Lambda} =  \frac{1}{a}
\end{equation}

\begin{equation}\label{eq: ndili13b}
\tau  =  b \ln \frac{Q}{\tilde \Lambda}  = \frac{1}{a}   +  c \ln \left | \frac{c a}{1 + c a} \right |  
\end{equation}

\begin{equation}\label{eq: ndili13c} 
\tau  =  b \ln \frac{Q}{\tilde \Lambda}  = \frac{1}{a}   + 
\frac{c}{2} \ln \frac{a^2}{\left | X \right |} +
\frac{c^2 - 2c_{2}}{2 a c_{2} + c}  - \frac{c^2 - 2c_{2}}{c}
\end{equation}
where $X = c_{2} a^2 +  c a + 1$.  
 
The main feature of these couplant equations is the existence of one couplant solution at any given  high
momentum Q, (or $\mu$), with this solution going to infinity when the momentum decreases to a critical value given by
$Q  = \Lambda_{\mathrm{QCD}}$, analogous to a Landau pole.  The behavior cuts off any access from the PQCD
region. into the infra-red region  $0 \le  Q  \le \Lambda_{\mathrm{QCD}}$ of QCD. We are  now to examine 
how  this picture of PQCD IR region changes when we replace the above truncated PQCD beta functions by 
their Pad\'{e} approximants and couplant equations.

\subsection{The Pad\'{e}  approximant QCD.}
According to the Pad\'{e} principle~\cite{Samuel95} to~\cite{Ellis98}, given a generic infinite series:

\begin{equation}\label{eq: ndili26}
S(x) = \sum_{n = 0}^{\infty} c_{n} x^n                                         
\end{equation}
whose terms are unknown except the first few  p terms, meaning the given series is in effect a truncated
series of only p usable terms, one can construct an  infinite series approximation for the same S(x) but
now of fully known terms, by solving the following identity equation :

\begin{equation}\label{eq: ndili27} 
S_{(N|M)} (x) =  \sum_{n = 0}^p c_{n}x^n                                              
\end{equation}

where explicitly :

\begin{equation}\label{eq: ndili28}
S_{(N|M)} (x) =  \frac{1 + \gamma_{11} x + \gamma_{12} x^2 + \cdots + \gamma_{1N} x^N}{1 + \gamma_{21} x + \gamma_{22} x^2 + \cdots + \gamma_{2M} x^M}   
\end{equation}

with  N and M being  integers  chosen  such  that  $N + M = p$.  The  quantities  $\gamma_{ij}$ are  unknown
coefficients which can all be determined in terms of the p known coefficients $c_{n}$ of eqn.~(\ref{eq: ndili26}).

The quantity  $S_{(N|M)}(x)$ as explicitly  given in eqn.~(\ref{eq: ndili28}) is what is called the Pad\'{e} 
infinite  series approximation (approximant) of the original infinite series $S(x)$ given in eqn.~(\ref{eq: ndili26}).
By inverting the  denominator  and expanding out with the numerator, the Pad\'{e} function is seen to be an
infinite series of fully known terms.  By  construction,  its first p terms are identical in value with
the first p terms of $S(x)$, but beyond these first p terms, other corresponding  terms in general differ
in their values were one to compute the unknown terms of $S(x)$ and compare with the fully known terms of
$S_{(N|M)} (x)$ as done by Ellis et.al.~\cite{Ellis97}.  One finds in general good agreement in the
various  studies~\cite{Samuel95} to~\cite{Jack97} carried out with the Pad\'{e} approximant.

Depending on our choice of values of N and M for a given p, the Pad\'{e} series  is
said to be of order (N, M).  Based directly on the only known terms of QCD beta function given in
eqns.~(\ref{eq: ndili3}) to~(\ref{eq: ndili5a}), the only non-trivial Pad\'{e} approximants we can consider,
are the (1,1), (0,2) at NNLO level, and the (1,2) (2,1) and (0,3) at NNNLO level of Pad\'{e} summation.
In this paper we have analyzed the (1,1), (1,2) and (2,1) Pad\'{e} approximants as representative enough to
show us the general features of Pad\'{e} QCD couplant equations. For the (1,1) NNLO Pad\'{e} approximant case,
we have also analyzed its optimized form, providing further information on the nature of the Pad\'{e}
couplant equation of evolution.  We give details below of these Pad\'{e}  approximants and couplant
equations. 
\begin {enumerate}
\item {\bf The $[1|1]$ Pad\'{e} approximant and couplant equation}\\
At the NNLO order the $[1|1]$ Pad\'{e} approximant is given by:

\begin{equation}\label{eq: ndili30} 
S_{(1|1)} (a)  =  \frac{ 1 + \gamma_{1} a}{ 1 + \gamma_{2} a}           
\end{equation}
with the corresponding Pad\'{e} beta function being :
\begin{equation}\label{eq: ndili31}
\beta^{(3P)} (a) = - ba^2 S_{(1|1)} (a) = - ba^2 \left ( \frac{ 1 + \gamma_{1} a}{ 1 + \gamma_{2} a}  \right )  
\end{equation}

where $\gamma_{1}$ and  $\gamma_{2}$ are the two unknown Pad\'{e} coefficients to be determined from the identity equation:

\begin{equation}\label{eq: ndili32}
-ba^2 (1 + a \gamma_{1}) (1 + a \gamma_{2})^{-1}  \equiv  -ba^2 (1 + c a  + c_{2} a^2) +  \cdots  \rightarrow \infty  
\end{equation}

We deduce that:

\begin{equation}\label{eq: ndili33}  
\gamma_{1}  = c  - \frac{c_{2}}{c}                                 
\end{equation}

\begin{equation}\label{eq: ndili34} 
\gamma_{2}  =  - \frac{c_{2}}{c}                                    
\end{equation}

Then eqn.~(\ref{eq: ndili8}) becomes replaced by:

\begin{eqnarray}
\beta^{(3)} (a)  \rightarrow  \beta^{(3P)} (a)   &=& -ba^2 \left ( \frac{1 + a \gamma_{1}}{1 + a \gamma_{2}} \right )     \nonumber\\ 
                                                 &=&  -ba^2 (1 + a \gamma_{1}) (1 + a \gamma_{2})^{-1}      \label{eq: ndili35}
\end{eqnarray}

Substituting  eqn~(\ref{eq: ndili31}) into  eqn.~(\ref{eq: ndili13}), we obtain the dynamical
relationship  between the Pad\'{e} QCD couplant $a$, and the dynamical variables Q and $N_{f}$ of QCD.  This  dynamical 
equation of evolution of $\alpha_{s} (Q,N_{f})$ becomes after explicit integration:

\begin{equation}\label{eq: ndili37}
\tau  =  b \ln \frac{Q}{\tilde \Lambda}  = \frac{1}{a}   +  c \ln \left | \frac{a \gamma_{1}}{1 + a \gamma_{1}} \right |  
\end{equation}

This (1,1) Pad\'{e} QCD couplant equation~(\ref{eq: ndili37}) can be analyzed for its features particularly
in the infra-red region and compared with the  Landau pole structure of truncated 
eqn.~(\ref{eq: ndili8}).

If we were to analyze this NNLO Pad\'{e} QCD by the method of Elias et. al.~\cite{Elias98,Elias99}, we
would focus on the Pad\'{e} beta function eqn.~(\ref{eq: ndili31}) and note the sequence of occurrence
of its denominator and numerator zeros.  The sequence  is shown in Table~\ref{tab: ndili1a}, 
and would lead one to expect, a Kogan-Shifman type ~\cite{Kogan95}  infra-red attractor behavior for 
$0 \le N_{f} \le 8$, while the Mattingly-Stevenson~\cite{Mattingly92,Mattingly94} infra-red frozen
couplant behavior may exist for $9 \le N_{f} \le 16$.  We will see later that analyzing eqn.~(\ref{eq: ndili37})
by our graphical method described below, provides a lot more information  on the above scenario, 
and alters the conclusion significantly.

\item {\bf The $[2|1]$ Pad\'{e} approximant and couplant equation}\\
We write down also the  $[2|1]$ Pad\'{e} approximant given by:
 
\begin{equation}\label{eq: ndili59a}
\beta_{[2|1]}^{(4P)} (a) = - ba^2 S_{(2|1)} (a) = - ba^2 \left ( \frac{ 1 + \gamma_{21} a + \gamma_{22} a^2}
{ 1 + \eta_{21} a}  \right )  
\end{equation}
where\\
\begin{eqnarray}
\gamma_{21}  &=&  c - \frac{c_{3}}{c_{2}}   \nonumber\\
\gamma_{22}  &=&  c_{2} - \frac{c c_{3}}{c_{2}}    \nonumber\\
\eta_{21}    &=&  - \frac{c_{3}}{c_{2}}          \label{eq: ndili59b}
\end{eqnarray}

Substituting this Pad\'{e} beta function into eqn.~(\ref{eq: ndili13}) we obtain the $[2|1]$
Pad\'{e} QCD couplant equation:

\begin{eqnarray}
\tau  =  b \ln \frac{Q}{\tilde \Lambda} &=& \frac{1}{a}   + \frac{1}{2}(\gamma_{21} - \eta_{21}) \ln \frac{a^2}{\left | X \right |}  \nonumber\\
        &+& \frac{\gamma_{21}^2 - 2 \gamma_{22} - \eta_{21}\gamma_{21}}{2 \gamma_{22} a + \gamma_{21}}  \nonumber\\
        &+& \eta_{21} - \gamma_{21} + \frac{2 \gamma_{22}}{\gamma_{21}}  \label{eq: ndili59e}
 \end{eqnarray}
where $X = a^2 \gamma_{22} + a \gamma_{21} + 1$.  
 
\item {\bf The $[1|2]$ Pad\'{e} approximant and couplant equation}\\
In the same way, the $[1|2]$ Pad\'{e} approximant is given by:
 
\begin{equation}\label{eq: ndili59c}
\beta_{[1|2]}^{(4P)} (a) = - ba^2 S_{(1|2)} (a) = - ba^2 \left ( \frac{ 1 + \gamma_{11} a}
{ 1 + \eta_{11} a  + \eta_{12}}  \right )  
\end{equation}
where\\
\begin{eqnarray}
\gamma_{11}  &=&  \frac{c^3 - 2 c c_{2} + c_{3}}{c^2 - c_{2}}   \nonumber\\
\eta_{11}  &=&  \frac{c_{3} - c c_{2}}{c^2 - c_{2}}    \nonumber\\
\eta_{12}    &=&   \frac{c_{2}^2 - c c_{3}}{c^2 - c_{2}}          \label{eq: ndili59d}
\end{eqnarray}

Substituting  this Pad\'{e} beta function into eqn.~(\ref{eq: ndili13}) we obtain the 
 $\tau$ equation for the $[1|2]$ Pad\'{e} QCD:
\begin{eqnarray}
\tau  =  b \ln \frac{Q}{\tilde \Lambda} &=& \frac{1}{a}   + (\eta_{11} - \gamma_{11}) \ln \left | a \gamma_{11} + 1 \right |  \nonumber\\
        & - &  (\eta_{11} - \gamma_{11}) \ln \left | a \gamma_{11} \right |  \nonumber\\
        & - &  \frac{\eta_{12}}{\gamma_{11}} \ln \left | a \gamma_{11} + 1 \right |    \label{eq: ndili59f}
\end{eqnarray}

\end{enumerate}

\subsection{Optimization of the NNLO $[1|1]$  Pad\'{e} Approximant}
Regarding the $[1|1]$  Pad\'{e} approximant, we can come closer to the work of Mattingly-Stevenson for
later comparison purposes,  by subjecting our $[1|1]$ Pade approximant to the PMS  optimization  
principle~\cite{Stevenson81, Mattingly92, Mattingly94}, and obtaining in the process a slightly
modified  eqn.~(\ref{eq: ndili37}). The details are as follows. 

The PMS principle  derives from the fact that a physical
observable such as the cross section for $(e^+e^-  \rightarrow  hadrons)$, parameterized as an infinite perturbative series,
has the same  physical  value  irrespective  of the RS used to compute its  perturbative  coefficients.
However  if the series is  truncated,  (with a  corresponding  truncation  of the beta  function),  the
physical  observable  becomes  RS  dependent.  The idea of  Stevenson  is that  one can  minimize  this
dependence of a truncated physical  observable on its RS variables by using only those values of the RS
variables that satisfy his equation (called optimization equation):

\begin{equation}\label{eq: ndili38}  
\frac{\partial R^{(n)} (Q)}{\partial (RS)}  =  0                                    
\end{equation}

where (RS) here stands for the complete set of RS variables on which the  truncated  physical  observable
$R^{(n)}(Q)$, (and its  correspondingly  truncated  beta  function  $\beta^{(n)}(a)$) depend.  By solving the set of
simultaneous  equations  implied by~(\ref{eq: ndili38}), one obtains the optimal or  optimized  set of values of the RS
variables  one should  use in the  truncated  equations  for $R^{(n)}(Q)$,  together  with a  corresponding
optimized  QCD couplant $a$ and its truncated  beta  function,  all involved in $R^{(n)} (Q)$.  In the case of
eqn.~(\ref{eq: ndili31}) taken along with some third order  truncated  physical  observable  $R^{(3)}_{e^+e^-} (Q)$, the  optimization
equations we need to set up are:

\begin{equation}\label{eq: ndili39}
R_{e^+e^-}^{(3)}(Q) = a(1 + r_{1} a + r_{2} a^2 )                                              
\end{equation}

\begin{equation}\label{eq: ndili40}
\beta^{(3P)} (a)   = -ba^2 \left ( \frac{1 + a \gamma_{1}}{1 + a \gamma_{2}}  \right )                         
\end{equation}

where $\gamma_{1}$ and  $\gamma_{2}$  are as already given in eqns~(\ref{eq: ndili33})~,~(\ref{eq: ndili34}).
Also we have from eqn.~(\ref{eq: ndili1}), in the form of eqns.~(\ref{eq: ndili13}),
the basic dynamical equation for $\alpha_{s}(Q)$:

\begin{equation}\label{eq: ndili41}
\tau = b \ln \frac{Q}{\tilde \Lambda}  = \frac{1}{a}  +  c \ln \left | \frac{a \gamma_{1}}{1 + a \gamma_{1}} \right |     
\end{equation}

Identifying $\tau$ and $c_{2}$ as the only RS variables present in
eqns.~(\ref{eq: ndili39})~,~(\ref{eq: ndili40})~,~(\ref{eq: ndili41}), the optimization
eqn.~(\ref{eq: ndili38}) takes the form of two equations:

\begin{equation}\label{eq: ndili42}
\left (\frac{\partial}{\partial \tau} \left |_{a} \right . +  \frac{\beta^{(3P)}}{b} \frac{\partial}{\partial a} \right ) R^{(3)}(Q) = 0    
\end{equation}

\begin{equation}\label{eq: ndili43}
\left (\frac{\partial}{\partial c_{2}} \left |_{a} \right . +  \beta_{(2)} (a) \frac{\partial}{\partial a} \right ) R^{(3)} (Q)  = 0 
\end{equation}

where  $\beta_{(2)}(a) = \frac{\partial a}{\partial c_{2}}$.

Substituting eqns.~(\ref{eq: ndili39})~,~(\ref{eq: ndili40}) into eqns.~(\ref{eq: ndili42})~,~(\ref{eq: ndili43}), we obtain
from~(\ref{eq: ndili42}):

\begin{equation}\label{eq: ndili44}
(1 + a \gamma_{2})  +  (1 + a \gamma_{2}) (2 a r_{1} + a c) - (1 + a \gamma_{1}) (1 + 2 a r_{1} + 3 a^2 r_{2})  = 0 
\end{equation}

Next, from eq.~(\ref{eq: ndili43}) we obtain:

\begin{equation}\label{eq: ndili45}
- ( 1 + a \gamma_{2})  +  (1 + 2 a r_{1}  + 3 a^2 r_{2})    =   0                           
\end{equation}

Equations~(\ref{eq: ndili41})~,~(\ref{eq: ndili44})~,~(\ref{eq: ndili45}) become the simultaneous  equations to solve for the 
optimized quantities: $a = \bar a ; \tau =  \bar \tau;  r_{2} = \bar r_{2};  c_{2} = \bar c_{2}$ ; 
being the optimized values for which our physical reference system $R^{(3)} (Q)$ would be minimally affected by RS changes.

To the  optimization  constraints in eqns.~(\ref{eq: ndili41})~,~(\ref{eq: ndili44})~,~(\ref{eq: ndili45}), one now adds the intrinsic 
constraints on RS variables  provided by the general  existence in PQCD of certain  combinations of RS variables that are
themselves RS invariants~\cite{Stevenson81,Stevenson86,Dhar83,Dhar84}.  In our case there are two such RS invariants,  
$\rho_{1}$ and $\rho_{2}$, given explicitly by:

\begin{equation}\label{eq: ndili46}
\rho_{1}  =  \tau - r_{1}  = \bar \tau  - \bar r_{1}   =  \mathrm{invariant}.                     
\end{equation}

\begin{equation}\label{eq: ndili47}
\rho_{2} = r_{2} + c_{2} - (r_{1} + \frac{1}{2} c)^2 = \bar r_{2} + \bar c_{2} - (\bar r_{1} + \frac{1}{2} c)^2 = \mathrm{invariant}.  
\end{equation}

where the RS  variables  on the right hand side are  individually  dependent on the RS used to evaluate
them  (except $b$ and $c$), but $\rho_{1}$ and $\rho_{2}$ have  numerical  values not  dependent on the RS used to evaluate
each combination of RS variables shown on the right hand side.  Because of this RS invariance  property
of $\rho_{1}$ and $\rho_{2}$, one can  obtain a  numerical  value for them in any  convenient  RS, and these  numerical
values  of $\rho_{1}$ and $\rho_{2}$  become  usable  as  input  data  into the main  optimization  eqns.~(\ref{eq: ndili41})~,
~(\ref{eq: ndili44})~,~(\ref{eq: ndili45}).  Specifically  we used the $\overline {MS}$ computed  values of $\rho_{1}$ and 
$\rho_{2}$ given by~\cite{Mattingly94,Stevenson99}.  Although $\rho_{1}$ and $\rho_{2}$ now become mere numbers, 
eqns.~(\ref{eq: ndili3})~,~(\ref{eq: ndili4})~,~(\ref{eq: ndili5})~,~(\ref{eq: ndili46})~,~(\ref{eq: ndili47}), taken  together,
show that these  numerical values of $\rho_{1}$ and $\rho_{2}$ still  depend on the values we assign
to momentum Q and flavor  number $N_{f}$ .  Denoting  these  numerical quantities  that  depend for their  value only on our choice 
of Q and $N_{f}$ by $\rho_{1}(Q, N_{f})$ and  $\rho_{2}(N_{f})$,  the complete set of our  optimization  equations  becomes
eqns.~(\ref{eq: ndili41})~,~(\ref{eq: ndili44})~,~(\ref{eq: ndili45}), together with the numerical value  constraints  provided by
$\rho_{1}(Q, N_{f})$ and $\rho_{2}(N_{f})$, for the combinations of RS variables shown on the
right hand side of eqns.~(\ref{eq: ndili48})~,~(\ref{eq: ndili49}) below:

\begin{equation}\label{eq: ndili48}
\rho_{1} (Q, N_{f})   =  \tau  -  r_{1}                                            
\end{equation}

\begin{equation}\label{eq: ndili49}
\rho_{2} (Q, N_{f})  =  r_{2}  +  c_{2}  -  (r_{1}  + \frac{1}{2} c)^2                         
\end{equation}

It is these five simultaneous equations (eqns.~(\ref{eq: ndili41})~,~(\ref{eq: ndili44})~,~(\ref{eq: ndili45})~,~(\ref{eq: ndili48})~,
~(\ref{eq: ndili49})) involving five RS variables $(\tau, a , r_{1} , r_{2} , c_{2})$, (but $\tau$ and $a$ are not  independent),  
that we now solve by a process of elimination to obtain one final  equation for our  optimized  couplant $a = \bar a$, in terms of 
optimized $\bar \tau$, as our best value (or our  minimally RS dependent)  solution  of eqn.~(\ref{eq: ndili1}) into  which  
the NNLO Pad\'{e} approximant eqn.~(\ref{eq: ndili31})  was  already  substituted  in the  course of the optimization
process.

The elimination process proceeds as follows:  we substitute eqns.~(\ref{eq: ndili33}) and~(\ref{eq: ndili34})
into eqns~(\ref{eq: ndili44}) and~(\ref{eq: ndili45}) obtaining:

\begin{equation}\label{eq: ndili50}
3r_{2}  + 2r_{1}c  + c_{2}  + 3r_{2}(c - \frac{c_{2}}{c})a   = 0                        
\end{equation}
and \\
\begin{equation}\label{eq: ndili51}
a \left ( \frac{c_{2}}{c} \right )  +  2 a r_{1} + 3 a^2 r_{2}     =  0                         
\end{equation}

We then solve eqns.~(\ref{eq: ndili50})~,~(\ref{eq: ndili51}) to obtain the optimized values of $r_{1} = \bar r_{1}$,  $r_{2} = \bar r_{2}$.  
The result is:
 \begin{eqnarray}
\bar r_{1}    &=&  - \frac{1}{2}\left (\frac{c_{2}}{c} \right)       \nonumber\\
\bar r_{2}    &=&   0            \label{eq: ndili49B}
\end{eqnarray}

Substituting these optimized values $\bar r_{1}$ and $\bar r_{2}$ into eqns.~(\ref{eq: ndili48})~,~(\ref{eq: ndili49})
we obtain:

\begin{equation}\label{eq: ndili52}
\rho_{1}(Q, N_{f})   =   \tau + \frac{1}{2}\left (\frac{c_{2}}{c} \right )                                 
\end{equation}

\begin{equation}\label{eq: ndili53}
\rho_{2}(N_{f})  = \frac{3}{2} c_{2}  -  \frac{1}{4}\frac{c_{2}^2}{c^2}  - \frac{c^2}{4}     
\end{equation}

as the  optimization  constraints  $\tau$ and $c_{2}$ have to satisfy,  where from eqns.~(\ref{eq: ndili1})~,~(\ref{eq: ndili31}),
$\tau$ is already given by eqn.~(\ref{eq: ndili41}).  We can solve for optimized $c_{2}$ by regarding  eqn.~(\ref{eq: ndili53})
as a quadratic equation in $c_{2}$ written as:

\begin{equation}\label{eq: ndili54}
c_{2}^2  -  (16c^2) c_{2}  + (4 c^2 \rho_{2}  + c^4)   =  0                          
\end{equation}

We get that :

\begin{equation}\label{eq: ndili55}
c_{2} = \bar c_{2} = 3c^2 \pm  c \sqrt D  = c(3c  \pm  \sqrt D ) = 2 c \left ( \frac{3}{2} c  \pm  \frac{1}{2} \sqrt D \right )  
\end{equation}

where $D = 9c^2 - (4 \rho_{2} + c^2 ) = 8c^2 - 4 \rho_{2}$.
If we now substitute eqn.~(\ref{eq: ndili54}) into eqn.~(\ref{eq: ndili52}) we get the optimized $\tau = \bar \tau$ given by

\begin{equation}\label{eq: ndili56} 
\bar \tau =  \rho_{1} (Q, N_{f})  - \frac{1}{2} \left [ 3c  \pm  \sqrt D  \right ]             
\end{equation}

It is this $\bar \tau$ we finally  substitute into eqn.~(\ref{eq: ndili41}) to obtain our optimized Pad\'{e}  couplant $a = \bar a$ that 
satisfies our universal eqn~(\ref{eq: ndili1}) with eqns.~(\ref{eq: ndili31})~,~(\ref{eq: ndili32})~,~(\ref{eq: ndili33})~,
~(\ref{eq: ndili34})~,~(\ref{eq: ndili35}).  The result is that:

\begin{equation}\label{eq: ndili57}
\bar \tau = \frac{1}{\bar a}  + c \ln \left | \frac{\bar \gamma_{1} \bar a}{1 + \bar a \bar \gamma_{1}} \right |  = \rho_{1} (Q, N_{f}) - 
\frac{1}{2} \left [ 3 c \pm \sqrt D \right ]   
\end{equation}

is our equation for determining $\bar a$ of our $[1|1]$ Pad\'{e} QCD.  Next noting from
 eqns.~(\ref{eq: ndili33}) and~(\ref{eq: ndili34}) that $\gamma_{1} = c - \frac{c_{2}}{c}$, meaning
$\bar \gamma_{1}  =  c - \frac{\bar c_{2}}{c}$ we can rewrite:

\begin{equation}\label{eq: ndili49C}
1 + \bar a  \bar \gamma_{1}  = c \bar a - \bar a (3 c \pm \sqrt D) +1  =  1 + c \bar a - \bar a (3 c  \pm  \sqrt D) = 1 + c \bar a  -2 \bar a P
\end{equation}

where $P  =  \frac{1}{2} (3 c  \pm  \sqrt D)$, giving :

\begin{equation}\label{eq: ndili58}
\bar \tau = \frac{1}{\bar a}  + c \ln \left | \frac{k \bar a}{1 + c \bar a  - 2 \bar a P} \right |    
\end{equation}
where $k = c - 2P$.

The final result is that our optimized $[1|1]$ Pad\'{e} QCD couplant equation is given by:

\begin{equation}\label{eq: ndili59}
\rho_{1}(Q, N_{f})  -  P  =  \frac{1}{\bar a}  + c \ln \left | \frac{k \bar a}{1 + c \bar a  -  2 \bar a P} \right  |  
\end{equation}

To simplify notation, we shall from here on, drop the bar over the optimized couplant in eqn.~(\ref{eq: ndili59}), and simply
write  $a$  in the remaining parts of this paper.   

\subsection{Pad\'{e} beta function denominator and numerator zero arguments applied to the optimization equations}
As in the case of eqn.~(\ref{eq: ndili37}) discussed earlier, before using eqn.~(\ref{eq: ndili59}) to
investigate the behavior of Pad\'{e} QCD couplant with momentum scale Q, especially near the infra-red region, 
one can use Elias et. al type argument~\cite{Elias98,Elias99}, to indicate what to expect  from
eqn.~(\ref{eq: ndili59}).  In this case, we apply the Pad\'{e} beta function denominator and numerator zero 
tests
to the complete set of the optimization  equations given as eqns.~(\ref{eq: ndili41})~,~(\ref{eq: ndili44})~,~(\ref{eq: ndili45})~,~(\ref{eq: ndili48})~,
~(\ref{eq: ndili49}).  The question is do the equations remain consistent and viable if we solve them under
the specific condition: $(1 + a \gamma_{2}) = 0$  with   $(1 + a \gamma_{1})  \ne 0$, which will yield
infra-red couplant behavior of the Kogan-Shifman type.  Correspondingly we ask if the same equations remain 
consistent and viable if we require to solve them under the condition that $(1 + a \gamma_{1}) = 0$, but
$(1 + a \gamma_{2}) \ne 0$ which will yield Mattingly-Stevenson type infra-red couplant behavior.  We find
 by looking at eqns.~(\ref{eq: ndili41})~,~(\ref{eq: ndili44})~,~(\ref{eq: ndili45})~,~(\ref{eq: ndili48})~,
~(\ref{eq: ndili49}), that while these equations remain viable under the denominator zero condition, they 
lose this viability under the numerator zero condition.  In particular eqn.~(\ref{eq: ndili41}) blows up for
$(1 + a \gamma_{1}) = 0$,  

The indication would then be that while eqn.~(\ref{eq: ndili59}) is likely to  give the Kogan-Shifman type infra-red 
behavior in most cases, the opposite Mattingly-Stevenson type behavior will probably not be found for
any flavor state of eqn.~(\ref{eq: ndili59}). However, totally  independent of the above 
Elias et. al type considerations, we now present  our own graphical and numeraical method of analyzing
the Pad\'{e} couplant eqns.~(\ref{eq: ndili37})~,~(\ref{eq: ndili59e})~,~(\ref{eq: ndili59f})
 and~(\ref{eq: ndili59}), the results of which we can compare later with
the above  denominator and numerator zero  previews.  We give details for the case of
eqn.~(\ref{eq: ndili59}) and  quote our results for the other couplant equations.

\section{NUMERICAL SOLUTION OF THE $[1|1]$ OPTIMIZED PAD\'{E} QCD COUPLANT EQUATION OF MOTION}

Equation~(\ref{eq: ndili59}) can be solved  numerically  for the couplant $a$, at any one chosen value of momentum Q and
flavor  number $N_{f}$, which  means at any one chosen  value of $\rho_{1}(Q, N_{f})$ and $\rho_{2}(N_{f})$.  Then the value of Q
and $N_{f}$ can be changed and the numerical solution process repeated. We chose  to keep $N_{f}$ fixed at
any one  value of $N_{f}$
in the  range :  $0 \le N_{f} \le 16$,  while we  varied  Q over a wide range from very small to very large Q,
and obtained a set of values
of the (optimized) couplant $a$, that are solutions of eqn.~(\ref{eq: ndili59}) at each  chosen  value of Q.  In the  process  one can 
plot out a graph of these  solutions  of eqn.~(\ref{eq: ndili59}) against Q, for any one fixed value of flavor number $N_{f}$, 
to see how Pad\'{e} QCD couplant, for a given flavor $N_{f}$,  behaves over a wide range of Q values, from  large
 $Q  \rightarrow \infty$, to  small $Q \rightarrow  0$.  Subsequently,  the value of flavor number $N_{f}$
can be changed and the entire process repeated to obtain a  separate  $(a, Q)$ plot at the  corresponding  new value of $N_{f}$.
Equation~(\ref{eq: ndili59}) was  solved separately in this way for all integer flavor numbers $N_{f} = 0, 1, 2, 3, 4,  \ldots, 16.$
We point out immediately that our varying Q over the wide range $Q  \rightarrow \infty$, to
$Q \rightarrow  0$ does not in any way imply that a Pad\'{e} couplant solution necessarily exists
in these momentum limits. As we explain further below, the procedure enables the system to pick its own
cut-off momentum below which or above which a solution exists or does not exist. This will become fully 
clear below.

We now give details of the exact method of numerical  solution used, the method being the same for any one of
the above $N_{f}$ values.  Our method  was to regard the  couplant $a$ in eqn.~(\ref{eq: ndili59}) as a floating  variable  that is allowed to
assume a wide range of values at any one fixed Q value (and also fixed $N_{f}$), such that for any one  floating value of $a$, the left
and the right hand sides of eqn.~(\ref{eq: ndili59}) are in general not equal, meaning that the 
floating value of $a$, is in general not a solution of eqn.~(\ref{eq: ndili59}), and therefore
not acceptable as a Pad\'{e} QCD couplant.

However,  there exists some unique value (or multiple values) of the floating variable $a$, (at a given Q and fixed 
$N_{f}$) for which the left and the  right  hand  sides of eqn.~(\ref{eq: ndili59}) are  exactly  equal,
measured by a curve crossing a solution line. That  unique  floating  value of $a$, becomes  identifiable  as our exact
solution  of  eqn.~(\ref{eq: ndili59}).  Thereafter,  we change  the value of Q and again  keep this Q fixed at its new
value,  while the  couplant $a$ floats again over a wide range (and in very fine steps we chose as: $\Delta a = 0.00001)$,
until we again find that  unique  value of  couplant $a$ (crossing point) , for which the left and the right hand sides of
eqn.~(\ref{eq: ndili59}) are exactly equal.  In this way it was possible to obtain a set of couplant values $a$, that are
exact  solutions  of eqn.~(\ref{eq: ndili59}) at given values of Q, all the time  keeping to one fixed $N_{f}$ value.  For a
different $N_{f}$ value the entire search process is repeated, always starting from very low Q values, and
moving up in suitable  steps to higher Q values, or vice versa.  The further  details of the above manner of floating
the couplant $a$, and constantly comparing when the left and the right hand sides of eqn.~(\ref{eq: ndili59}) are exactly
equal, are what we describe next.

First, we note that the  double-valuedness of the optimized variable $c_{2}$ in eqn~(\ref{eq: ndili55}) implies that
eqn.~(\ref{eq: ndili59}) can be written out as two separate equations, each one of which can be separately solved numerically to
yield its own (optimized) Pad\'{e} couplant solution, together with a corresponding  separate $(a, Q)$ plot.  The
two separate equations we get out of eqn.~(\ref{eq: ndili59}) are:

\begin{equation}\label{eq: ndili60}
\rho_{1}(Q, N_{f})  -  P_{1}  =  \frac{1}{\bar a}  + c \ln \left | \frac{k \bar a}{1 + c \bar a  -  2 \bar a P_{1}} \right  |  
\end{equation}

and\\

\begin{equation}\label{eq: ndili61}
\rho_{1}(Q, N_{f})  -  P_{2}  =  \frac{1}{\bar a}  + c \ln \left | \frac{k \bar a}{1 + c \bar a  -  2 \bar a P_{2}} \right  |  
\end{equation}

where:

\begin{equation}\label{eq: ndili62}
P_{1}  = \frac{3c}{2}  + \frac{1}{2} \sqrt D   = \frac{1}{2c} \bar c_{2} (+)                
\end{equation}

and\\

\begin{equation}\label{eq: ndili63}
P_{2}  = \frac{3c}{2}  - \frac{1}{2} \sqrt D   = \frac{1}{2c} \bar c_{2} (-)                   
\end{equation}

with $\bar c_{2}(+)$  and  $\bar c_{2}(-)$,  being the two possible values of the optimized $c_{2}$ of eqn.~(\ref{eq: ndili55}).

Next let us denote the left hand sides of eqns.~(\ref{eq: ndili60})~and~(\ref{eq: ndili61}) by:

\[
X_{1} = \rho_{1}(Q, N_{f}) - P_{1} 
\] 

\[
X_{2} = \rho_{1}(Q, N_{f}) - P_{2}
\]

Similarly, we denote the right hand sides of the same  eqns.~(\ref{eq: ndili60})~and~(\ref{eq: ndili61}) by:

\[
X_{3} = \frac{1}{\bar a}  + c \ln \left | \frac{k \bar a}{1 + c \bar a  -  2 \bar a P_{1}} \right  |    
\]

\[
X_{4} = \frac{1}{\bar a}  + c \ln \left | \frac{k \bar a}{1 + c \bar a  -  2 \bar a P_{2}} \right  |     
\]

\begin{figure}
\scalebox{1.0}{\includegraphics{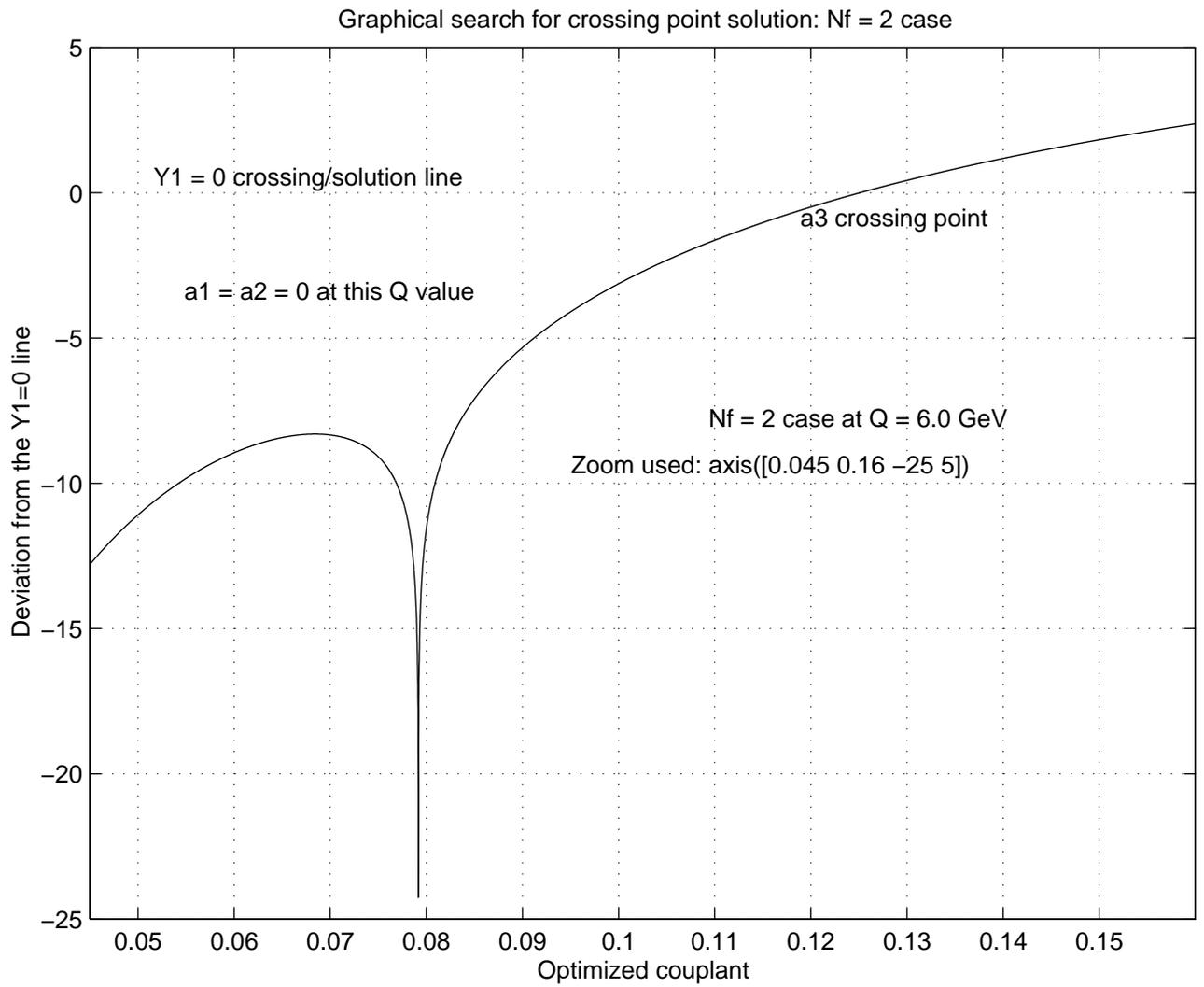}}
\caption{Sample crossing point profile on the Y1 = 0 line, Nf = 2 case}
\label{fig: ndili2}
\centering
\end{figure}

\begin{figure}
\scalebox{1.0}{\includegraphics{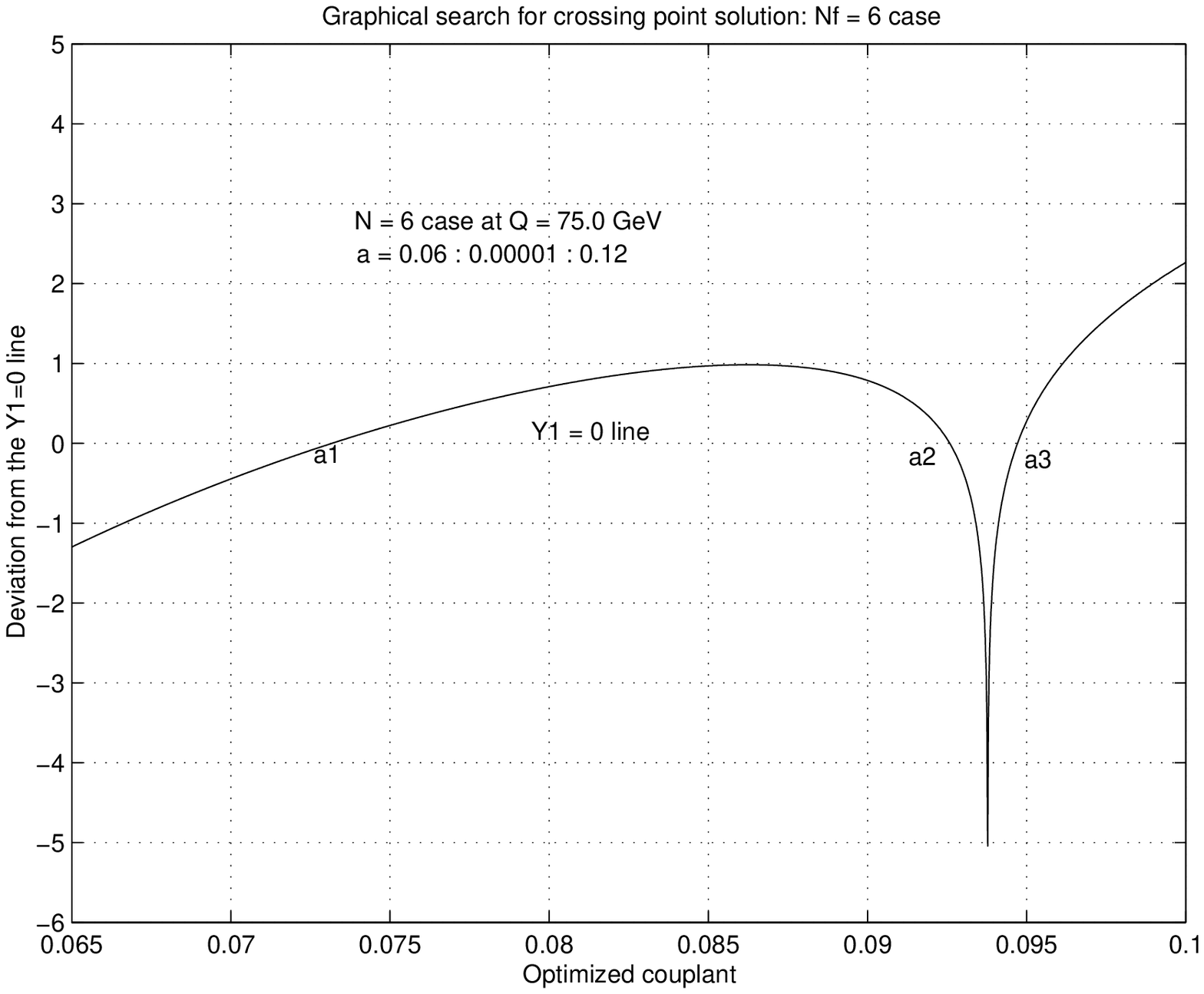}}
\caption{Sample crossing point profile on the Y1 = 0 line, Nf = 6 case}
\label{fig: ndili3}
\centering
\end{figure}

\begin{figure}
\scalebox{1.0}{\includegraphics{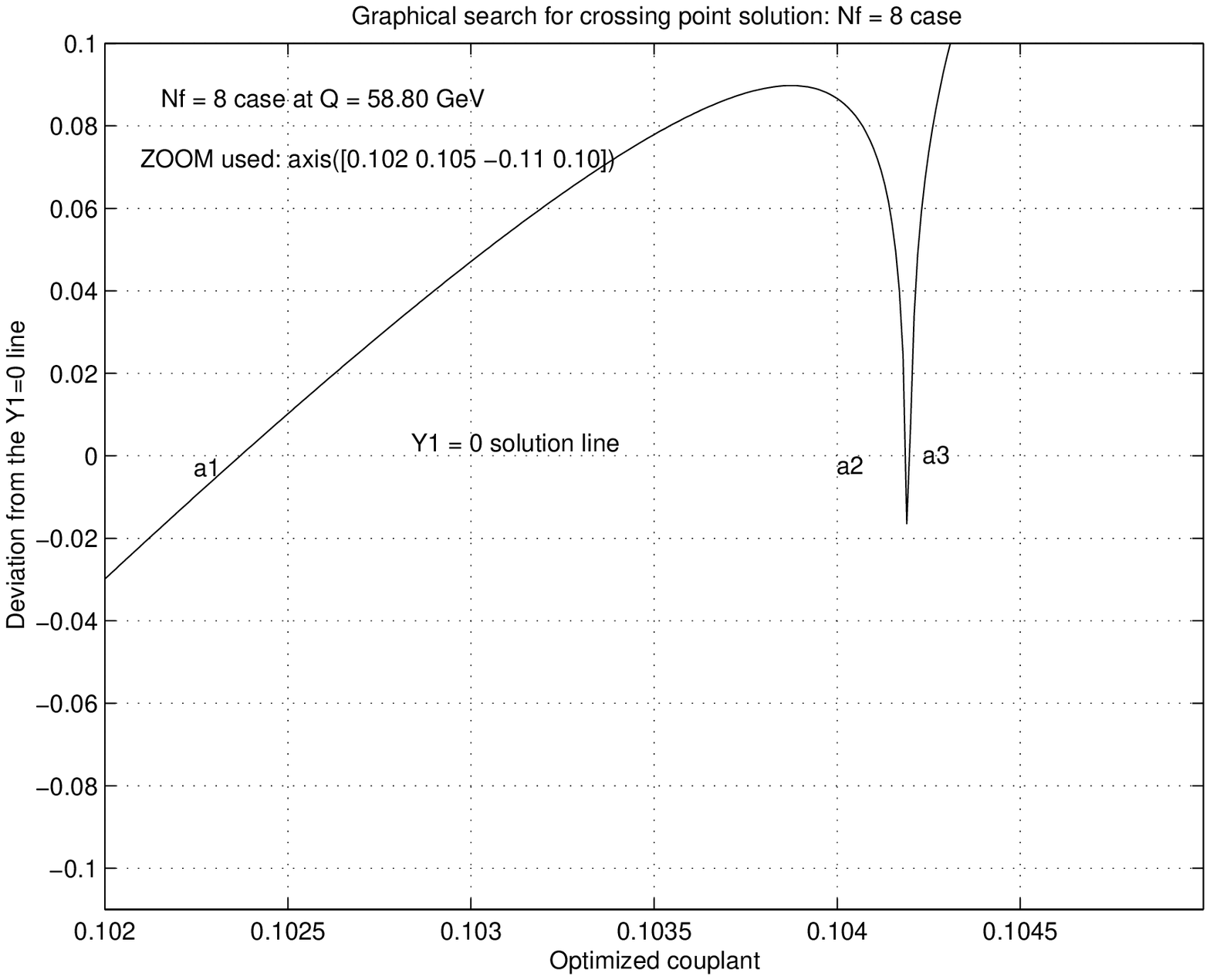}}
\caption{Sample triple point crossing profile on the Y1 = 0 line, Nf = 8 case}
\label{fig: ndili4}
\centering
\end{figure}

\begin{figure}
\scalebox{1.0}{\includegraphics{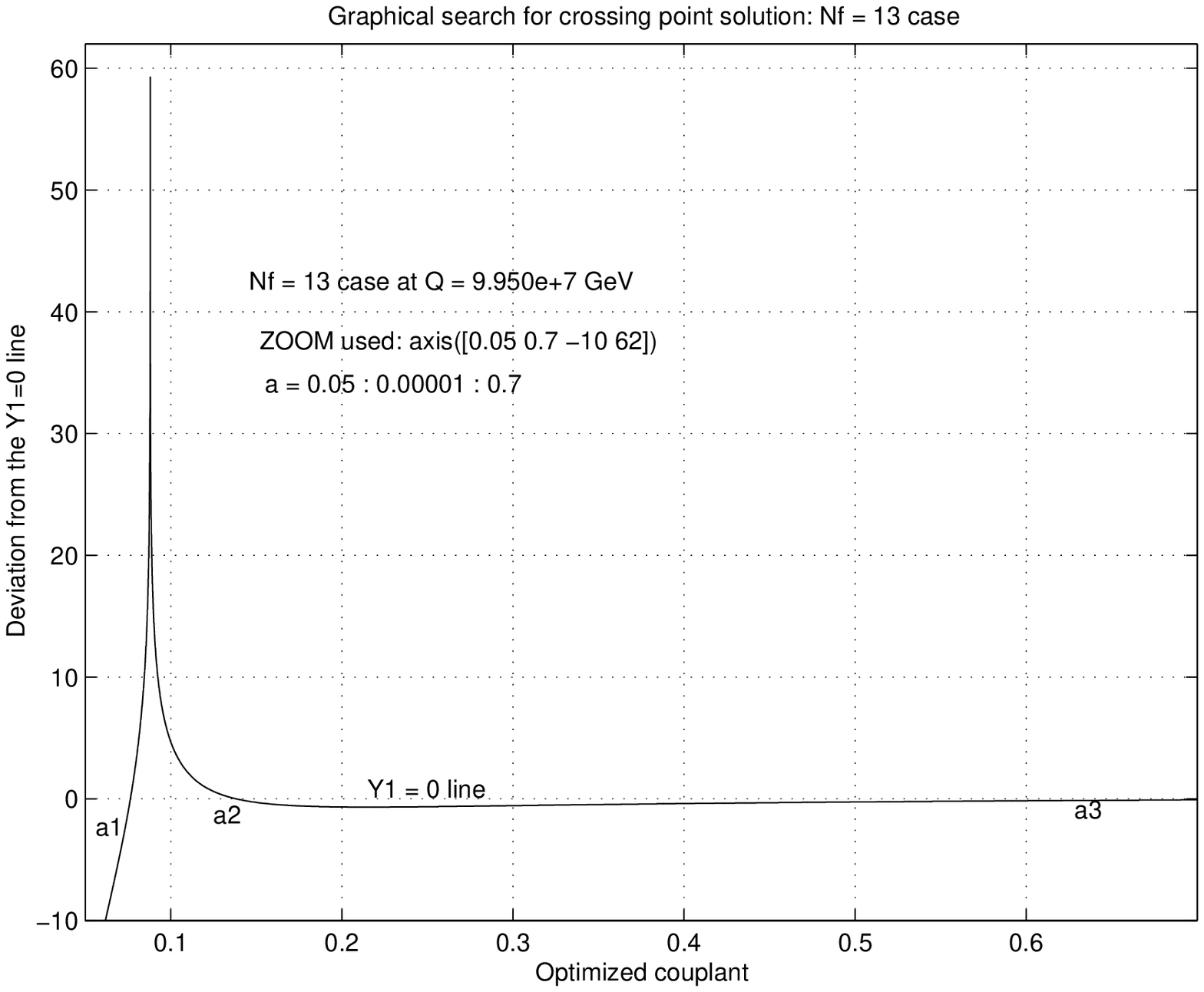}}
\caption{Sample crossing point profile on the Y1 = 0 line, Nf = 13 case}
\label{fig: ndili6}
\centering
\end{figure}

\begin{figure}
\scalebox{1.0}{\includegraphics{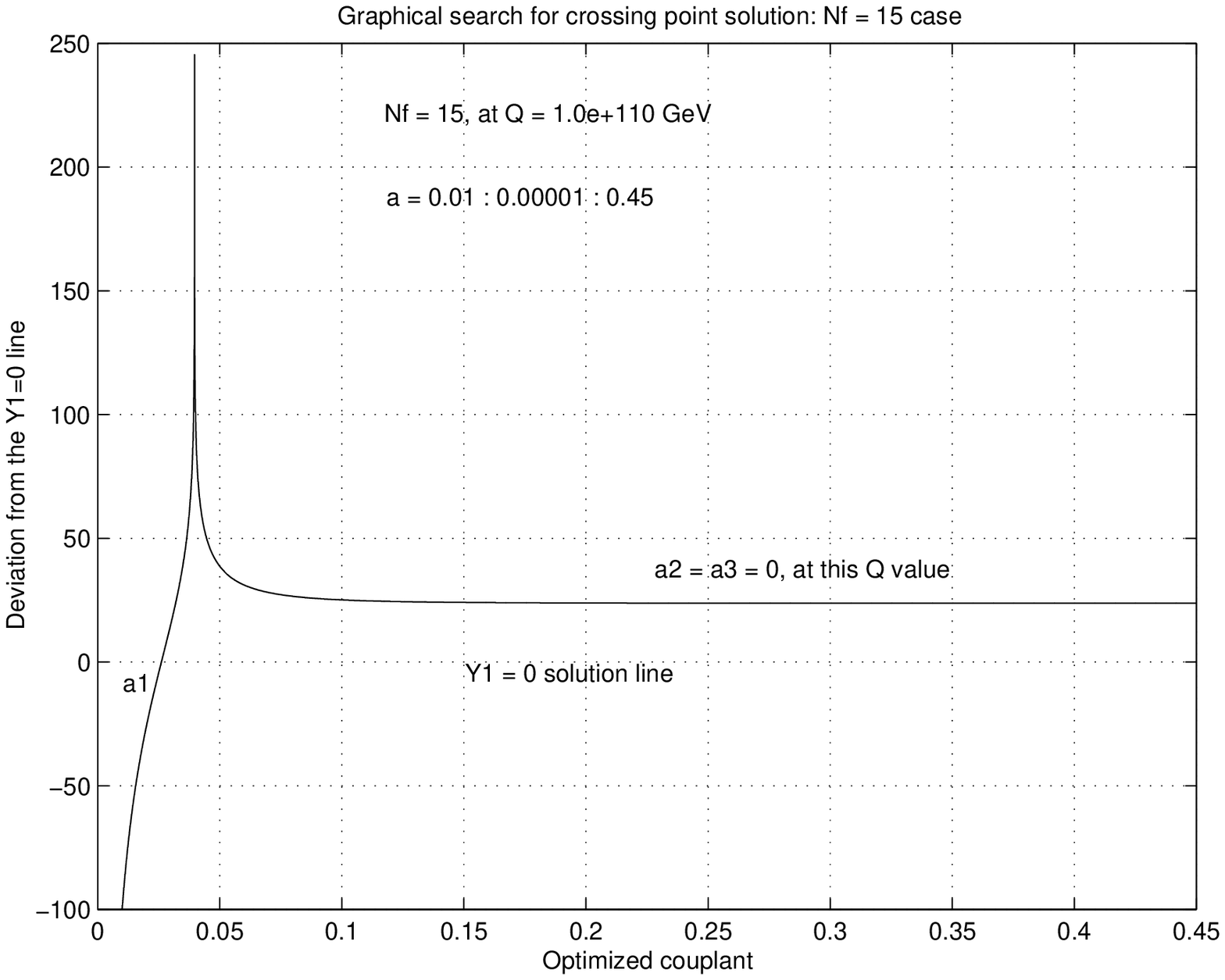}}
\caption{Sample crossing point profile on the Y1 = 0 line, Nf = 15 case}
\label{fig: ndili7}
\centering
\end{figure}

\begin{figure}
\scalebox{1.0}{\includegraphics{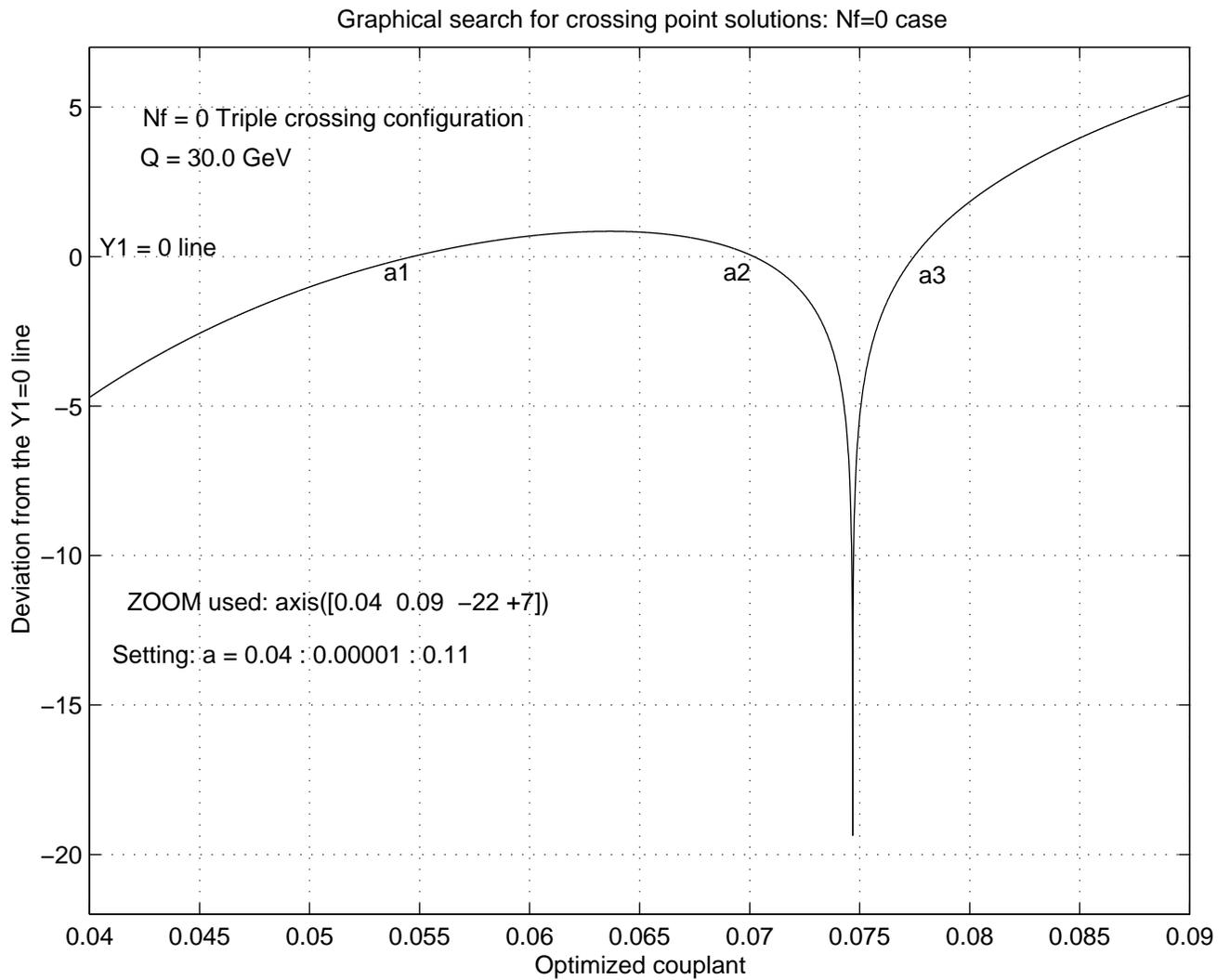}}
\caption{Triple point crossing configuration on the Y1 = 0 line, Nf = 0 case}
\label{fig: ndili12}
\centering
\end{figure}

\begin{figure}
\scalebox{1.0}{\includegraphics{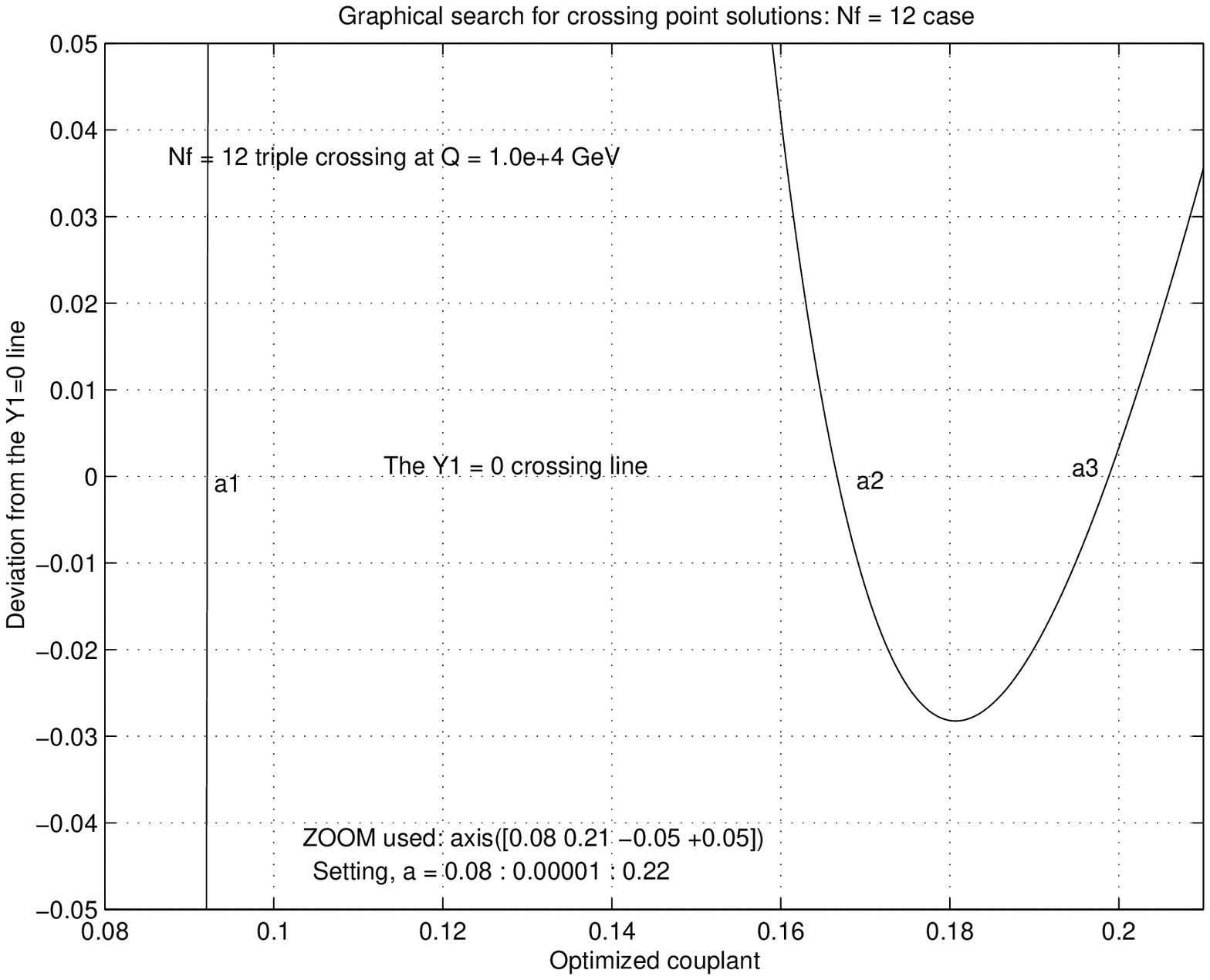}}
\caption{Sample (triple) point crossing  profile on the Y1 = 0 line, Nf = 12 case}
\label{fig: ndili15}
\centering
\end{figure}

\begin{figure}
\scalebox{1.0}{\includegraphics{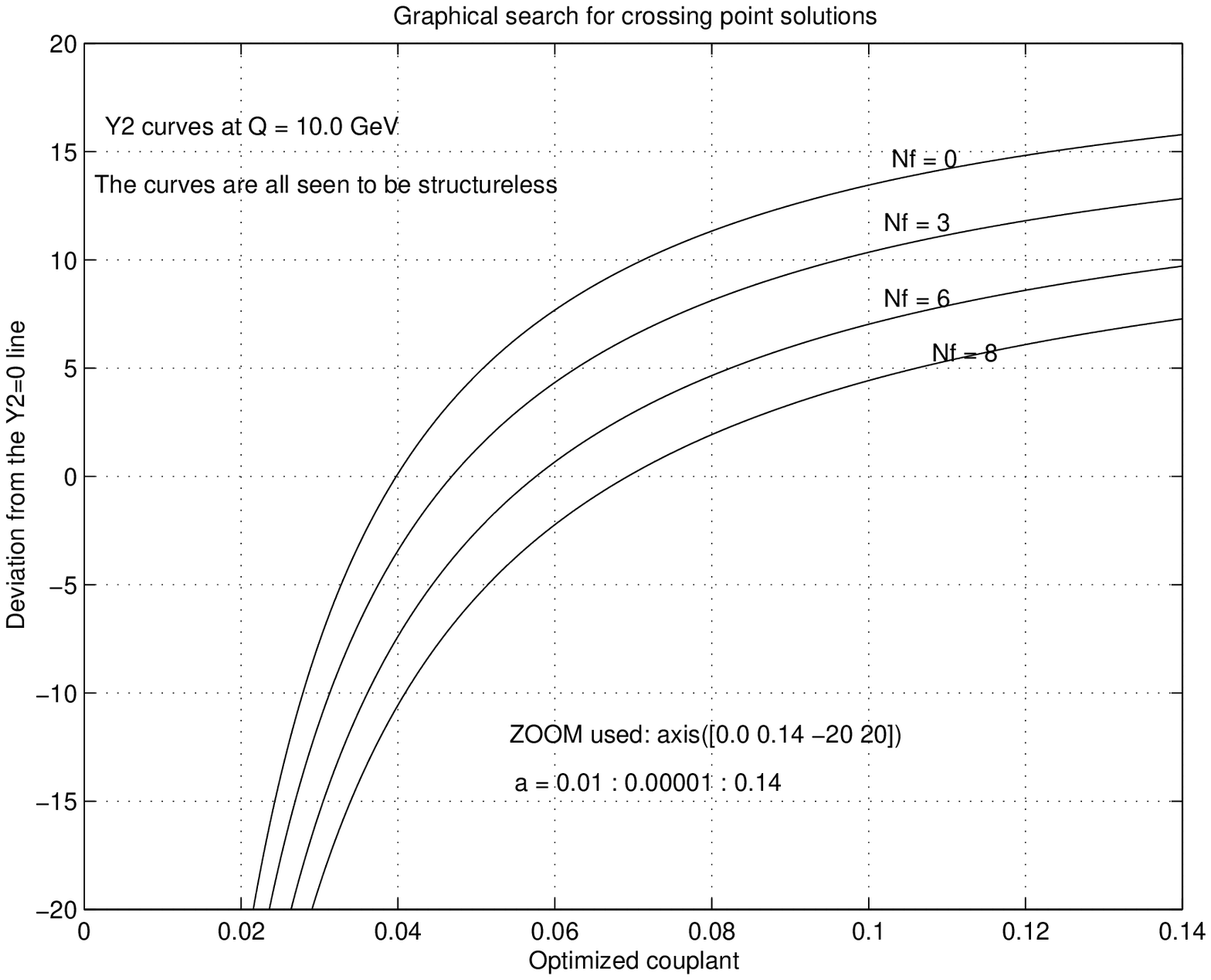}}
\caption{Crossing point profiles on the Y2 = 0 line}
\label{fig: ndili10}
\centering
\end{figure}

\begin{figure}
\scalebox{1.0}{\includegraphics{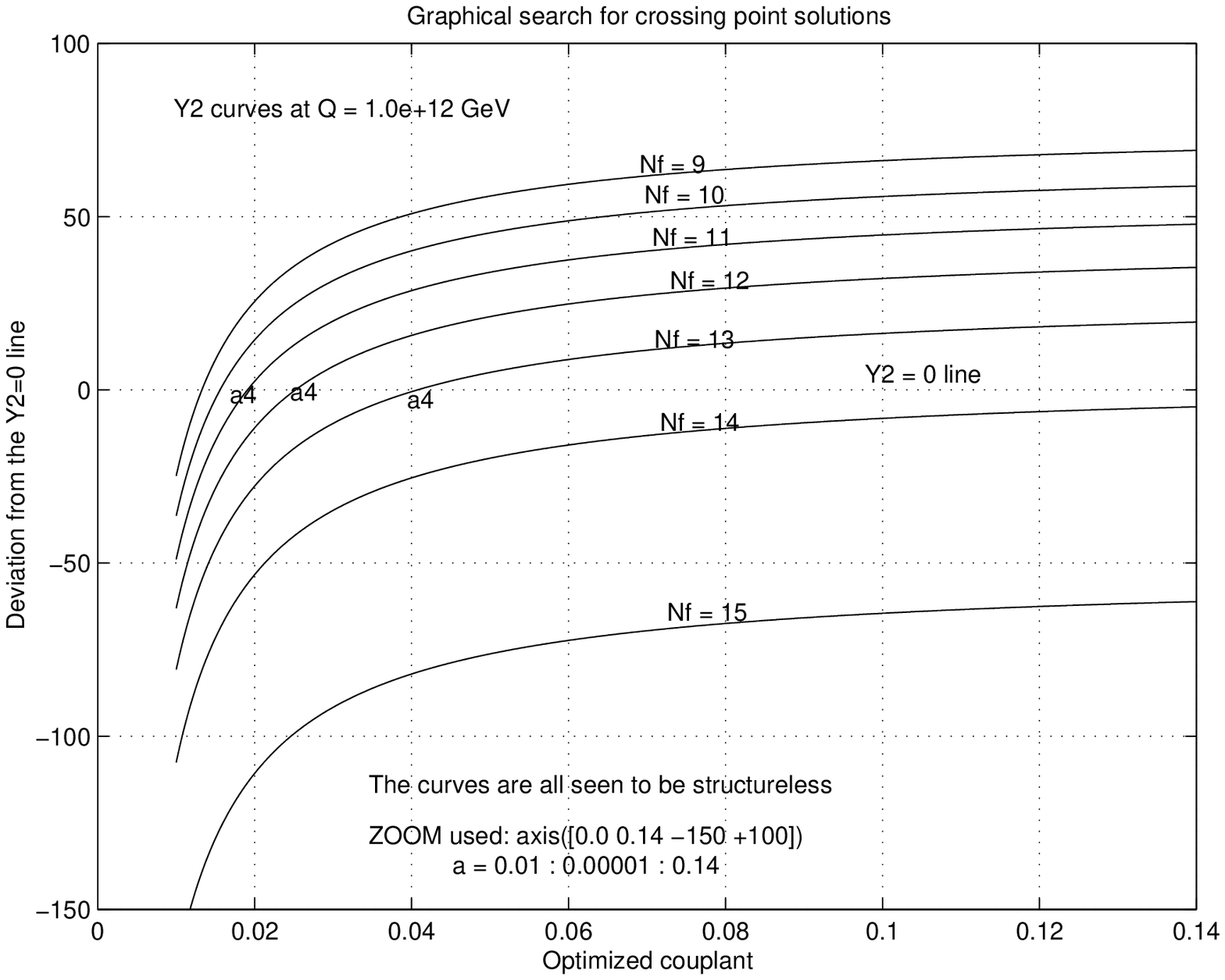}}
\caption{Crossing point profiles on the Y2 = 0 line}
\label{fig: ndili11}
\centering
\end{figure}

Then the two equations~(\ref{eq: ndili60})~and~(\ref{eq: ndili61}) we want to solve numerically for $a$, can be written simply as
 
\begin{equation}\label{eq: ndili64}
X_{1} - X_{3} = Y_{1}(a) = 0                                                           
\end{equation}

\begin{equation}\label{eq: ndili65}
X_{2} - X_{4} = Y_{2}(a) = 0                                                           
\end{equation}

so that what we have to look for as the exact  solutions of eqn.~(\ref{eq: ndili59}) are the zeros of the two functions,
$Y_{1}(a)$ and $Y_{2}(a)$.  We do this by  assigning a wide range of values, in small  incremental  steps $(\Delta a = 0.00001)$, 
to the  floating  couplant $a$, all at one fixed value of Q (and fixed $N_{f}$ ).  At any  assigned
value of $a$, we  compute  $Y_{1}(a)$  and $Y_{2}(a)$.  Then over the  chosen  wide range of $a$, (and at one fixed Q
value and fixed $N_{f}$),  we plot  the two  separate  graphs  of  $(Y_{1}(a), a)$, and $(Y_{2}(a), a)$.  The  exact  numerical
solutions  of eqns~(\ref{eq: ndili60})~and~(\ref{eq: ndili61}) we are looking for at a fixed Q, can now be read off these 
$(Y_{1}(a), a)$, and $(Y_{2}(a), a)$  plots, as the points on the couplant axis, where each curve crosses the couplant axis.
Typical  such plots and  crossing  points are shown in figs.~\ref{fig: ndili2} to~\ref{fig: ndili15}
for the $(Y_{1}(a), a)$ solution, and figs.~\ref{fig: ndili10} and~\ref{fig: ndili11} for the $(Y_{2}(a), a)$ solution.

A simple MATLAB program we used for plotting out these $(Y_{1}(a), a)$, and $(Y_{2}(a), a)$ curves, and reading
off the crossing point values of $a$, on the couplant axis, is reproduced as
Appendix A.  The crossing point readings from the $(Y_{1}(a), a)$ plot, being  solutions of eqn.~(\ref{eq: ndili60}), we have
called the $Y_{1}$ component of Pad\'{e} QCD.  Similarly, we can call the crossing point  readings from the $(Y_{2}(a), a)$ 
plot, being solutions of eqn.~(\ref{eq: ndili61}),  the $Y_{2}$  component of Pad\'{e} QCD.  We find that the two  
solutions  or components of Pad\'{e} QCD are in general different and distinct from each other, implying immediately,
that the $[1|1]$ optimized Pad\'{e} QCD couplant has at least two separate  components  (solutions), we shall from here onwards  refer to as the
$Y_{1}$ and $Y_{2}$ component couplants of Pad\'{e}  QCD.

\section{FEATURES FOUND IN THE $[1|1]$ OPTIMIZED PAD\'{E} COUPLANT EQUATION} 
We now state the features we found with the above graphical and computational analysis of the
optimized $[1|1]$ Pad\'{e} QCD couplant eqn.~(\ref{eq: ndili59}). Later we point out which of these 
features persist as intrinsic features of the other  Pad\'{e} QCDs we also analysed by the same 
graphical method.

\subsection{The Multiplicity Structure in the $Y_{1}$  Pad\'{e} Component Solution}
One striking feature we found in the course of studying the above $(Y_{1}(a) , a)$, and
$(Y_{2}(a), a)$ Pad\'{e} QCD plots, is that  while the $(Y_{2}(a), a)$  plot is simple  and  structureless  for all  flavor
numbers  $N_{f}$, eqn.~(\ref{eq: ndili60})  and its $(Y_{1}(a) , a)$  plots  have in  general  three  different  but  simultaneous
crossing  points on the couplant axis, for a given Q value,  implying that eqn.~(\ref{eq: ndili60}) has in general three
different  solutions for the same Q value and the same flavor  number, and this feature  holds for each
flavor number :  $0 \le N_{f} \le 16$.  Typical such (simultaneous) triple crossing points can be seen in 
figs.~\ref{fig: ndili3} to~\ref{fig: ndili15}. The feature exists for every flavor number $0 \le N_{f} \le 16$.

\begin{figure}
\scalebox{1.0}{\includegraphics{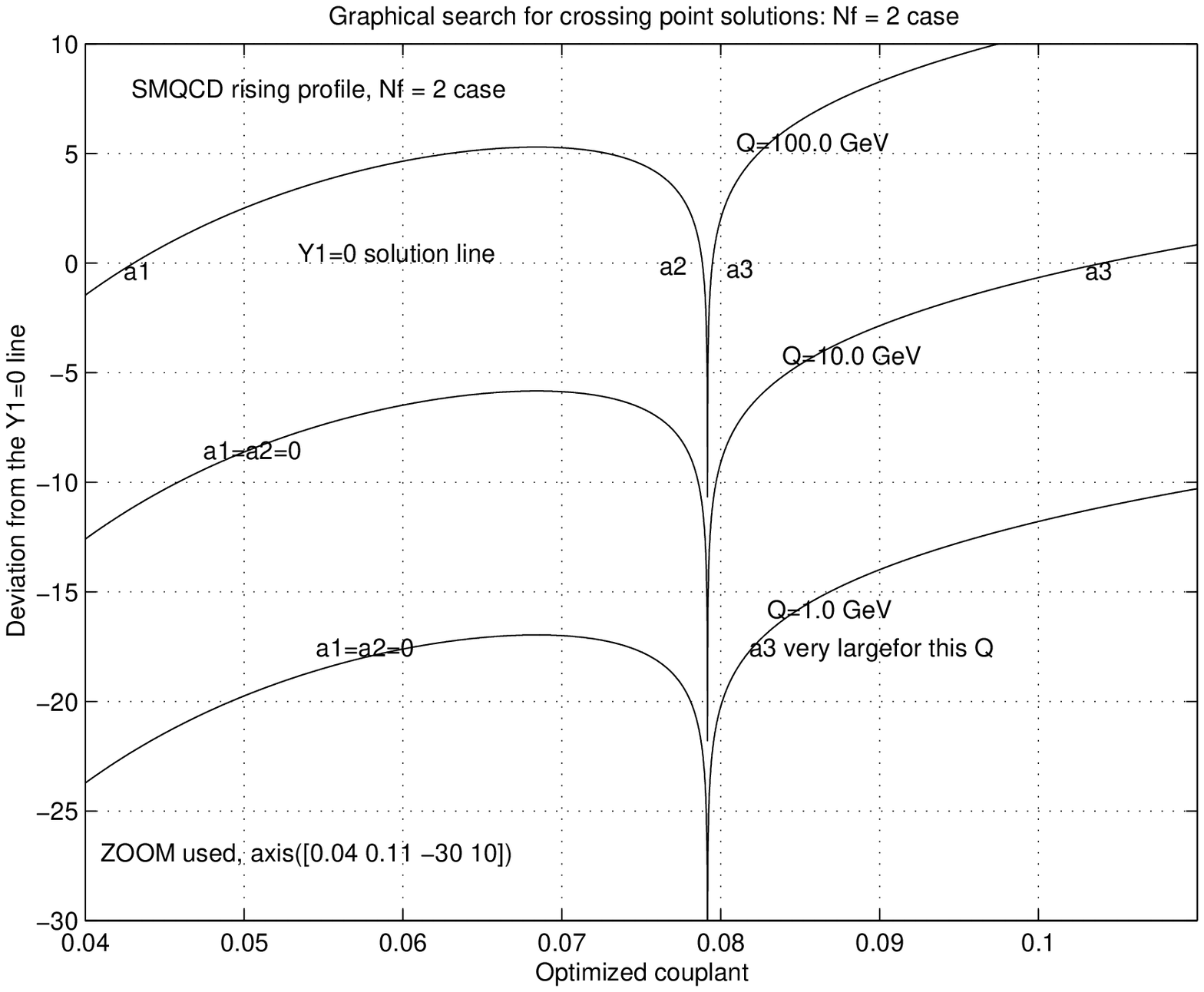}}
\caption{Y1 rising profiles, Nf = 2 case}
\label{fig: ndili18}
\centering
\end{figure}

\begin{figure}
\scalebox{1.0}{\includegraphics{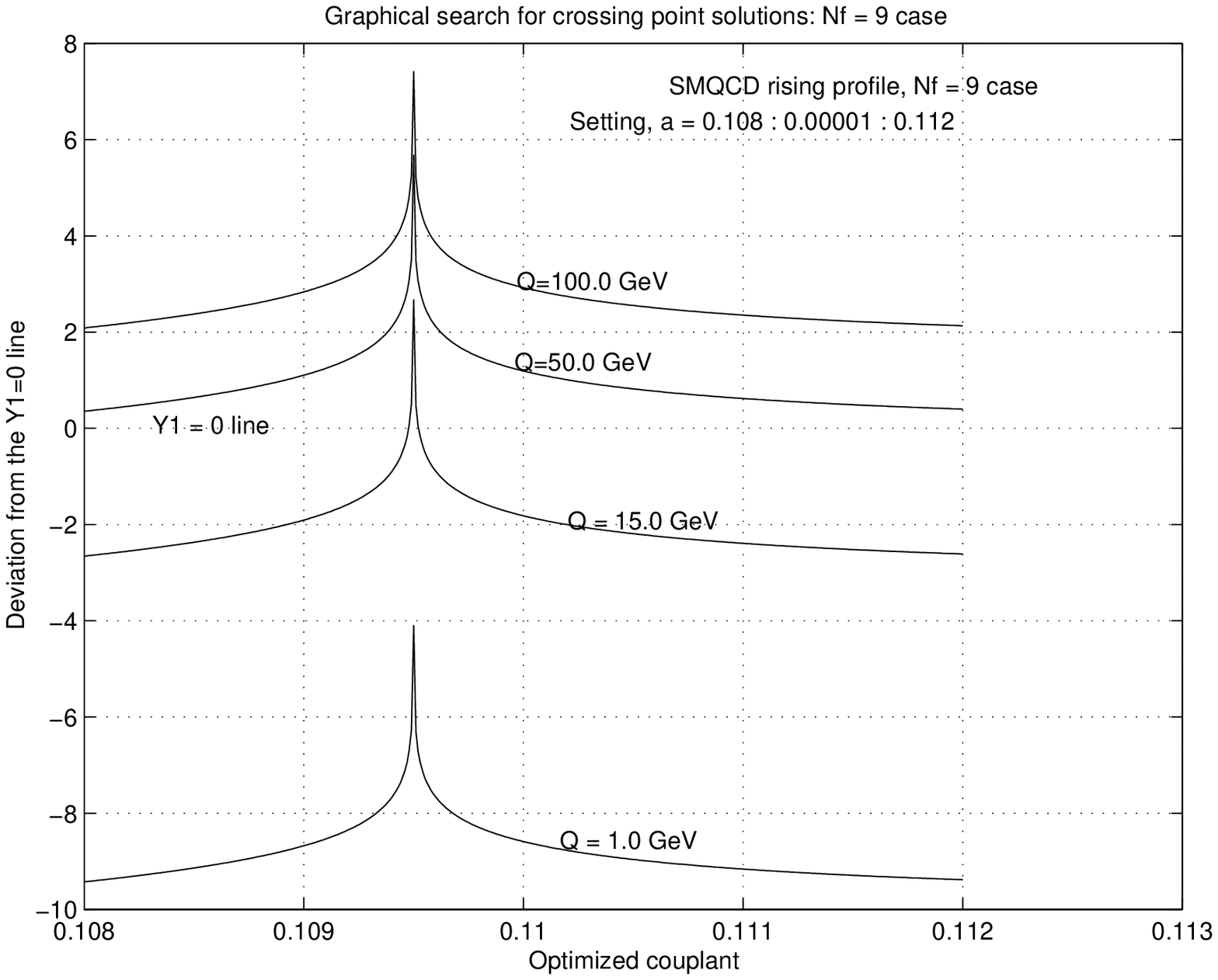}}
\caption{Y1 rising profiles, Nf = 9 case}
\label{fig: ndili20}
\centering
\end{figure}

\begin{figure}
\scalebox{1.0}{\includegraphics{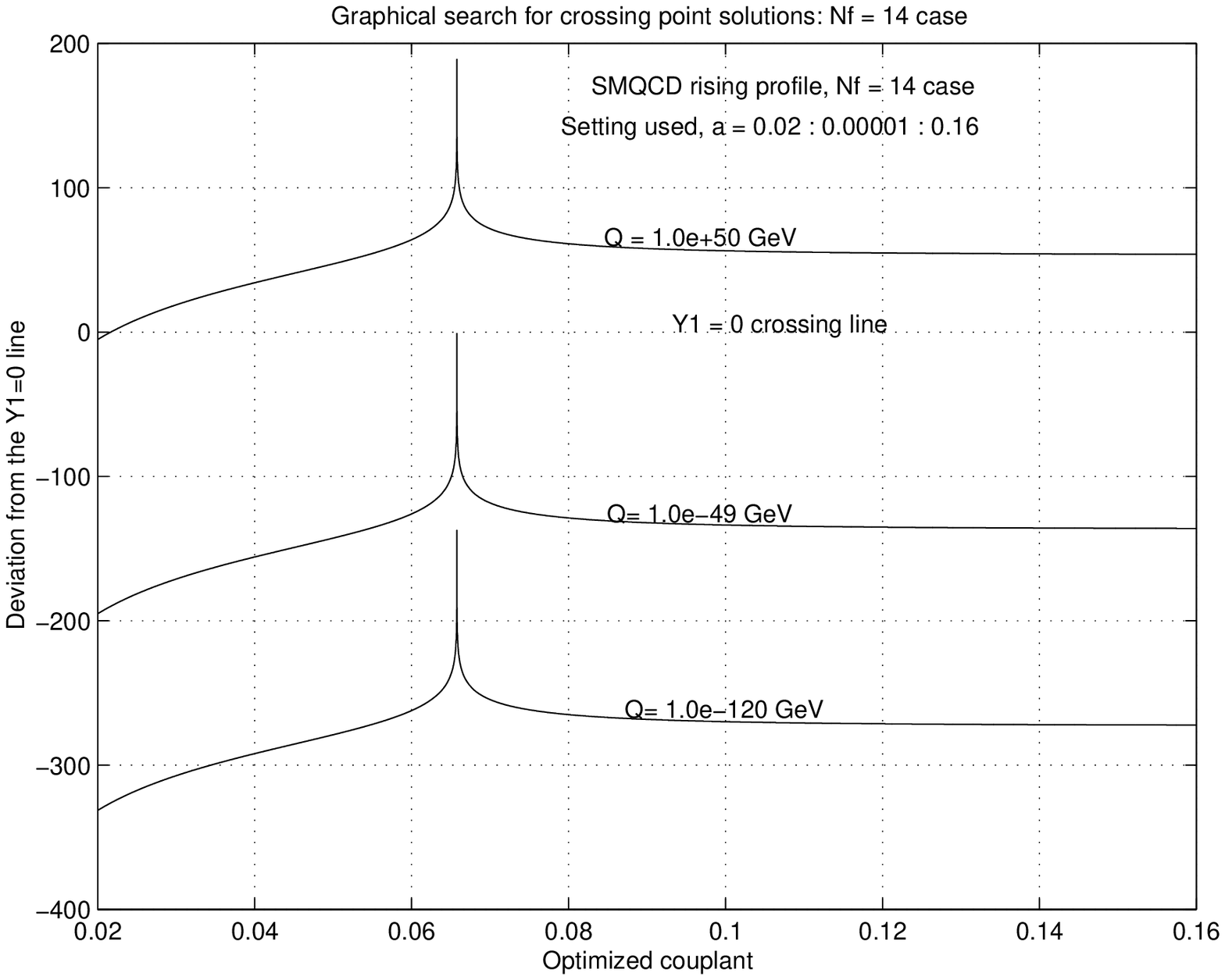}}
\caption{Y1 rising profiles, Nf = 14 case}
\label{fig: ndili21}
\centering
\end{figure}

\begin{figure}
\scalebox{1.0}{\includegraphics{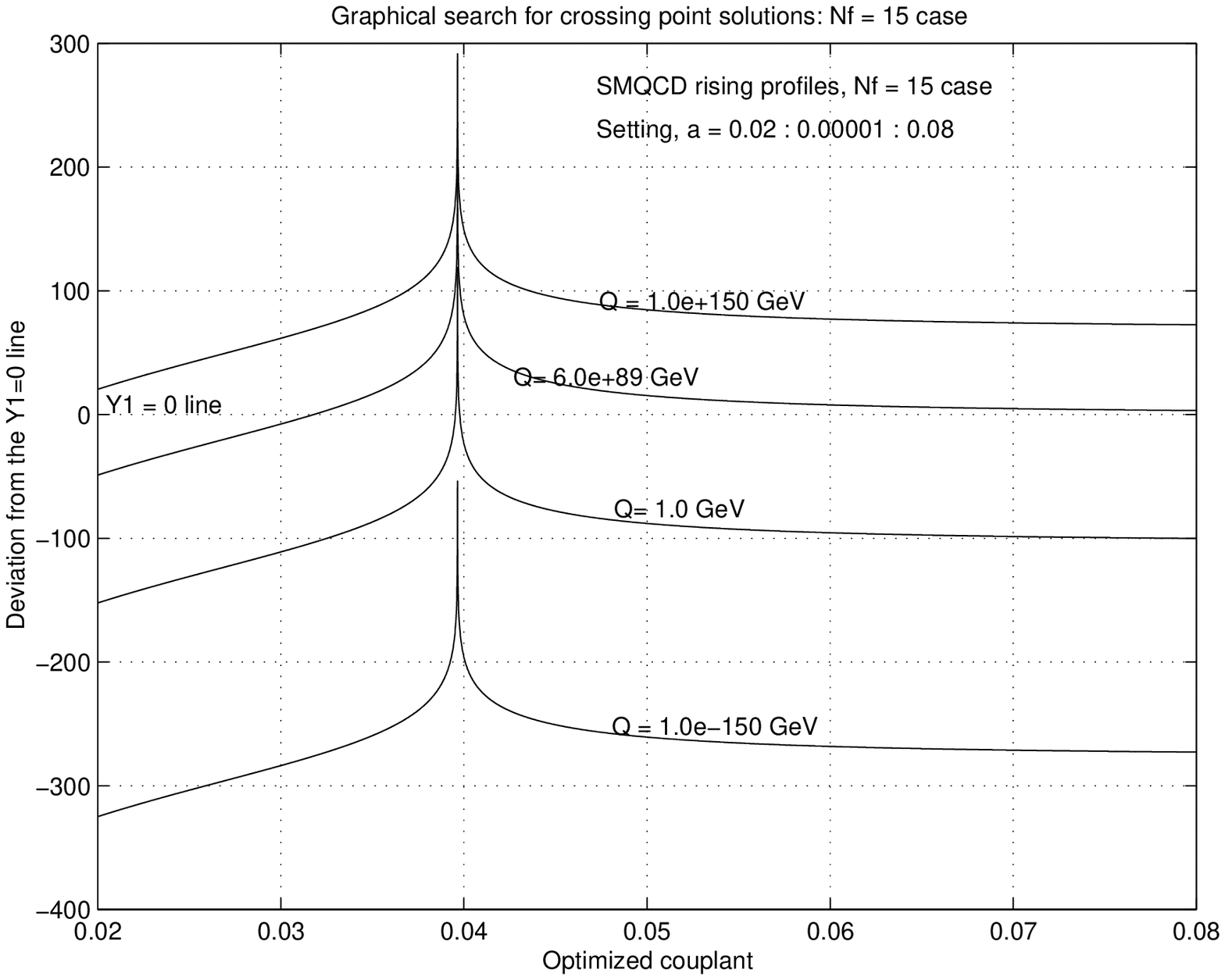}}
\caption{Y1 rising profiles, Nf = 15 case}
\label{fig: ndili22}
\centering
\end{figure}

\begin{figure}
\scalebox{1.0}{\includegraphics{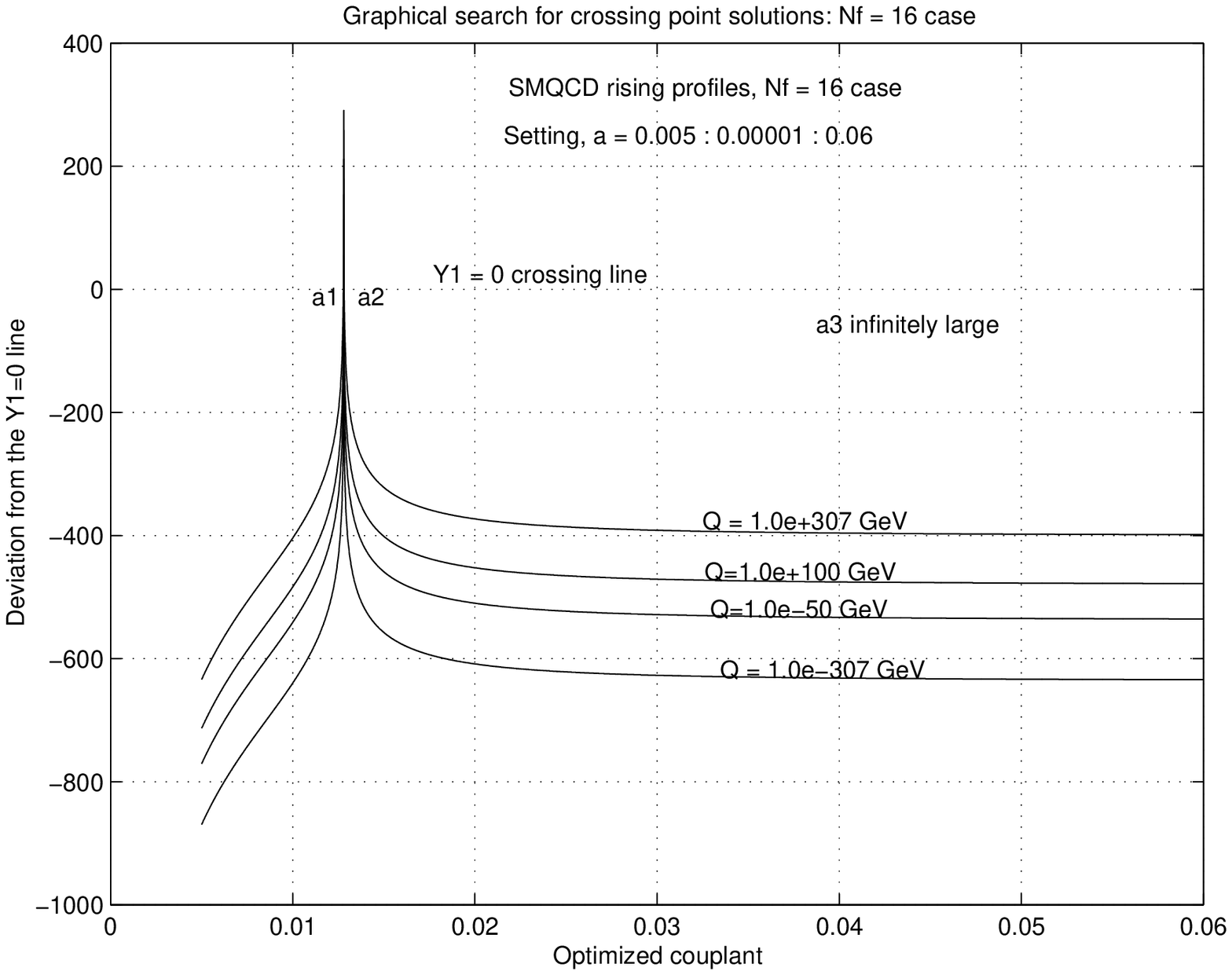}}
\caption{Y1 rising profiles, Nf = 16 case}
\label{fig: ndili23}
\centering
\end{figure}

\begin{figure}
\scalebox{1.0}{\includegraphics{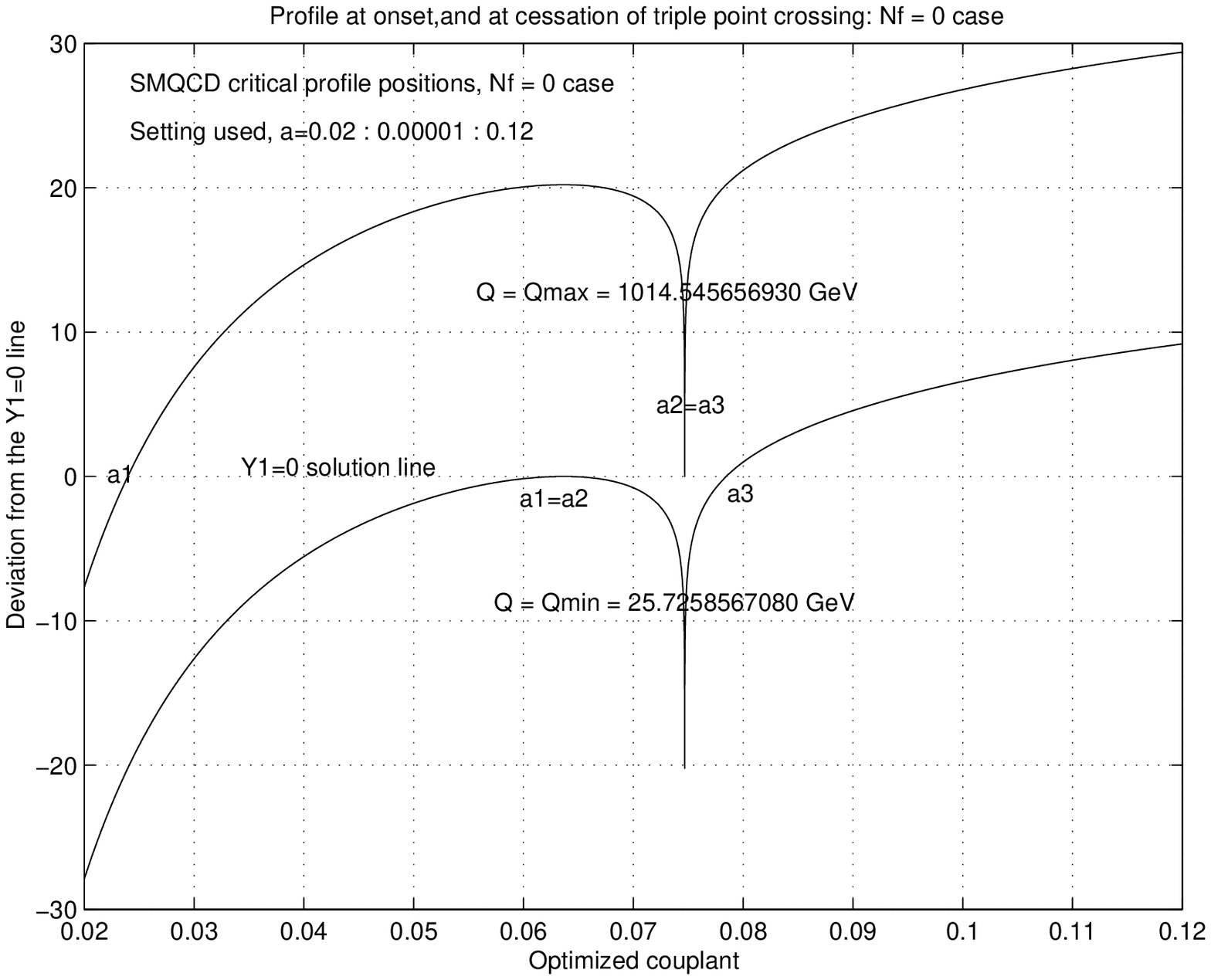}}
\caption{Critical momentum profile configurations on the Y1 = 0 line, Nf = 0 case}
\label{fig: ndili24}
\centering
\end{figure}

\begin{figure}
\scalebox{1.0}{\includegraphics{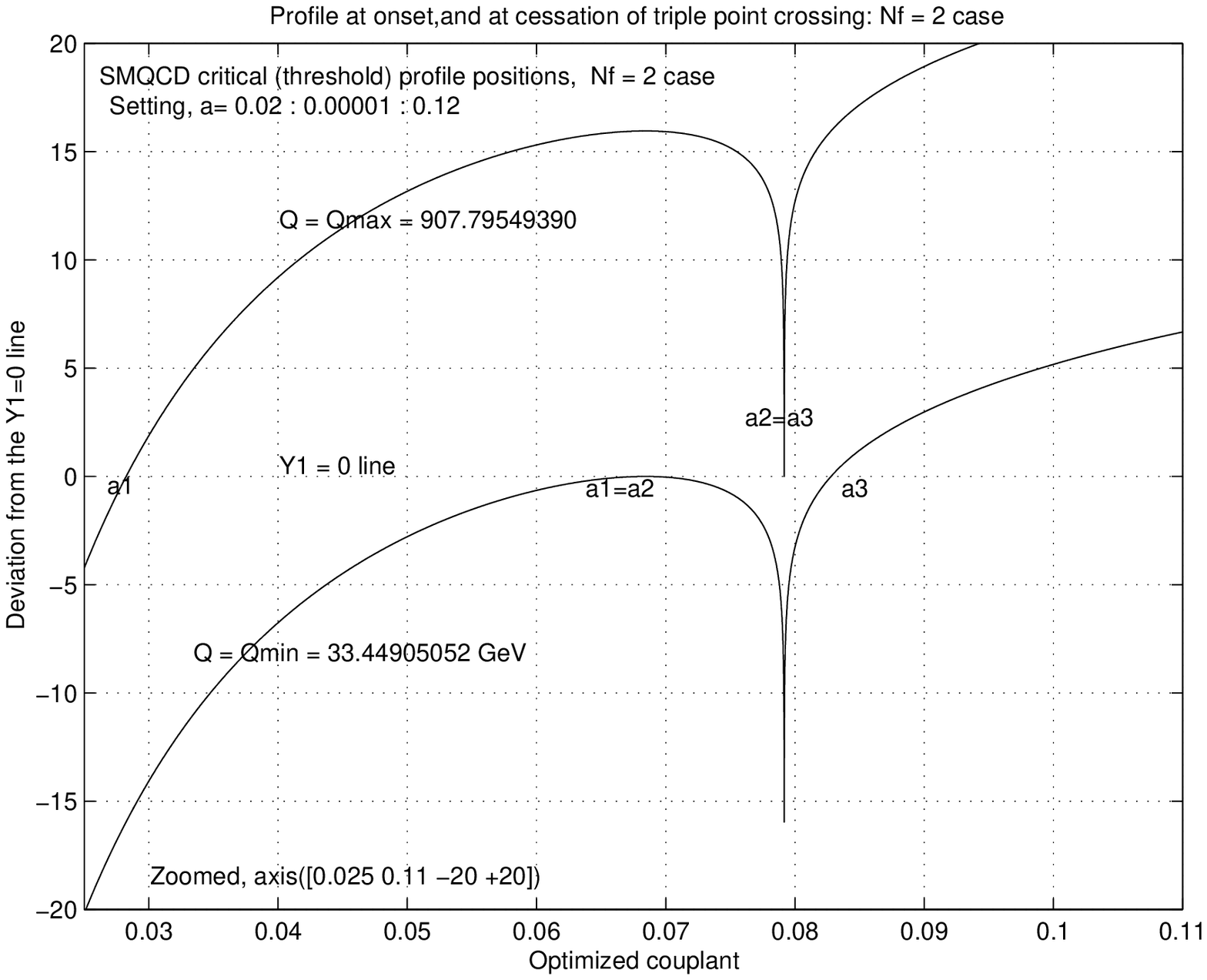}}
\caption{Critical momentum profile configurations on the Y1 = 0 line, Nf = 2 case}
\label{fig: ndili25}
\centering
\end{figure}

\begin{figure}
\scalebox{1.0}{\includegraphics{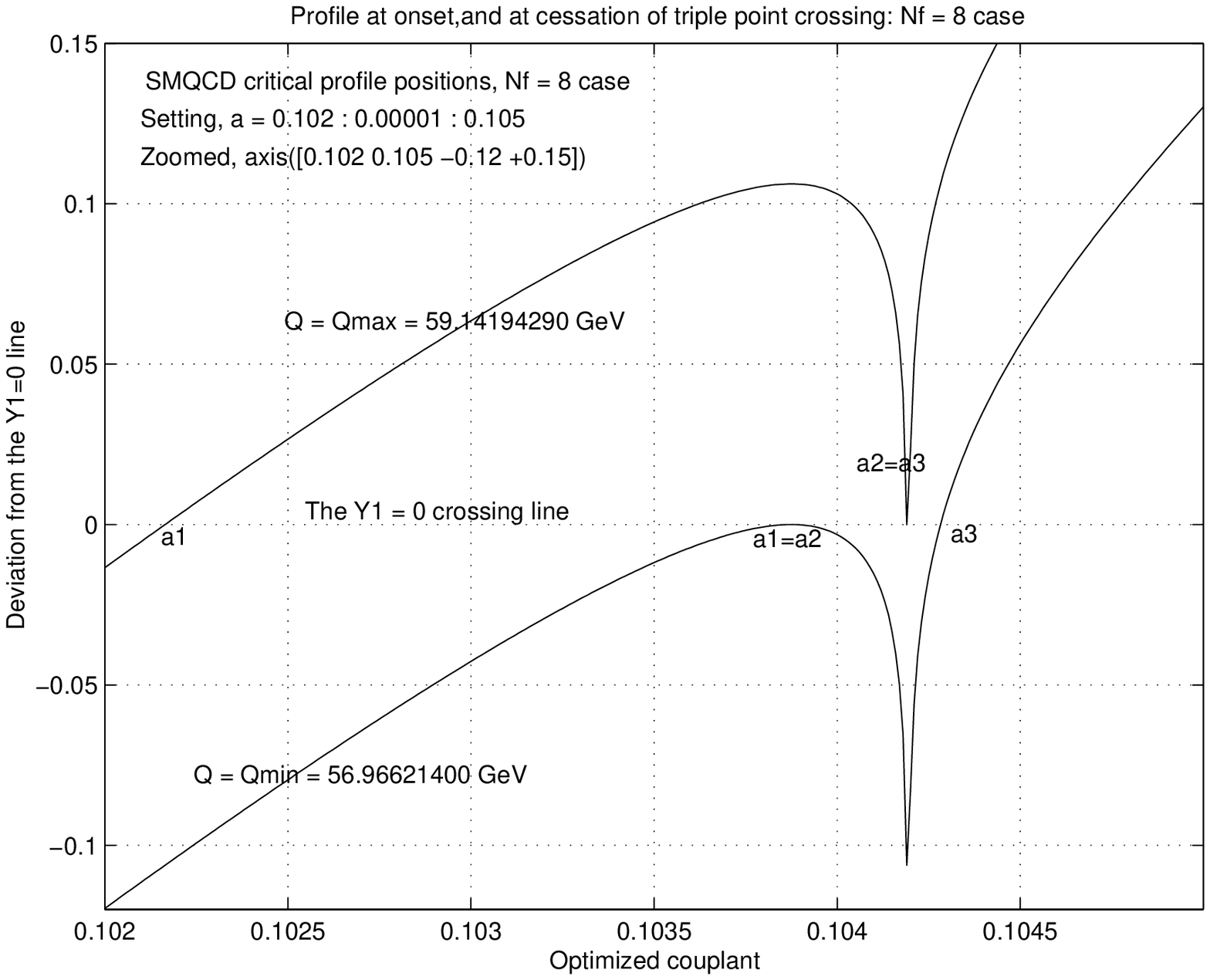}}
\caption{Critical momentum profile configurations on the Y1 = 0 line, Nf = 8 case}
\label{fig: ndili26}
\centering
\end{figure}

\begin{figure}
\scalebox{1.0}{\includegraphics{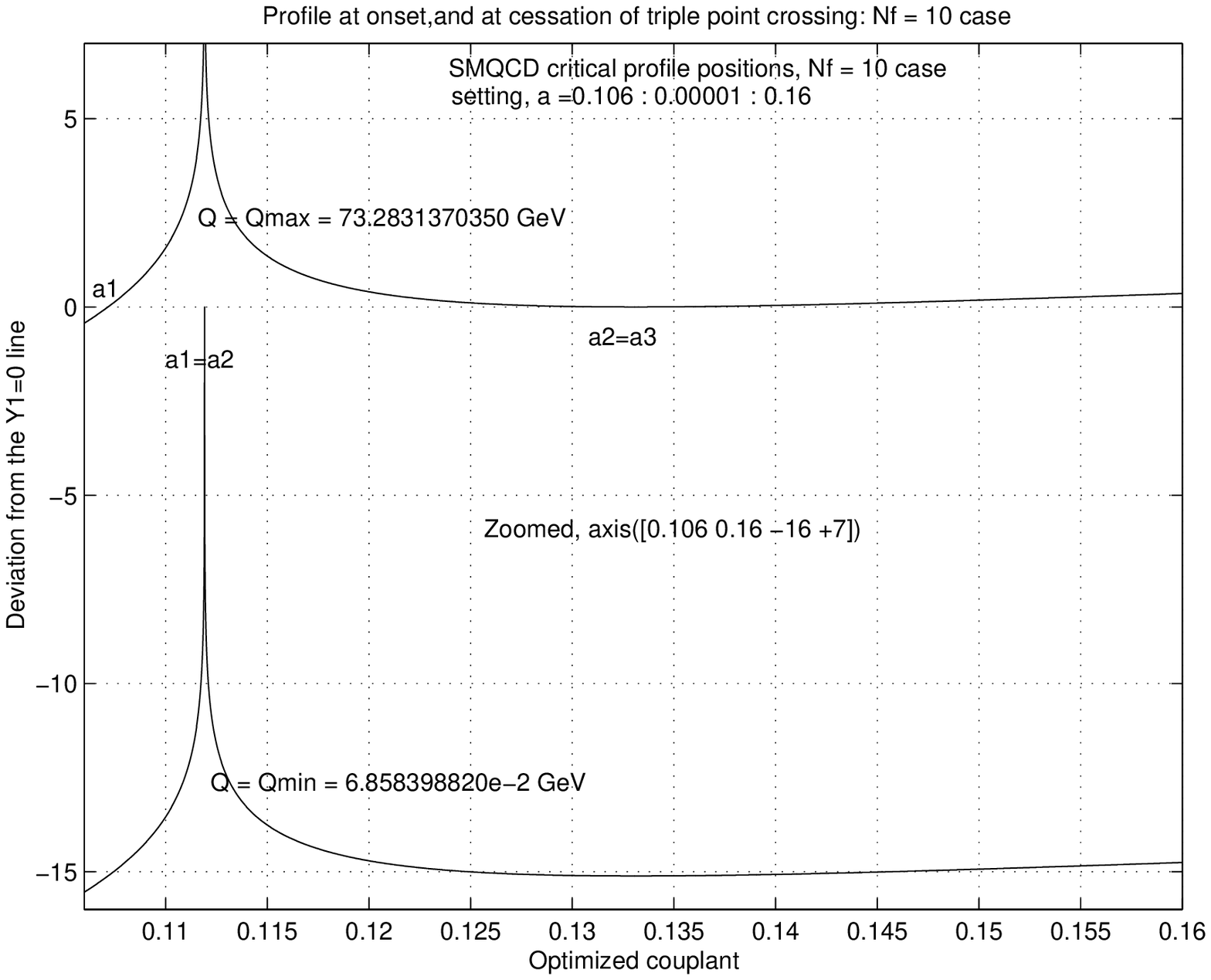}}
\caption{Critical momentum profile configurations on the Y1 = 0 line, Nf = 10 case}
\label{fig: ndili27}
\centering
\end{figure}

The first graphical indication of the existence of this triple  multiplicity  solution of
eqn.~(\ref{eq: ndili60}) is the observation  that as Q increases from below, the entire profile of the 
$(Y_{1}(a) , a)$ plot rises  upwards,
such that beginning with the plot (at low Q) cutting the couplant axis at only one distant point to the far
right hand side of the couplant axis, the curve later rises enough (at higher Q values) to begin to cut
and cross the  couplant  axis  simultaneously  at three  distinct  points.  We found that this begins to
occur when Q has attained a certain  minimum or threshold  value we have denoted by $Q_{\mathrm{min}}$, and
that this minimum Q value exists as a sharply defined  threshold point for each flavor number $N_{f}$ . 
Significantly however,  the actual value of $Q_{\mathrm{min}}$, (see Table~\ref{tab: ndili2}),  differed 
from one flavor  number  system to another.  These rising profiles of the $(Y_{1}(a) , a)$ plots and 
their triple crossings for $Q \geq  Q_{\mathrm{min}}$ are shown in
figs.~\ref{fig: ndili18} to~\ref{fig: ndili23} for some flavor  numbers, the pattern being however 
the same for all flavor numbers.  Figures~\ref{fig: ndili24} to~\ref{fig: ndili27} show sample  profile
positions when one is at exactly the threshold (critical) point $Q = Q_{\mathrm{min}}$ for these
rising profiles.

\begin{figure}
\scalebox{1.0}{\includegraphics{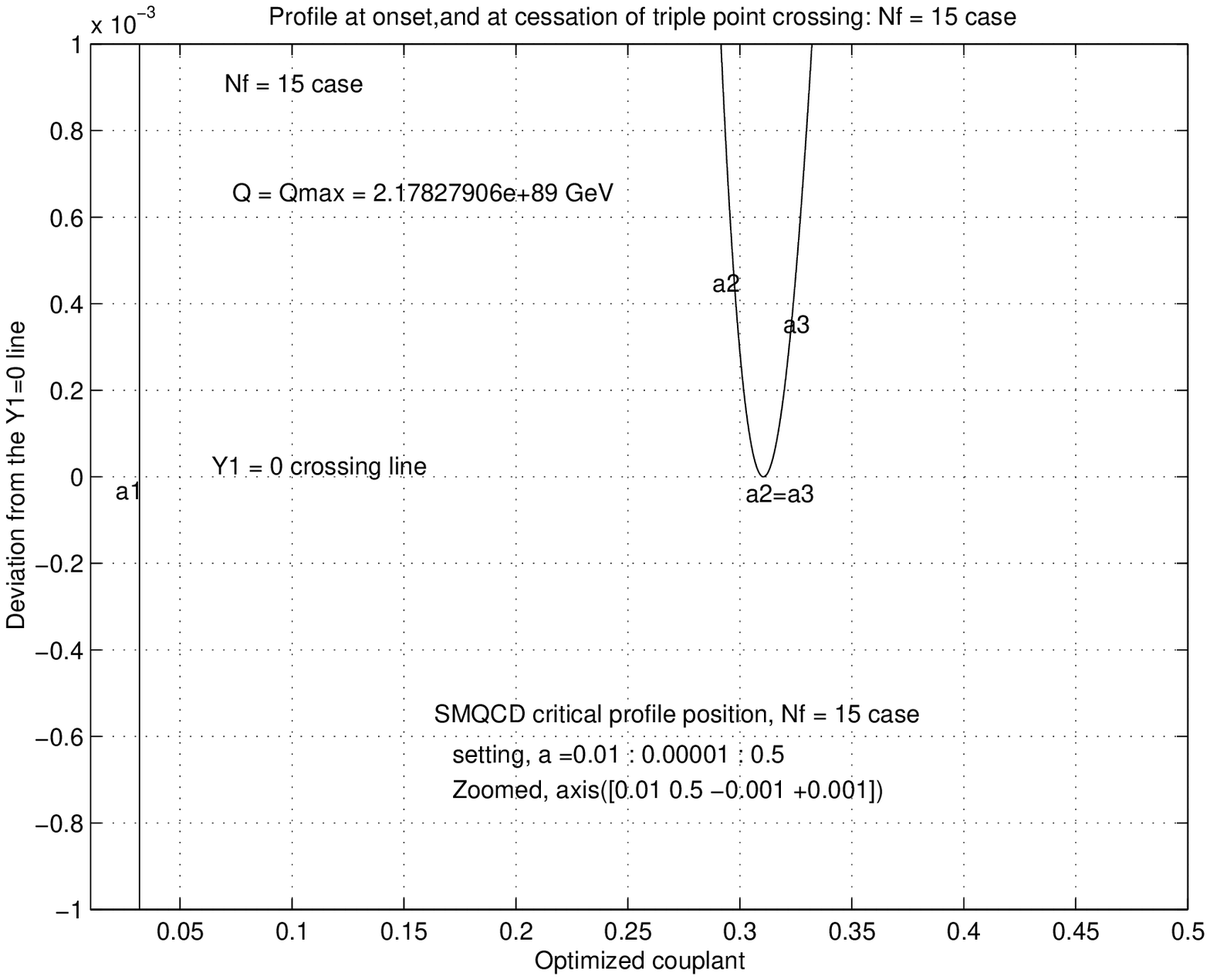}}
\caption{Critical momentum profile configurations on the Y1 = 0 line, Nf = 15 case}
\label{fig: ndili28}
\centering
\end{figure}

We observed also from the same $(Y_{1}(a) , a)$  profile plots that the triple crossing point feature
does not  persist  for all $Q  \geq  Q_{\mathrm{min}}$.  Rather, as Q rises  higher and higher  above  $Q_{\mathrm{min}}$, 
the  triple  crossing suddenly  ceases and we noted that this occurs when Q has attained some upper cut-off value we denoted by
$Q = Q_{\mathrm{max}}$.  This upper cut off value of Q was found to be as precise and sharply  defined for each $N_{f}$ flavor
system as the lower threshold point $Q = Q_{\mathrm{min}}$.  But, again as in the $Q_{\mathrm{min}}$ case, 
the value of $Q_{\mathrm{max}}$, (see Table~\ref{tab: ndili2}),
differed from one $N_{f}$  flavor  system to  another.  The upper parts of Figures~\ref{fig: ndili24}
to~\ref{fig: ndili28} show  sample  profiles  and  configurations  at the cessation point of triple
crossings when $Q = Q_{\mathrm{max}}$.

\begin{figure}
\scalebox{1.0}{\includegraphics{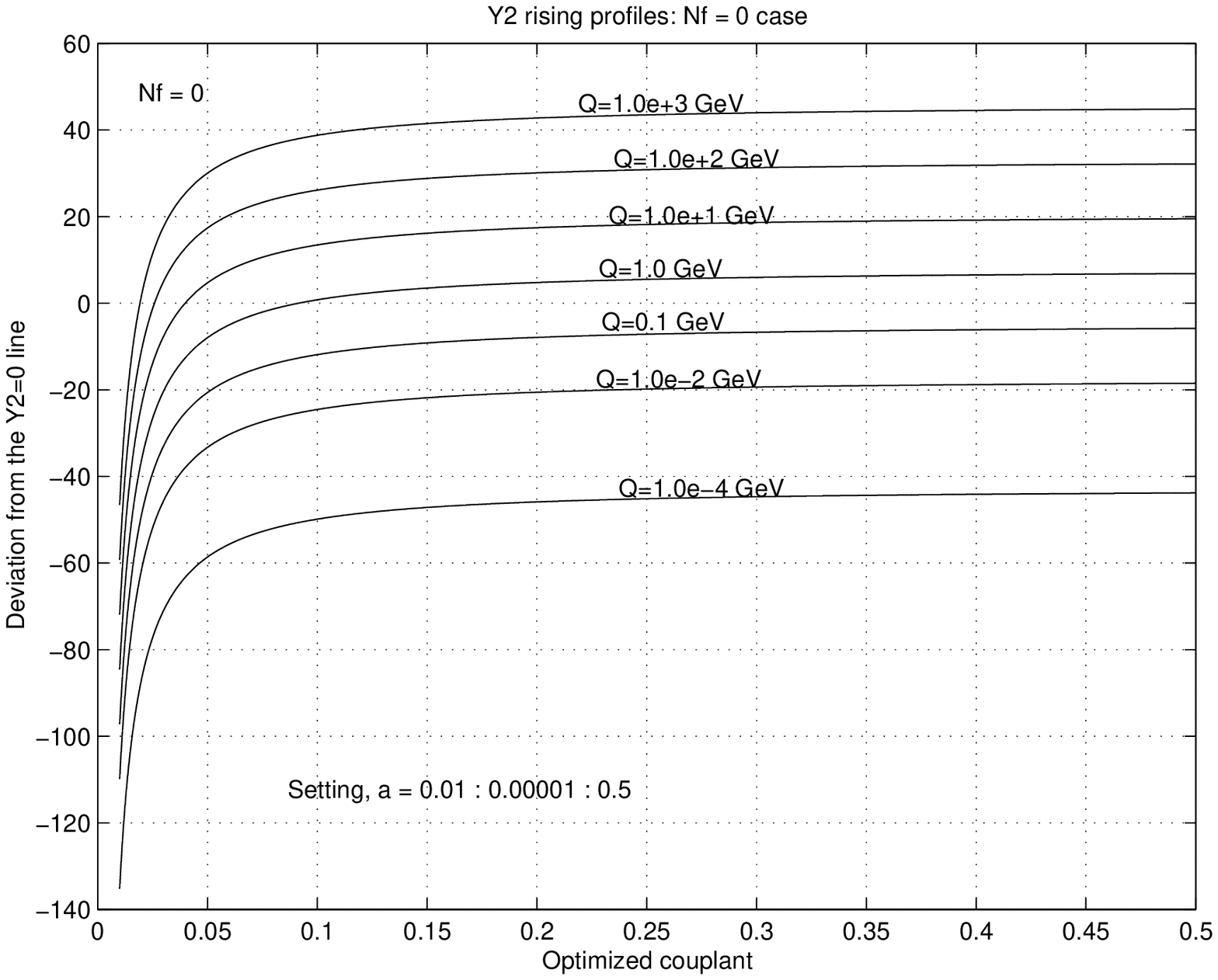}}
\caption{Y2 rising profiles, Nf = 0 case}
\label{fig: ndili29}
\centering
\end{figure}

\begin{figure}
\scalebox{1.0}{\includegraphics{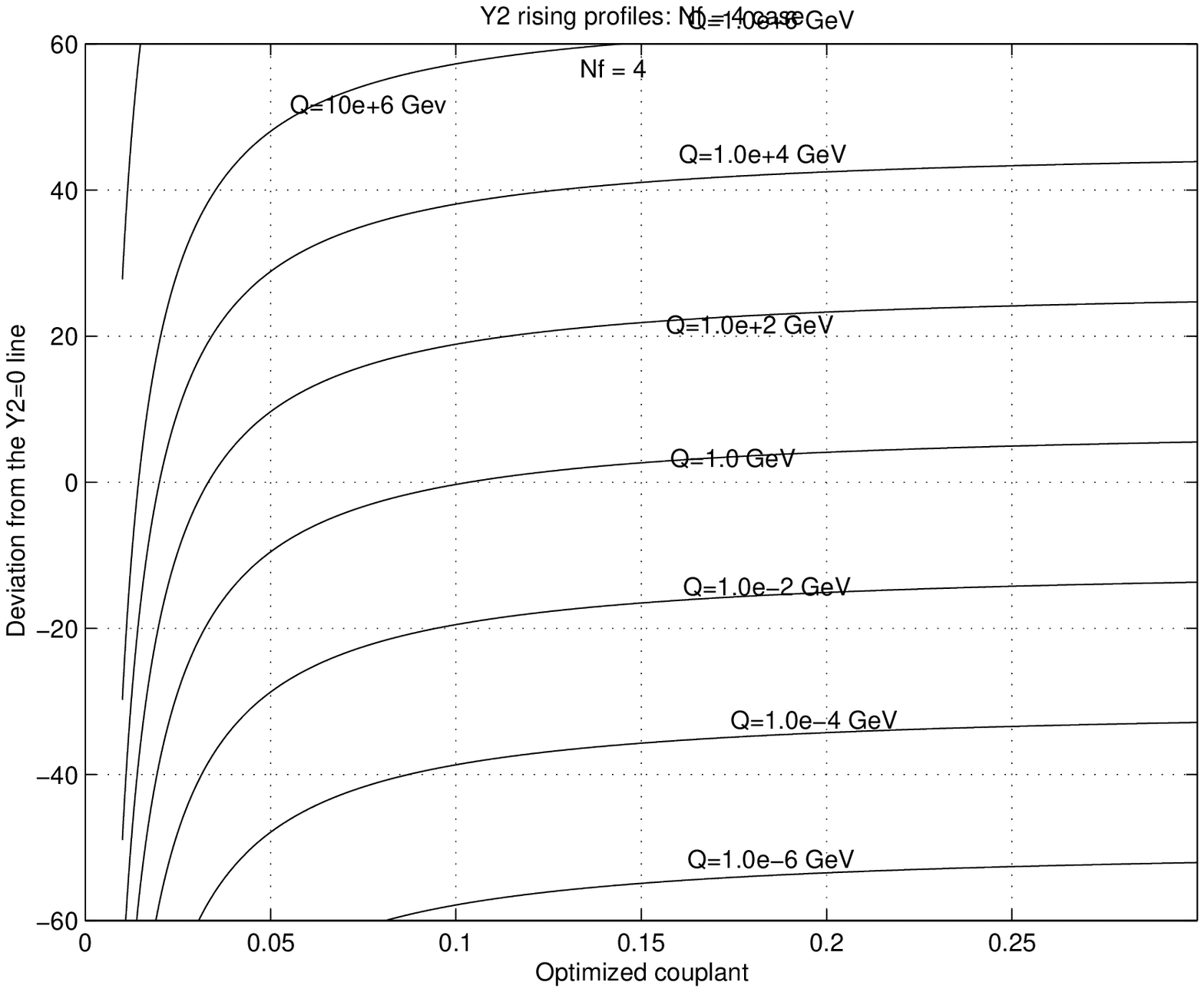}}
\caption{Y2 rising profiles, Nf = 4 case}
\label{fig: ndili30}
\centering
\end{figure}

\begin{figure}
\scalebox{1.0}{\includegraphics{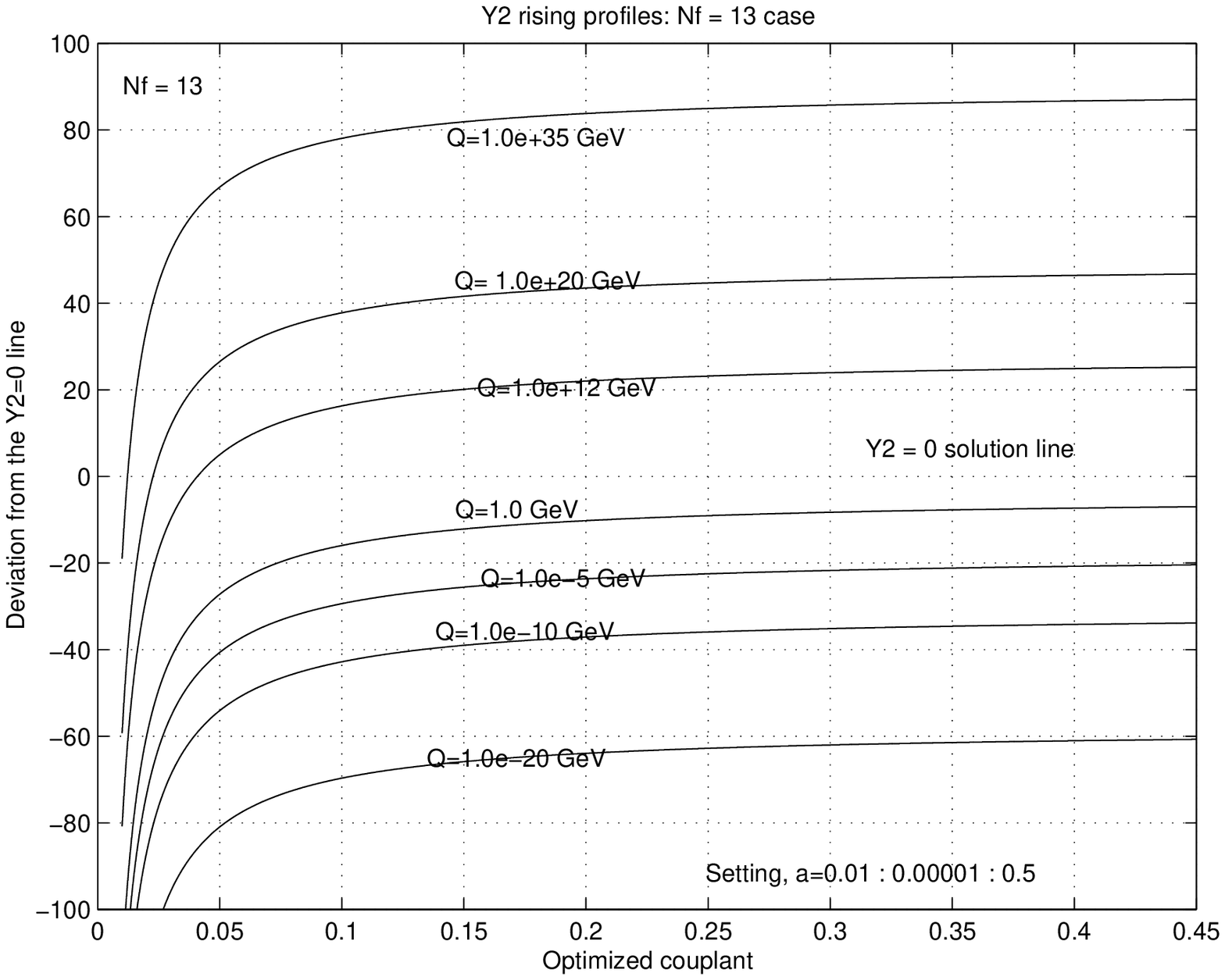}}
\caption{Y2 rising profiles, Nf = 13 case}
\label{fig: ndili31}
\centering
\end{figure}

By way of  comparison,  we show in figs.~\ref{fig: ndili29} to~\ref{fig: ndili31} the  profiles  of the
$(Y_{2}(a), a)$ plots  for some  flavor numbers.  None of these has a triple  crossing  structure, but only one
solution or crossing  point for eqn.~(\ref{eq: ndili61}) at any one value of Q, and for all flavor numbers.

Focusing now on these crossing point  solutions of eqns.~(\ref{eq: ndili60})~and~(\ref{eq: ndili61}), we 
can denote by $a_{1}, a_{2}, a_{3}$, the general triple point crossing solutions  of
eqn.~(\ref{eq: ndili60}), and by $a_{4}$ the one solution  (crossing point) of eqn.~(\ref{eq: ndili61}). 
Here $a_{1}$ is the  extreme  left  crossing  point in the  $(Y_{1}(a) , a)$
plot,  $a_{2}$, is the  middle crossing  point, while $a_{3}$, is the extreme right crossing point,
all in the same $(Y_{1}(a) , a)$ plot of any given (fixed) $N_{f}$ system.  Together the four solutions 
$a_{1}, a_{2}, a_{3}, a_{4}$  imply that  the  Pad\'{e} approximant QCD couplant 
$a$ of eqn.~(\ref{eq: ndili59}) or eqn.~(\ref{eq: ndili1}),  has a multiplicity  structure of four
distinct  component couplants or color force solutions, we can read off as crossing points on our
Y1 = 0 and Y2 = 0 solution lines. This multiplicity of solutions  of the Pad\'{e} couplant
equation~(\ref{eq: ndili59}), exhibited in  the momentum band  
$Q_{\mathrm{min}} \le Q \le Q_{\mathrm{max}}$, represents our first explicit finding 
concerning the features of Pad\'{e} QCD couplant analyzed by our above computational 
and graphical method.

\subsection{Features of the $a_{1}(Y_{1}), a_{2}(Y_{1}), a_{3}(Y_{1})$, and $a_{4}(Y_{2})$ 
Pad\'{e} couplant Solutions.}
We investigated next  the behavior of each Pad\'{e} component couplant solution, by plotting its 
graphical Pad\'{e} crossing point values against the momentum Q of the profile plot at which  the
crossing point value was read off. We found the following features.

\subsubsection{Features of the $a_{3}$  Pad\'{e} component couplant solution}
\begin{figure}
\scalebox{1.0}{\includegraphics{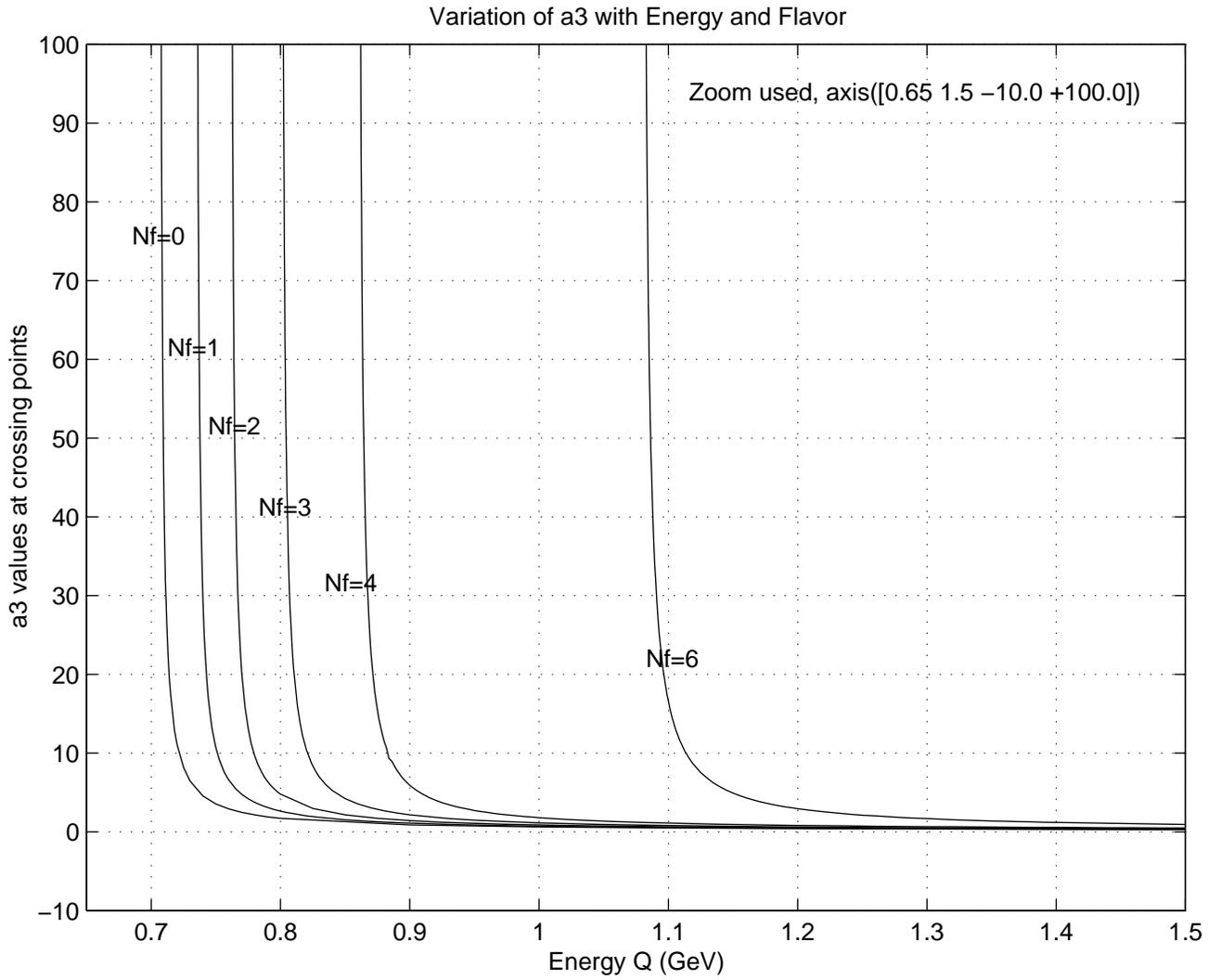}}
\caption{Pattern of variation with momentum and flavor, of isolated Pad\'{e}  $a_{3}$ component color force solution} 
\label{fig: ndili33}
\centering
\end{figure}

From fig.~\ref{fig: ndili2} we see that the $a_{3}$ is the only solution or  
crossing point of  eqn.~(\ref{eq: ndili60}) in the low energy  momentum  region $Q < Q_{\mathrm{min}}$, 
and for any given flavor  number $N_{f}$,  particularly in the range, $0 \le N_{f} \le 8$.  For
$9 \le N_{f} \le 16$, the indication from figs.~\ref{fig: ndili20} to~\ref{fig: ndili23} 
as well as figs.~\ref{fig: ndili45}~,~\ref{fig: ndili47} and~\ref{fig: ndili51}, 
is that the $a_{3}$ is already infinitely large even before we reach the  point $Q = Q_{\mathrm{min}}$.
If we combine this with the features of the $a_{3}$ shown in fig.~\ref{fig: ndili33} for the
$0 \le N_{f} \le 8$ cases, the indication is that while the $a_{3}$ does exist in general in the region
$Q < Q_{\mathrm{min}}$, it has the form of a Landau type pole singularity for a Pad\'{e} beta
function or couplant. This is re-inforced by the fact seen in fig.~\ref{fig: ndili33} that even
for the lowest $N_{f} = 0$ case, the  $a_{3}$ already rises sharply towards very large
values at $Q = 0.705$ GeV, in the process still cutting off access into the infra-red region
$Q < 0.705$ Gev. 
  
The $a_{3}$  exists  in the  medium  energy region $Q_{\mathrm{min}} \le  Q \le  Q_{\mathrm{max}}$  
but with a value that decreases  progressively  with increasing momentumQ. 
It finally cuts off at the characteristic  higher momentum $Q = Q_{\mathrm{max}}$ such that for all
$Q > Q_{\mathrm{max}}$,  $a_{3} = 0$. Because this second Pad\'{e}
couplant bifurcation point occurs well inside the normal PQCD region of QCD, 
$Q \geq \Lambda_{\mathrm{QCD}}$,
with  $Q_{\mathrm{max}} \gg \Lambda_{\mathrm{QCD}}$, the possibility arises that if the above
Pad\'{e} couplant structure represents physical  QCD reality,  the Pad\'{e} $(a_{3}, a_{2})$ structure
can affect  substantially even relatively high energy QCD processes, by way of  distinctively
non-perturbative contributions.  

\subsubsection{Features of the $a_{1}$   Pad\'{e} component couplant solution}
\begin{figure}
\scalebox{1.0}{\includegraphics{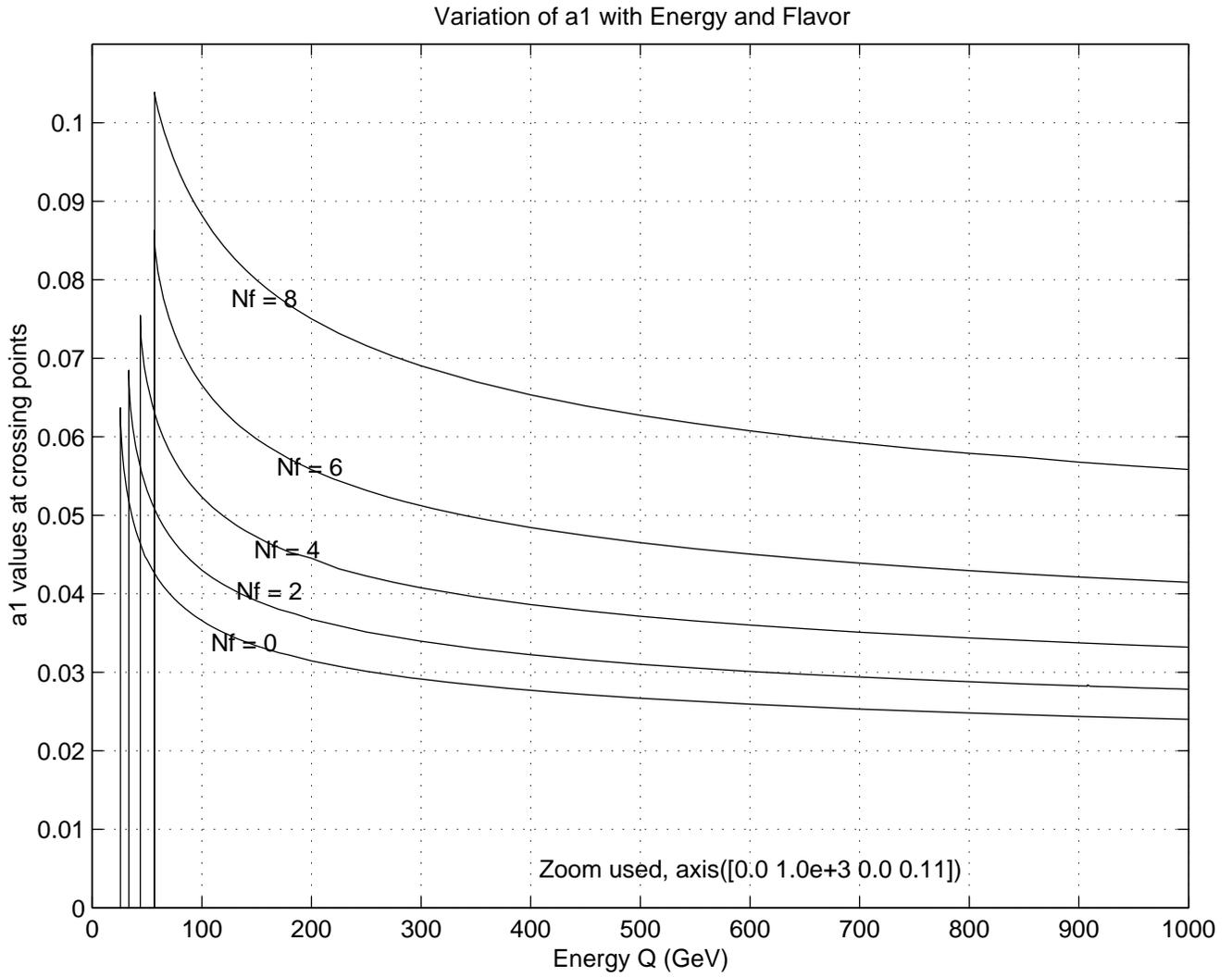}}
\caption{Pattern of variation with momentum and flavor, of isolated Pad\'{e}  $a_{1}$ component  couplant solution}
\label{fig: ndili32}
\centering
\end{figure}

Examining the $a_{1}$ Pad\'{e} component couplant solution in the same way,  we observe from 
fig.~\ref{fig: ndili7} as well as  the upper halfs of figs.~\ref{fig: ndili24} to~\ref{fig: ndili28}, 
that the $a_{1}$ solution  (crossing point) is the only  Pad\'{e} QCD present in the very high energy 
region $Q > Q_{\mathrm{max}}$ where it also has very small values for any flavor  number, such that for
$Q \rightarrow \infty$,  $a_{1} \rightarrow  0$, shown in fig.~\ref{fig: ndili32}.

The $a_{1}$  couplant solution exists in the intermediate 
energy region:  $Q_{\mathrm{min}} \le Q  \le Q_{\mathrm{max}}$ attaining its highest but still moderate
value at $Q = Q_{\mathrm{min}}$ from where it decreases progressively  towards  zero as 
$Q \rightarrow \infty$.  The $a_{1}$  solution  does not exist in the low energy region 
$0 \le Q < Q_{\mathrm{min}}$.

The indication from these features is  that the Pad\'{e}  $a_{1}$ component solution can be identified 
as the  asymptotically  free  purely PQCD  component color force of QCD we started with in
eqns.~(\ref{eq: ndili1}) and~(\ref{eq: ndili8}).  We shall so identify the  $a_{1}$ component
Pad\'{e} solution.  Then the fact that the $a_{1}$ solution does not exist in the region
$0 \le Q < Q_{\mathrm{min}}$ is seen as a correct reflection of the denominator zero singularity of 
the Pad\'{e} beta function eqn.~(\ref{eq: ndili31}) and its attendant Kogan-Shifman~\cite{Kogan95} type
behavior, with our $Q_{\mathrm{min}}$ directly  identifiable with the $\mu_{c}$ cut off momentum of 
Kogan and Shifman, and of Elias et. al.~\cite{Elias98,Elias99}.

Based on these identifications, the question of whether infra-red scenario I (frozen couplant), or
scenario II (bifurcated infra-red attractor point) holds at any flavor  in Pad\'{e} QCD and perhaps in 
real QCD, boils down in our approach to the question  of the extent the momentum gap 
$0 \le Q < Q_{\mathrm{min}}$ remains finite and free of our $(a_{1}, a_{2})$ bifurcating components, 
in various flavor states of QCD. We will find that our method allows us to answer this question
explicitly later.

\subsubsection{The $a_{2}$ Pad\'{e} component couplant solution}
\begin{figure}
\scalebox{1.0}{\includegraphics{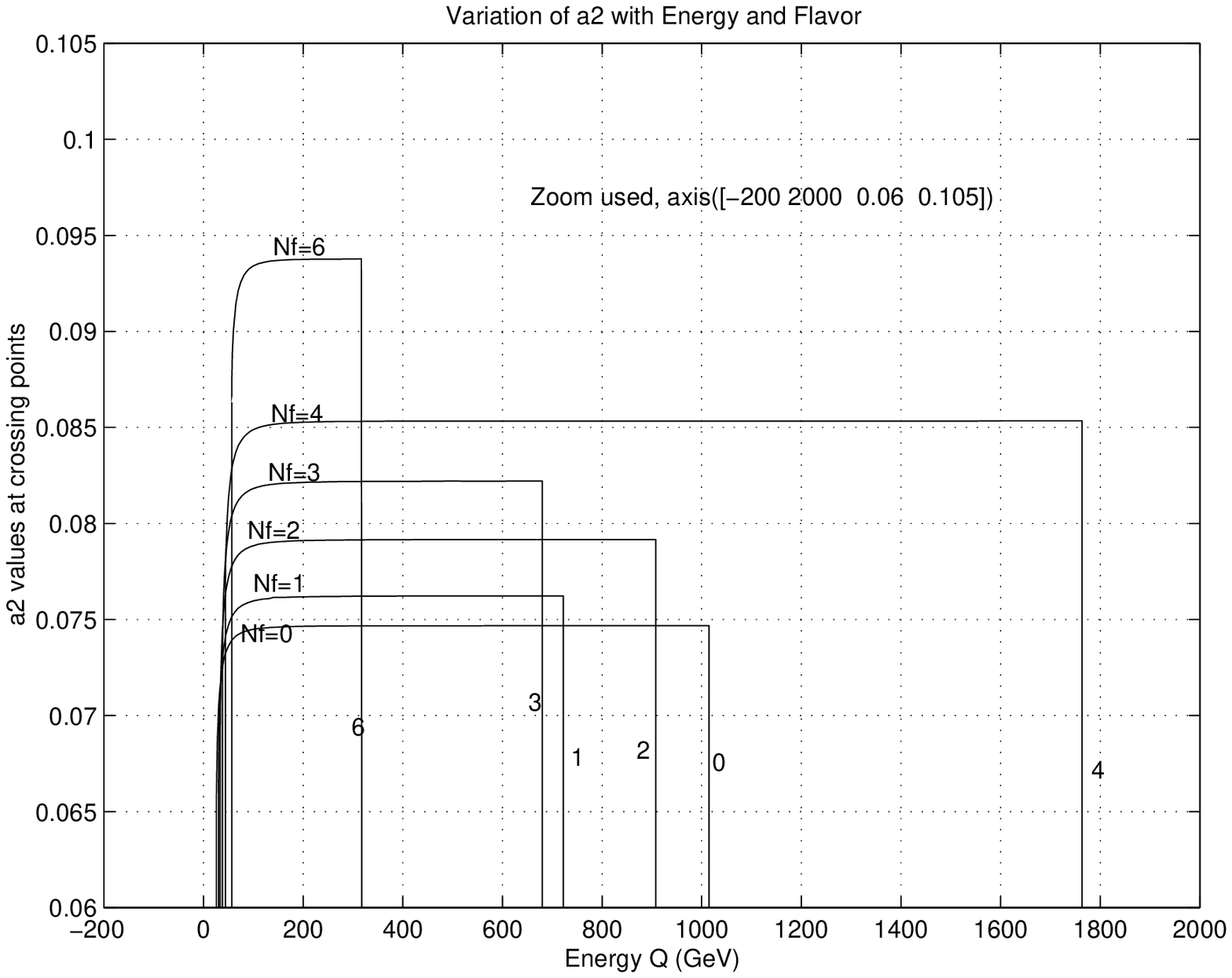}}
\caption{Pattern of variation with momentum and flavor, of isolated Pad\`{e} $a_{2}$ component couplant solution}
\label{fig: ndili34}
\centering
\end{figure}

Plotting the $a_{2}$ crossing point values against momentum Q, we find the behavior shown in fig.~\ref{fig: ndili34}.
That is,  the $a_{2}$ component color force solution is found to exist only in the intermediate energy region
$Q_{\mathrm{min}} \le Q \le Q_{\mathrm{max}}$ for any flavor number $N_{f}$, but vanishes or  cuts off
sharply at the two critical upper and lower momentum points $Q = Q_{\mathrm{max}}$ and
$Q = Q_{\mathrm{min}}$.

From fig~\ref{fig: ndili34}, we see that inside  its intermediate  energy domain of operation and existence, the $a_{2}$ 
couplant rises in value from a lowest value at $Q = Q_{\mathrm{min}}$ to a highest  value at
$Q = Q_{\mathrm{max}}$.  This manner of behavior suggests we identify our Pad\'{e}  $a_{2}$ component 
with  the upper branch of the bifurcated structure found by Elias et. al.~\cite{Elias98,Elias99} and
 by Kogan and Shifman~\cite{Kogan95}. This is reinforced by the observation from
our profile plots shown variously in figs.~\ref{fig: ndili2} to~\ref{fig: ndili28} that
the $a_{2}$ crossing point solution merges with and coincides exactly in value with the $a_{1}$ 
couplant solution at the point $Q = Q_{\mathrm{min}}$ which can be regarded as a bifurcation point 
of the $a_{1}$ into $a_{2}$. Then the infra-red region $0 \le Q < Q_{\mathrm{min}}$, is totally free
of the two bifurcated component couplants, $a_{1}$ and $a_{2}$, again agreeing fully with the behavior 
found by Elias et. al.~\cite{Elias98,Elias99}, and  by Kogan-Shifman~\cite{Kogan95}.

However, unlike the Elias et. al. structure which did not determine the progression of this 
$a_{2}$ upper branch, our  graphical crossing point solution  method shows the $a_{2}$ clearly as
terminating and merging  exactly with the $a_{3}$ crossing  point  solution
at the point $Q = Q_{\mathrm{max}}$, where $a_{2} = a_{3}$, and both couplants thereafter
disappear together for all $Q > Q_{\mathrm{max}}$. One may say that the $a_{2}$ bifurcates further
into the $a_{3}$ which then runs towards the infra-red region but soon rises sharply to very large values
in a manner suggestive of a Landau pole behavior. This $a_{3}$ presence was not discernible from the
numerator and denominator zero analysis of Elias et. al.~\cite{Elias98,Elias99}.

\subsubsection {The $a_{4}(Y_{2})$ Pad\'{e} component couplant solution}
\begin{figure}
\scalebox{1.0}{\includegraphics{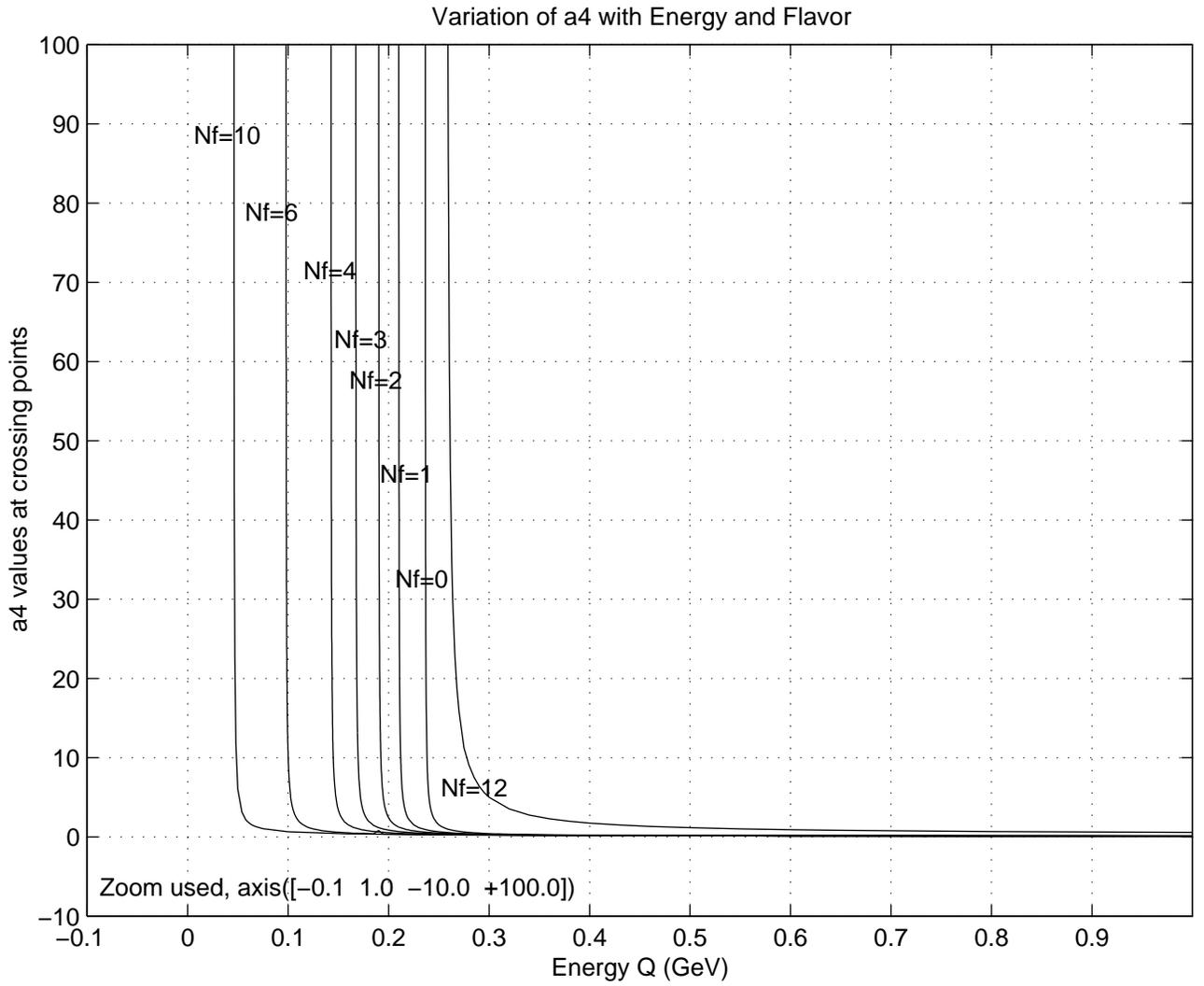}}
\caption{Pattern of variation with momentum and flavor  of  Pad\'{e} $a_{4}$ component couplant solution} 
\label{fig: ndili35}
\centering
\end{figure}

The  $a_{4}(Y_{2})$  crossing  point  solution, arose from the separate eqn.~(\ref{eq: ndili61}). As 
such it emerges for now as a lone star Pad\'{e} component color force, not connected to the 
$a_{1}, a_{2}, a_{3}$  components of eqn.~(\ref{eq: ndili60}).  Its features  are shown 
in fig.~\ref{fig: ndili35}, where it is seen that it has no structure or bifurcation but behaves 
smoothly, falling asymptotically in value from a large value $(a_{4} \rightarrow \infty)$ near
$Q \rightarrow  0$, to the value $(a_{4} \rightarrow  0)$ as $(Q \rightarrow \infty)$. 
The exact meaning of the $a_{4}$ couplant solution  is not yet clear to us, and is being investigated.

\subsection{Spiral Chain-like Structure found in Pad\'{e} QCD}

\begin{figure}
\scalebox{1.0}{\includegraphics{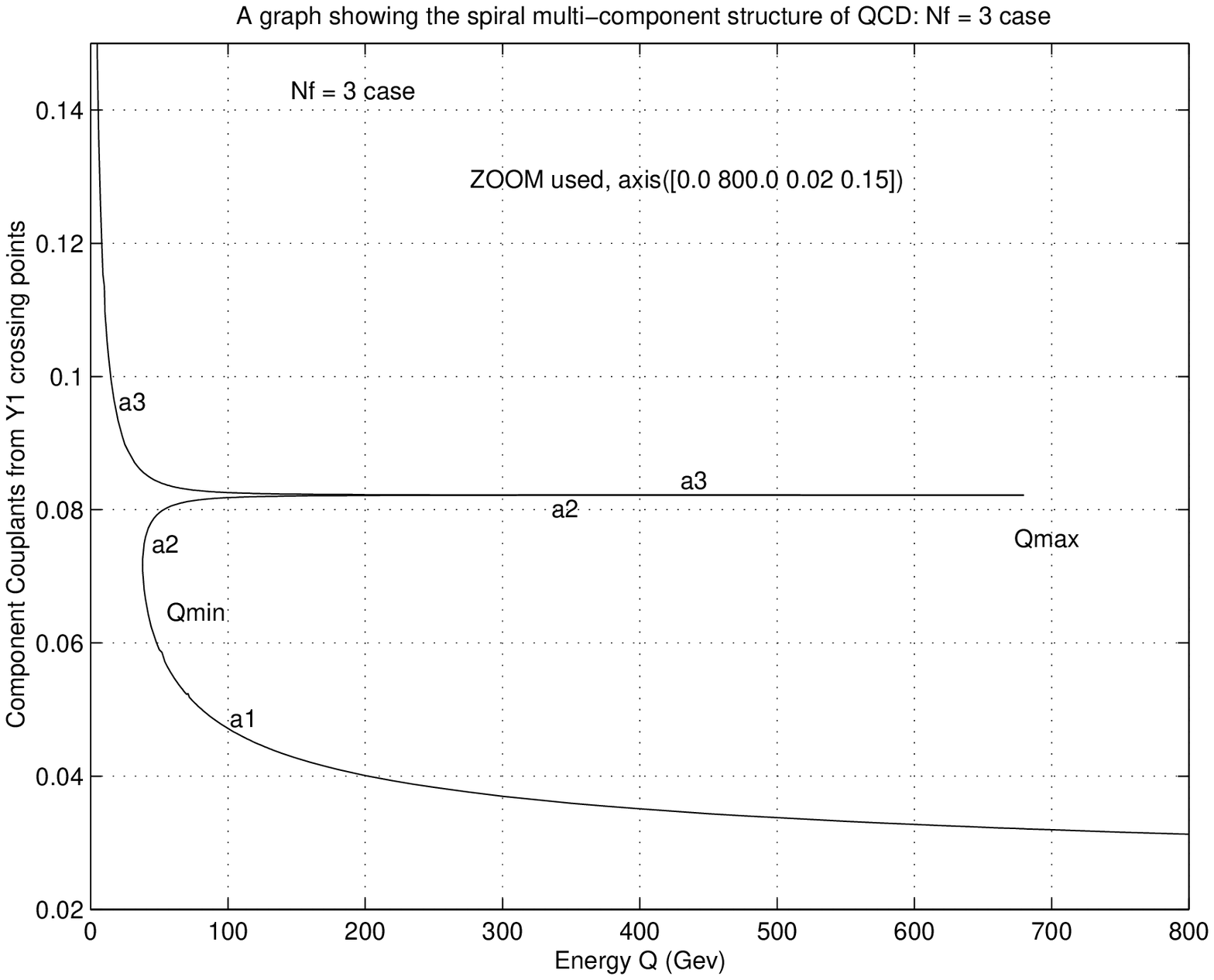}}
\caption{A plot showing a spiral multi-component structure of Pad\'{e} QCD: Nf = 3 case}
\label{fig: ndili37}
\centering
\end{figure}

\begin{figure}
\scalebox{1.0}{\includegraphics{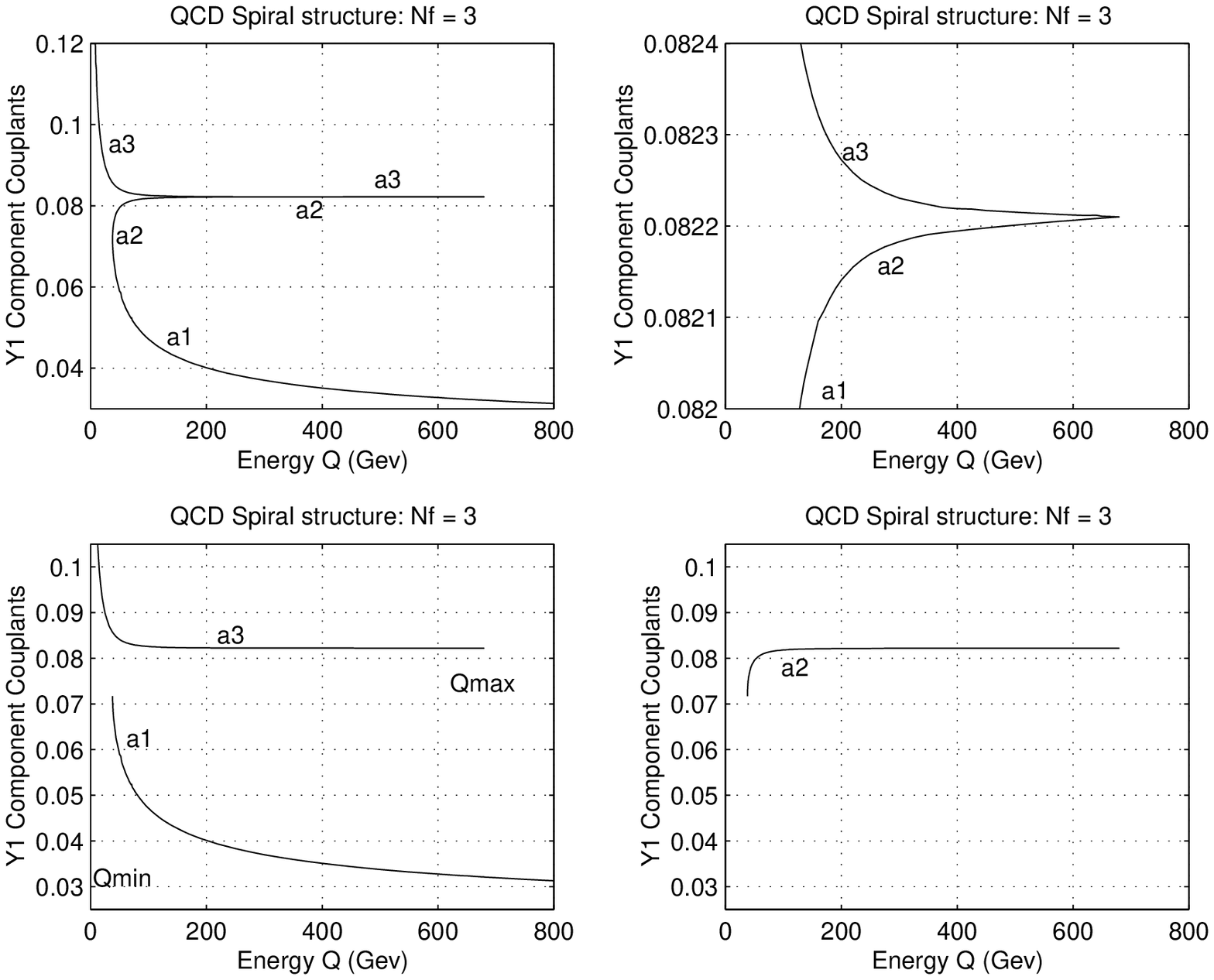}}
\caption{A plot showing a  multi-component couplant structure in Pad\'{e} QCD: Nf = 3 case}
\label{fig: ndili38}
\centering
\end{figure}

\begin{figure}
\scalebox{1.0}{\includegraphics{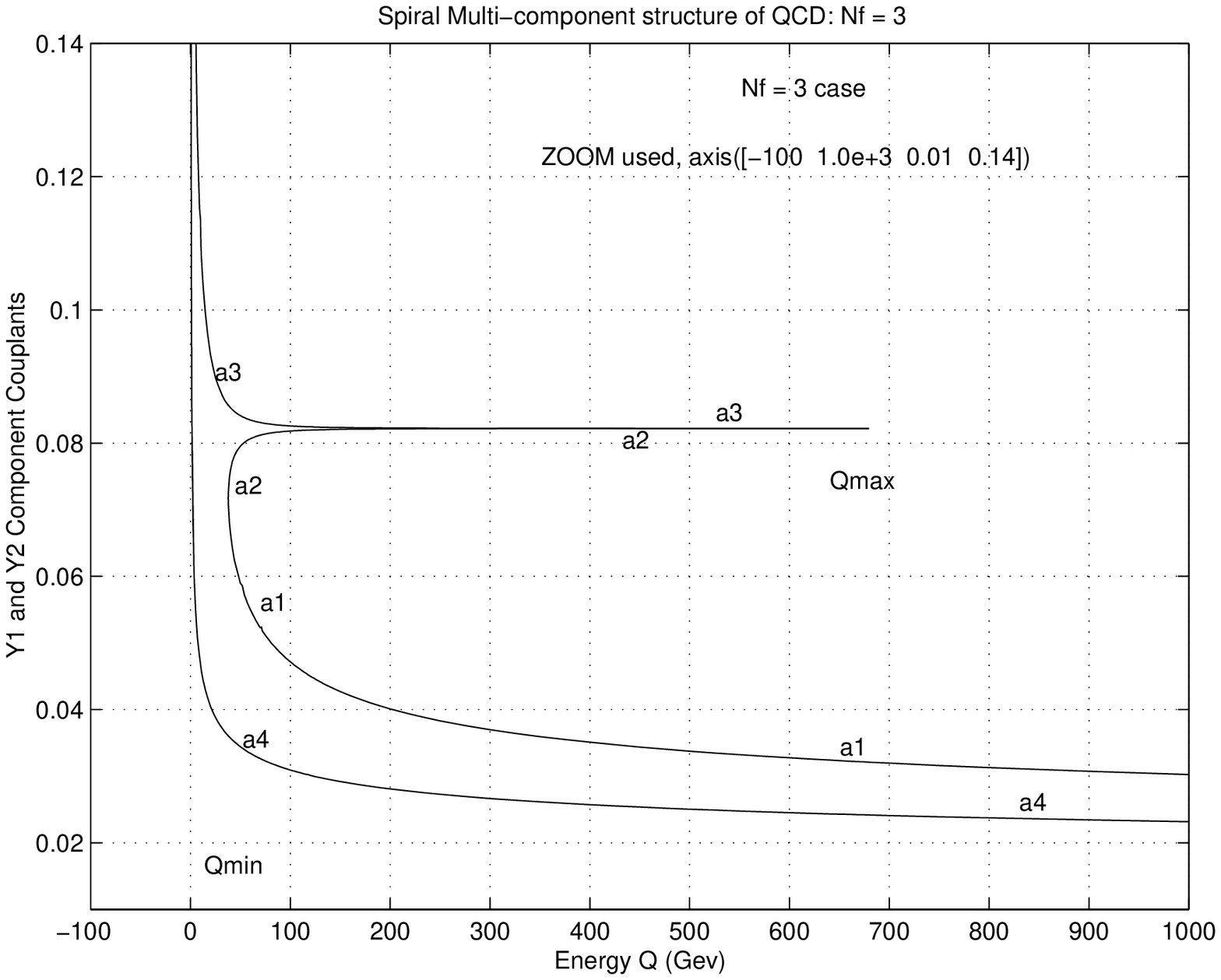}}
\caption{A plot showing a spiral multi-component structure in Pad\'{e} QCD: Nf = 3 case}
\label{fig: ndili39}
\centering
\end{figure}

\begin{figure}
\scalebox{1.0}{\includegraphics{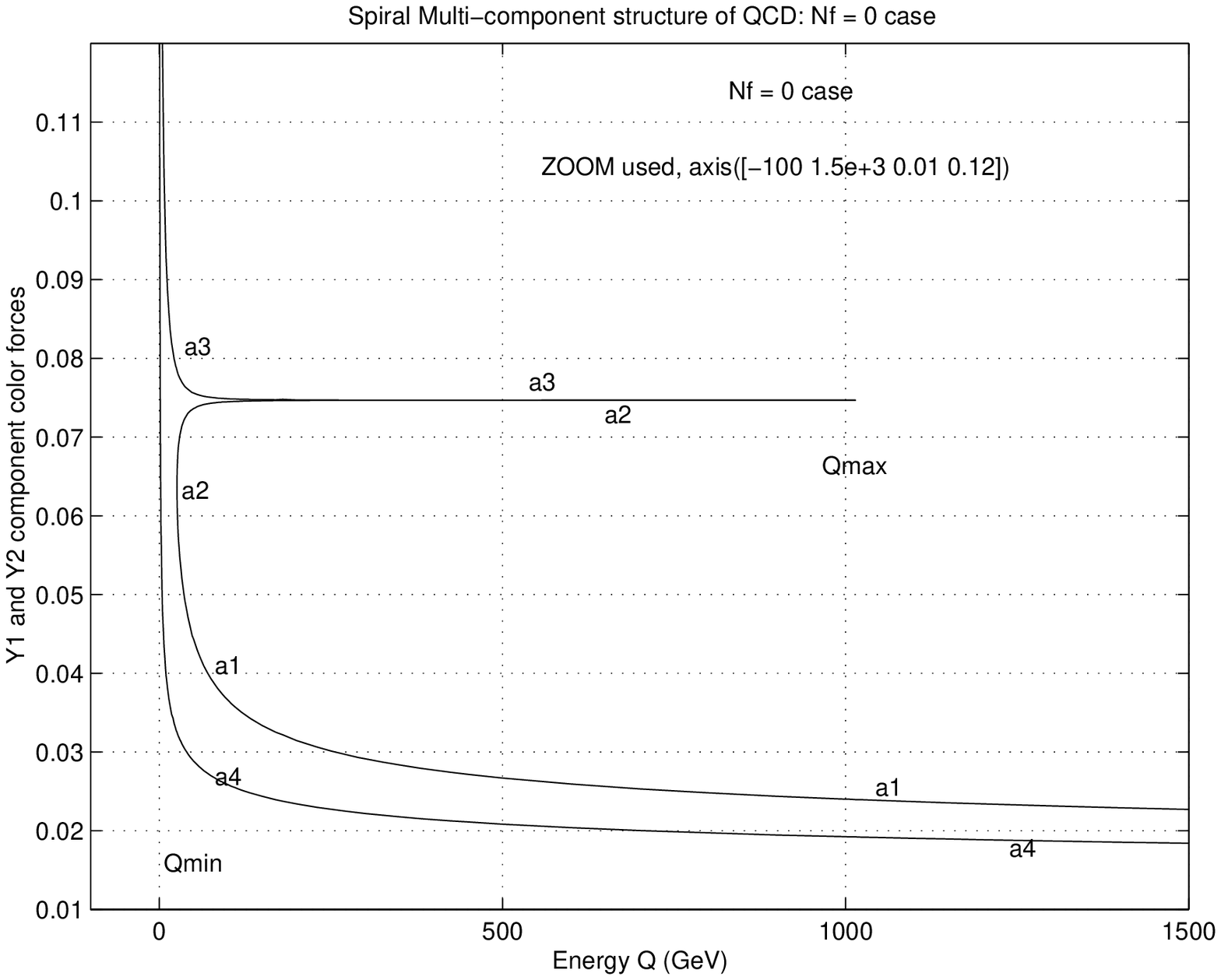}}
\caption{A plot showing a spiral multi-component structure in Pad\'{e}  QCD: Nf = 0 case}
\label{fig: ndili40}
\centering
\end{figure}

\begin{figure}
\scalebox{1.0}{\includegraphics{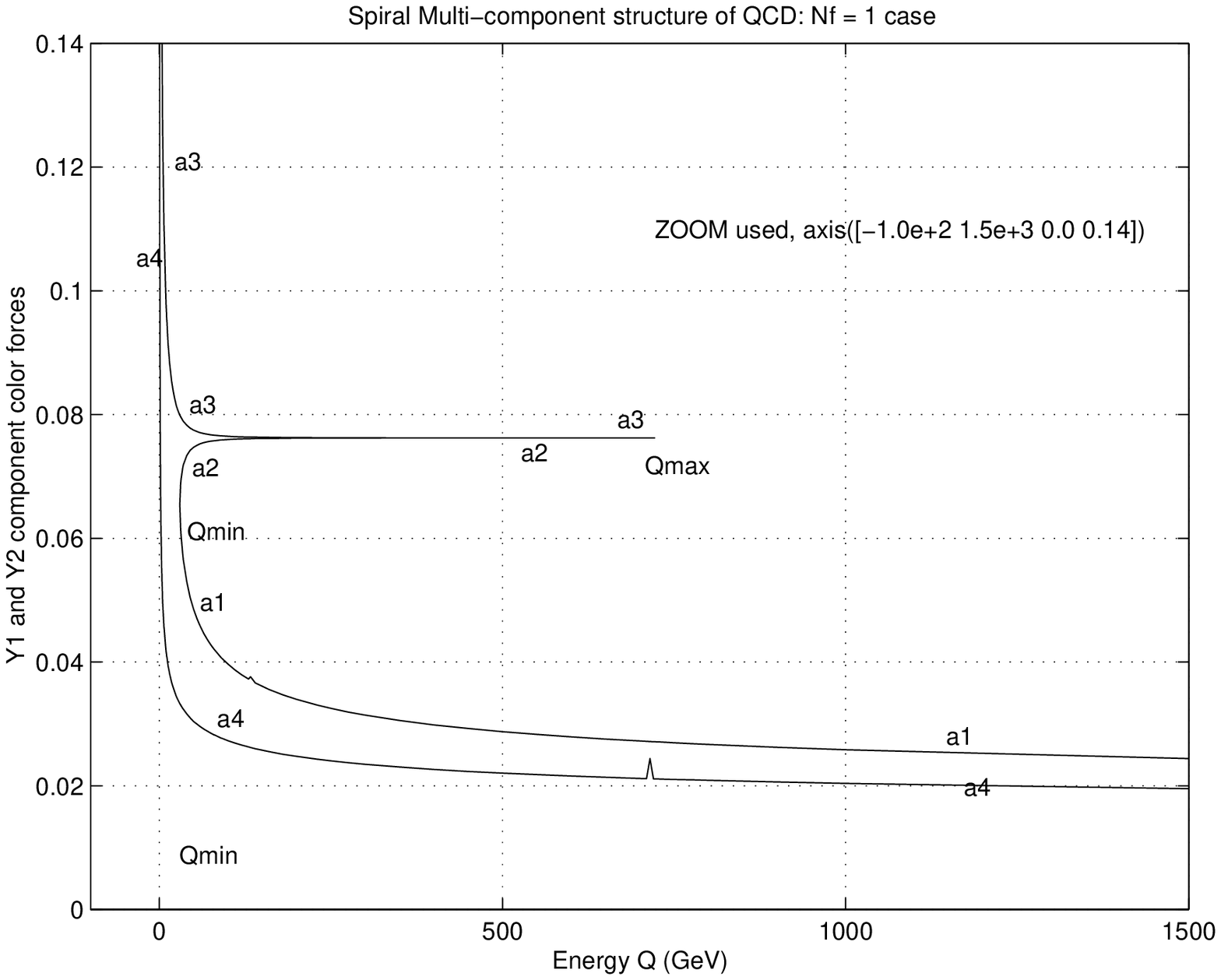}}
\caption{A plot showing a  spiral multi-component structure in Pad\'{e} QCD: Nf = 1 case}
\label{fig: ndili40a}
\centering
\end{figure}

\begin{figure}
\scalebox{1.0}{\includegraphics{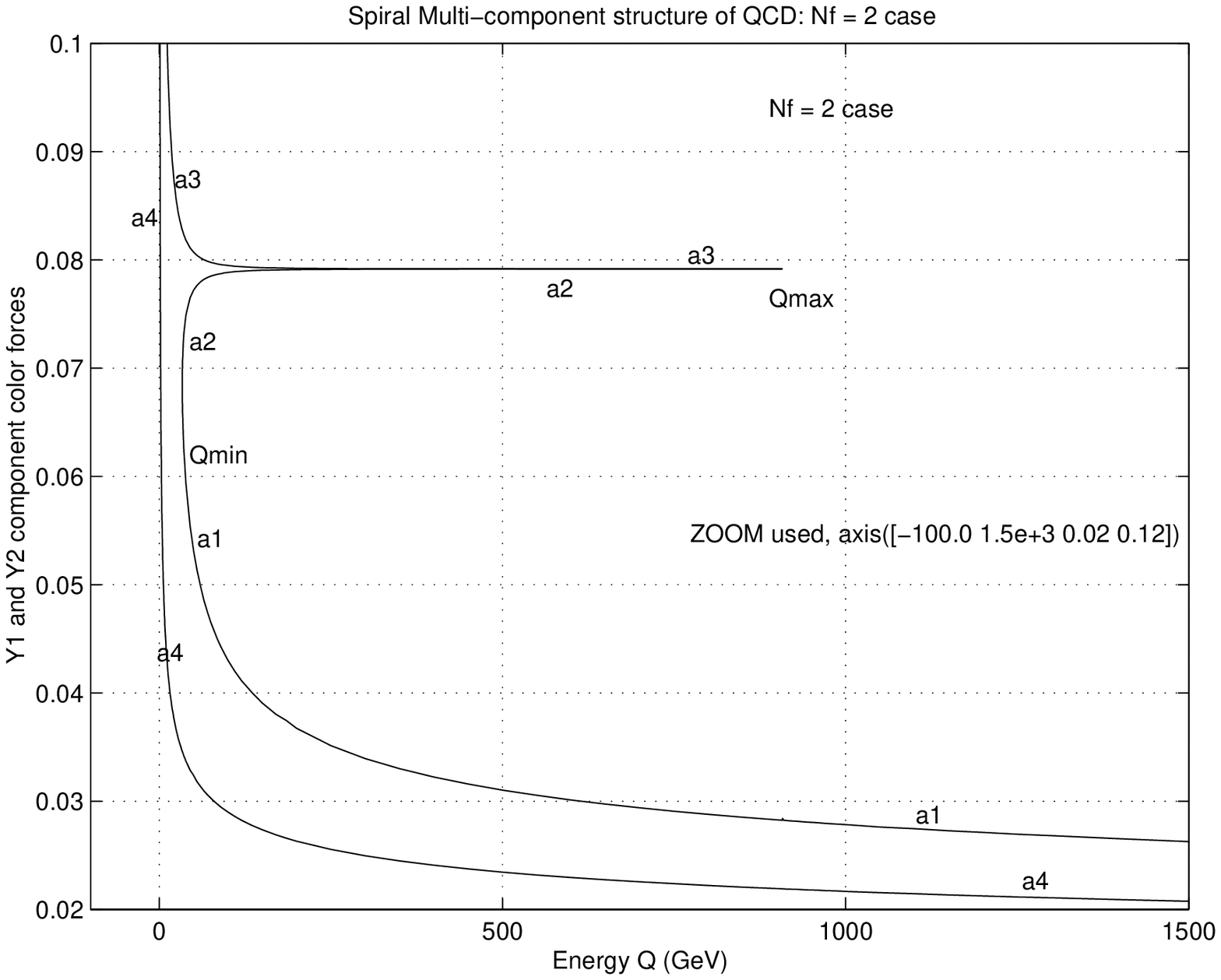}}
\caption{A plot showing a  spiral multi-component structure in Pad\'{e} QCD: Nf = 2 case}
\label{fig: ndili42}
\centering
\end{figure}

\begin{figure}
\scalebox{1.0}{\includegraphics{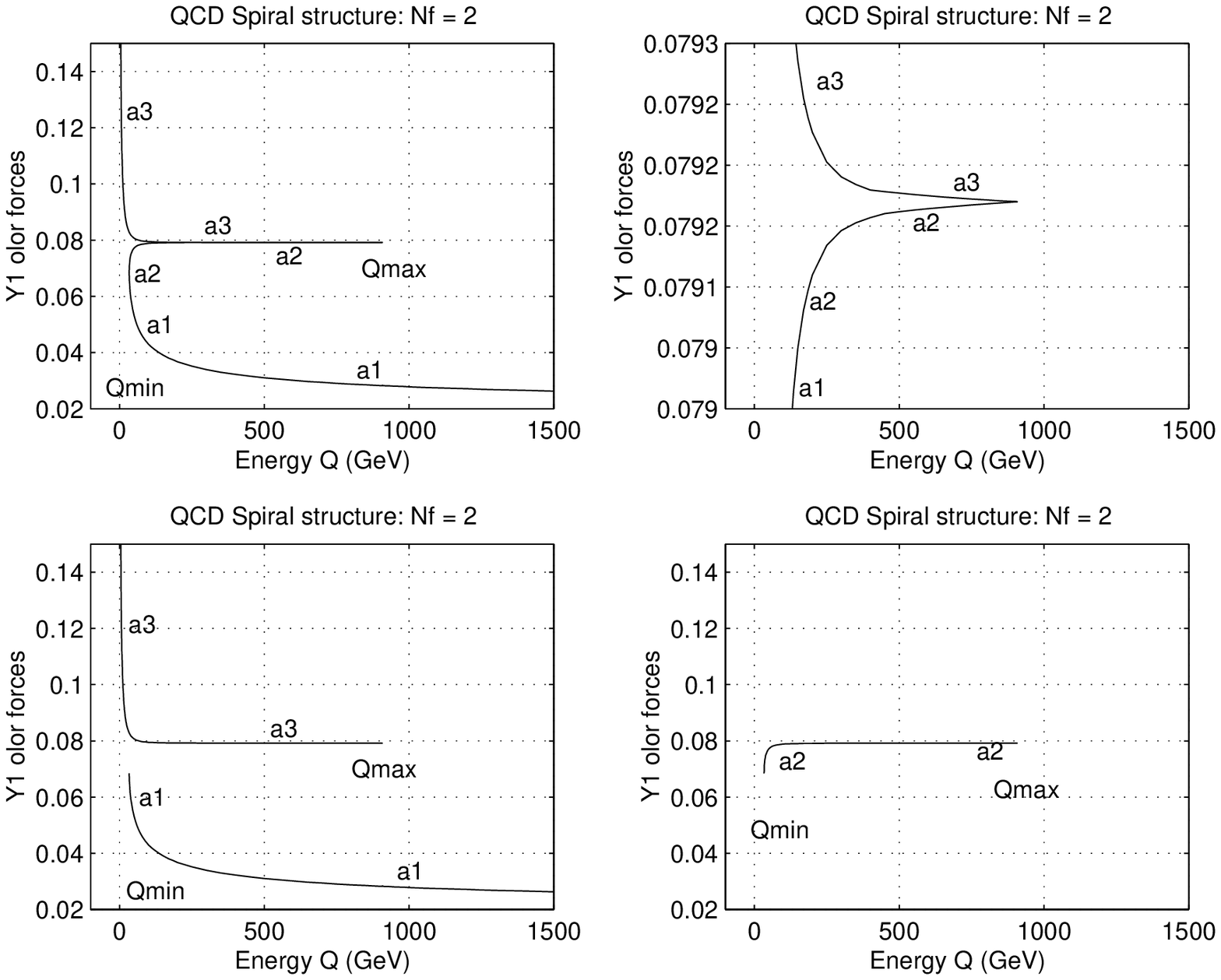}}
\caption{A plot showing a multi-component couplant structure in Pad\'{e}  QCD: Nf = 2 case}
\label{fig: ndili43}
\centering
\end{figure}

\begin{figure}
\scalebox{1.0}{\includegraphics{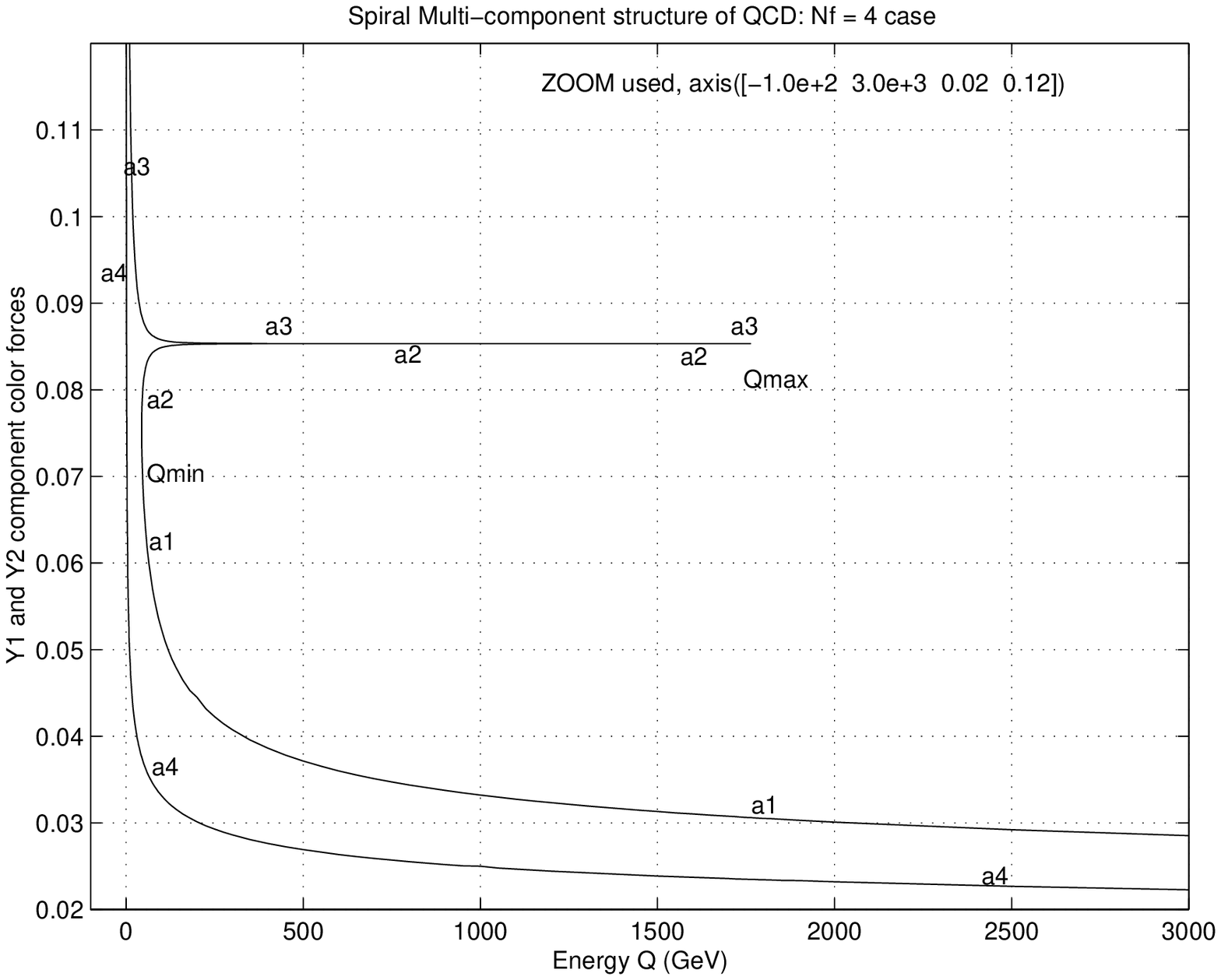}}
\caption{A plot showing a spiral multi-component structure in Pad\'{e} QCD: Nf = 4 case}
\label{fig: ndili43a}
\centering
\end{figure}

\begin{figure}
\scalebox{1.0}{\includegraphics{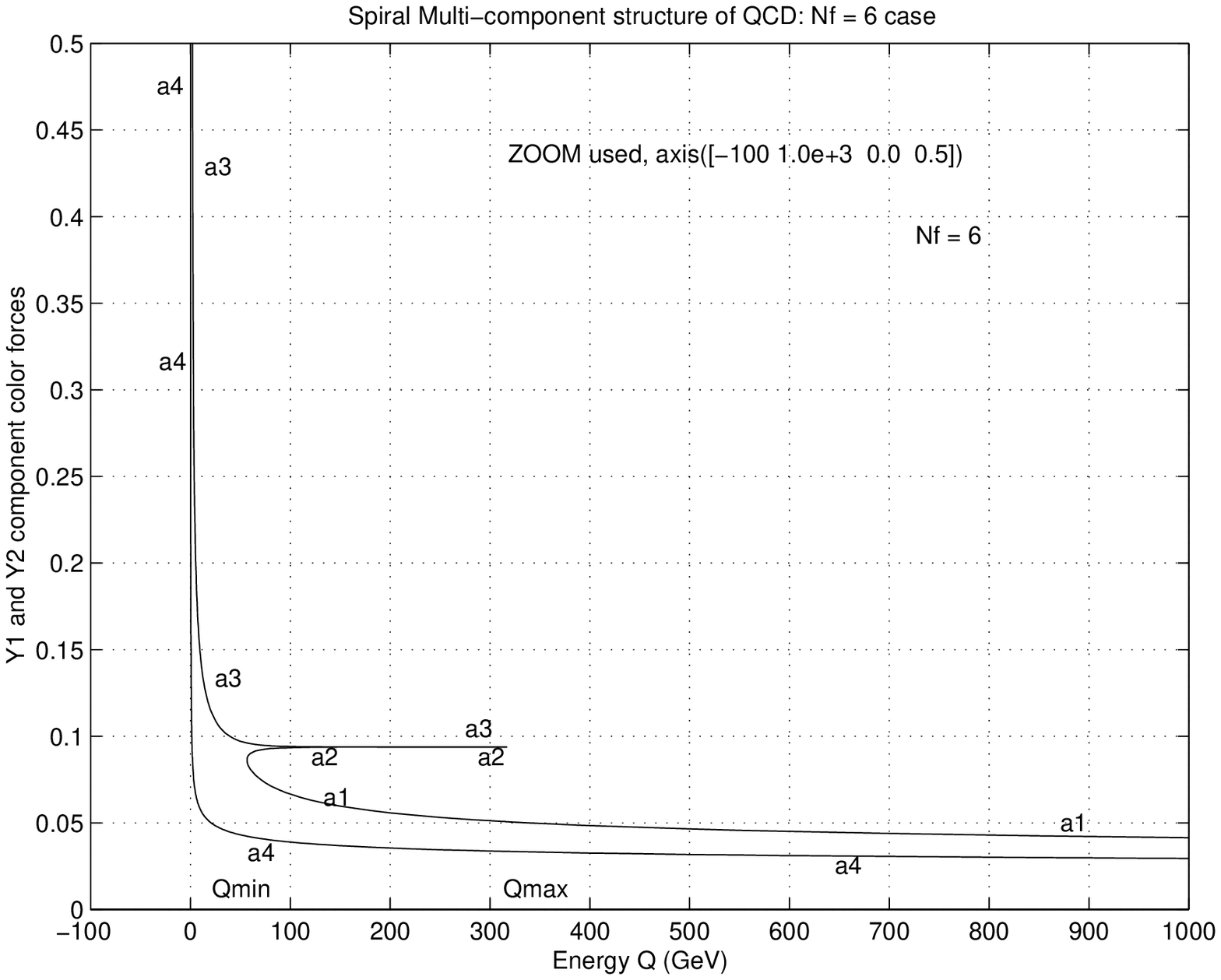}}
\caption{A plot showing a  spiral multi-component structure in Pad\'{e} QCD: Nf = 6 case}
\label{fig: ndili43b}
\centering
\end{figure}

\begin{figure}
\scalebox{1.0}{\includegraphics{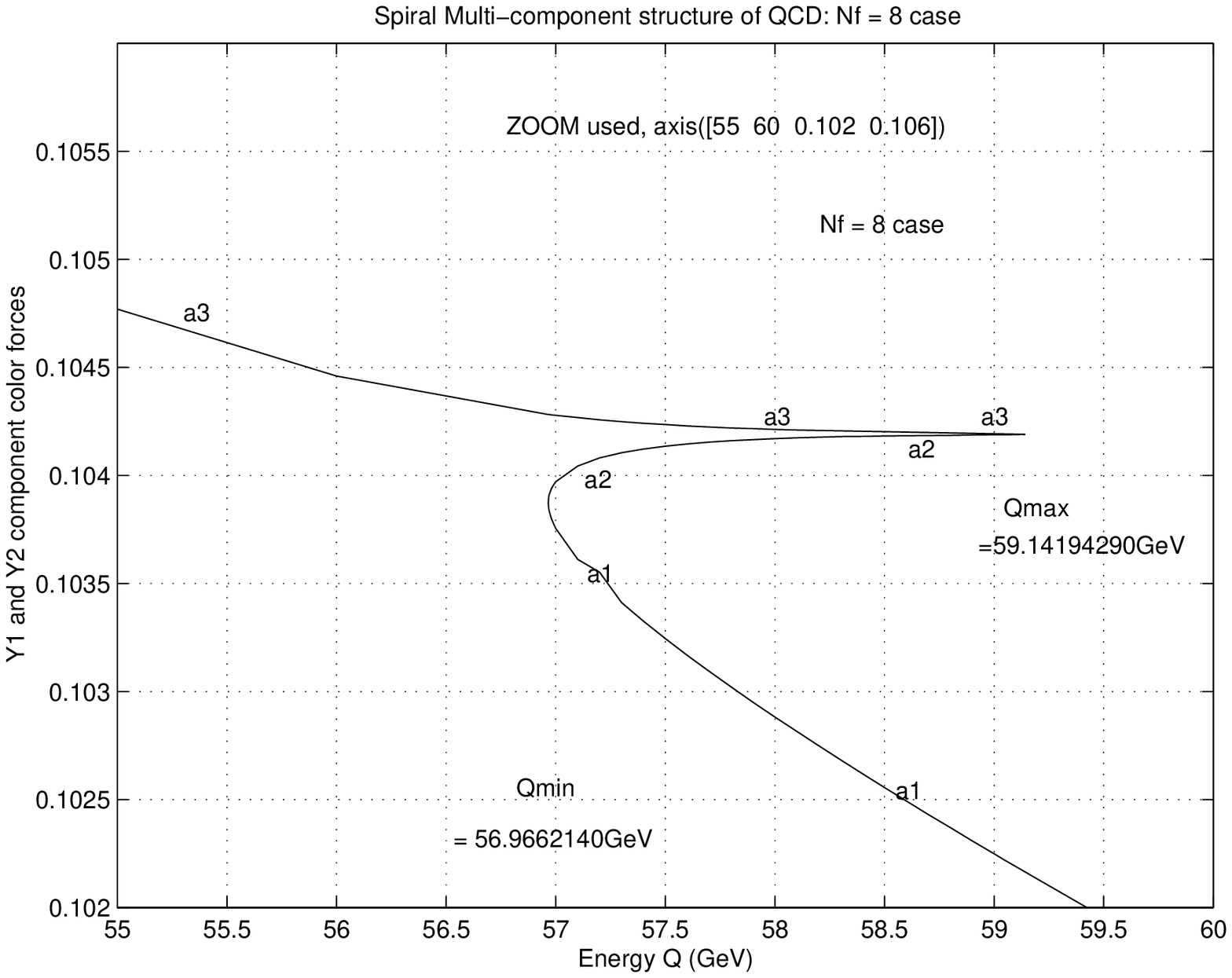}}
\caption{A plot showing a  spiral multi-component structure in Pad\'{e} QCD: Nf = 8 case}
\label{fig: ndili44a}
\centering
\end{figure}

\begin{figure}
\scalebox{1.0}{\includegraphics{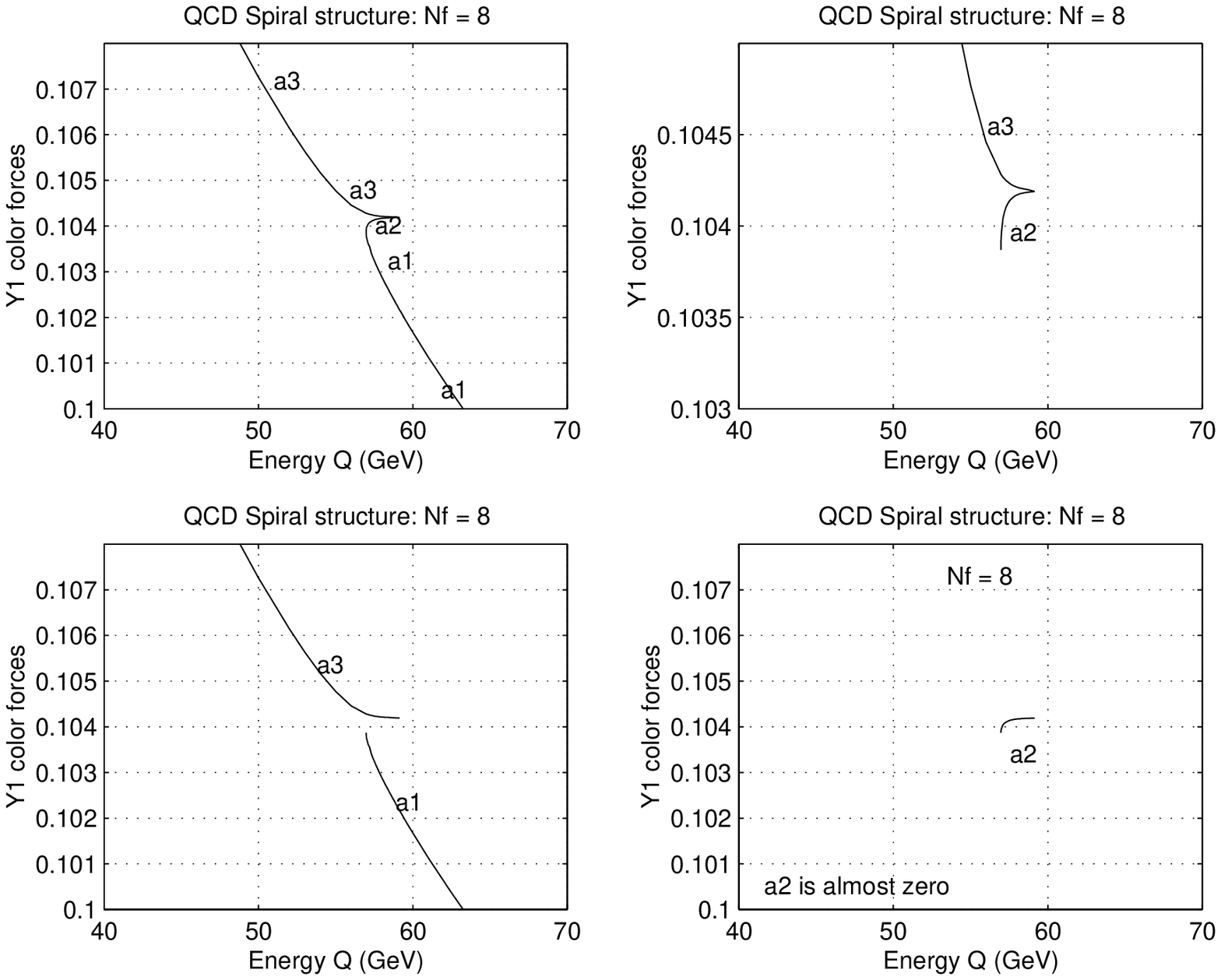}}
\caption{A plot showing a  multi-component couplant structure of Pad\'{e} QCD: Nf = 8 case}
\label{fig: ndili44}
\centering
\end{figure}

\begin{figure}
\scalebox{1.0}{\includegraphics{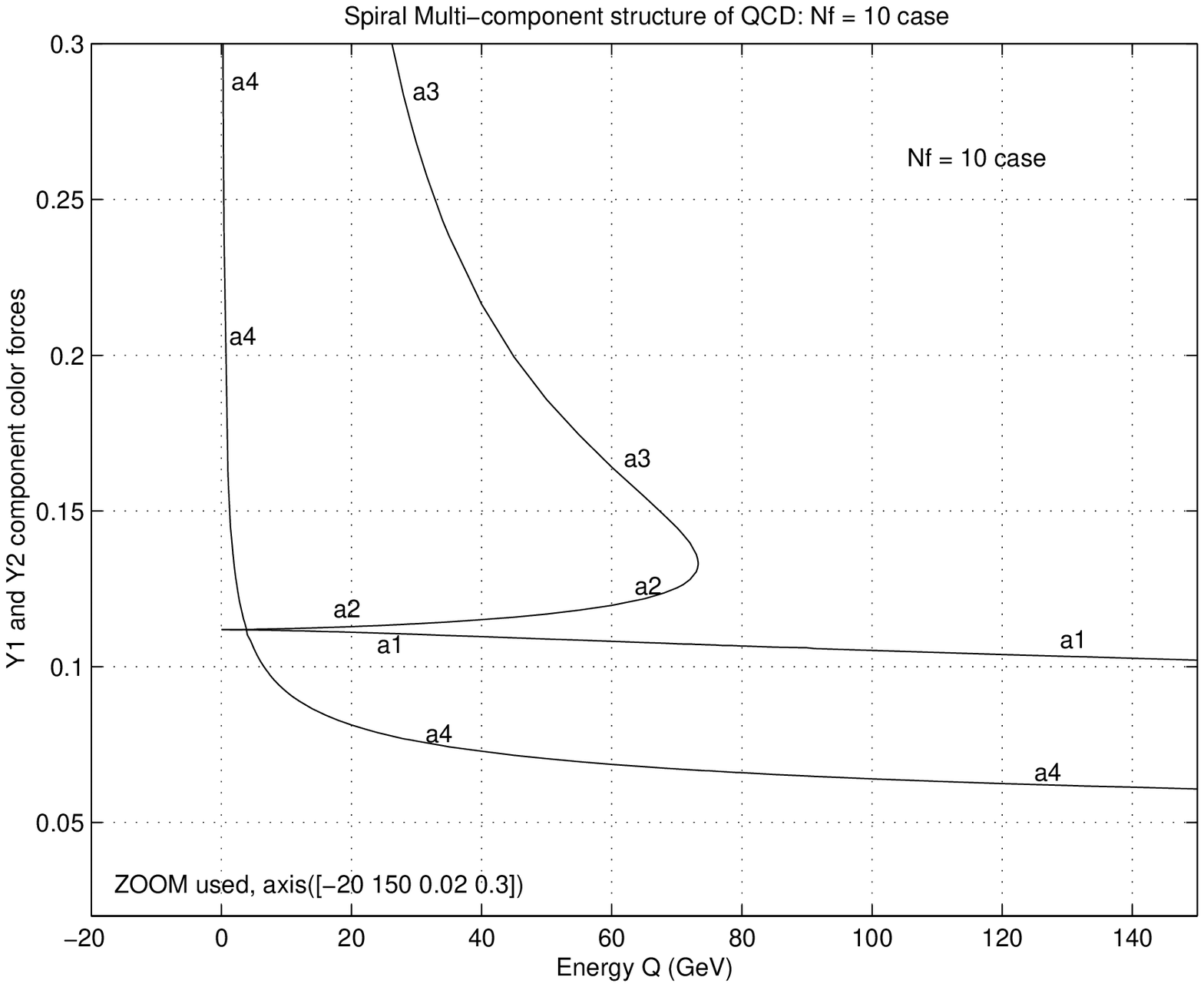}}
\caption{A plot showing a spiral multi-component structure in Pad\'{e} QCD: Nf = 10 case}
\label{fig: ndili45}
\centering
\end{figure}

\begin{figure}
\scalebox{1.0}{\includegraphics{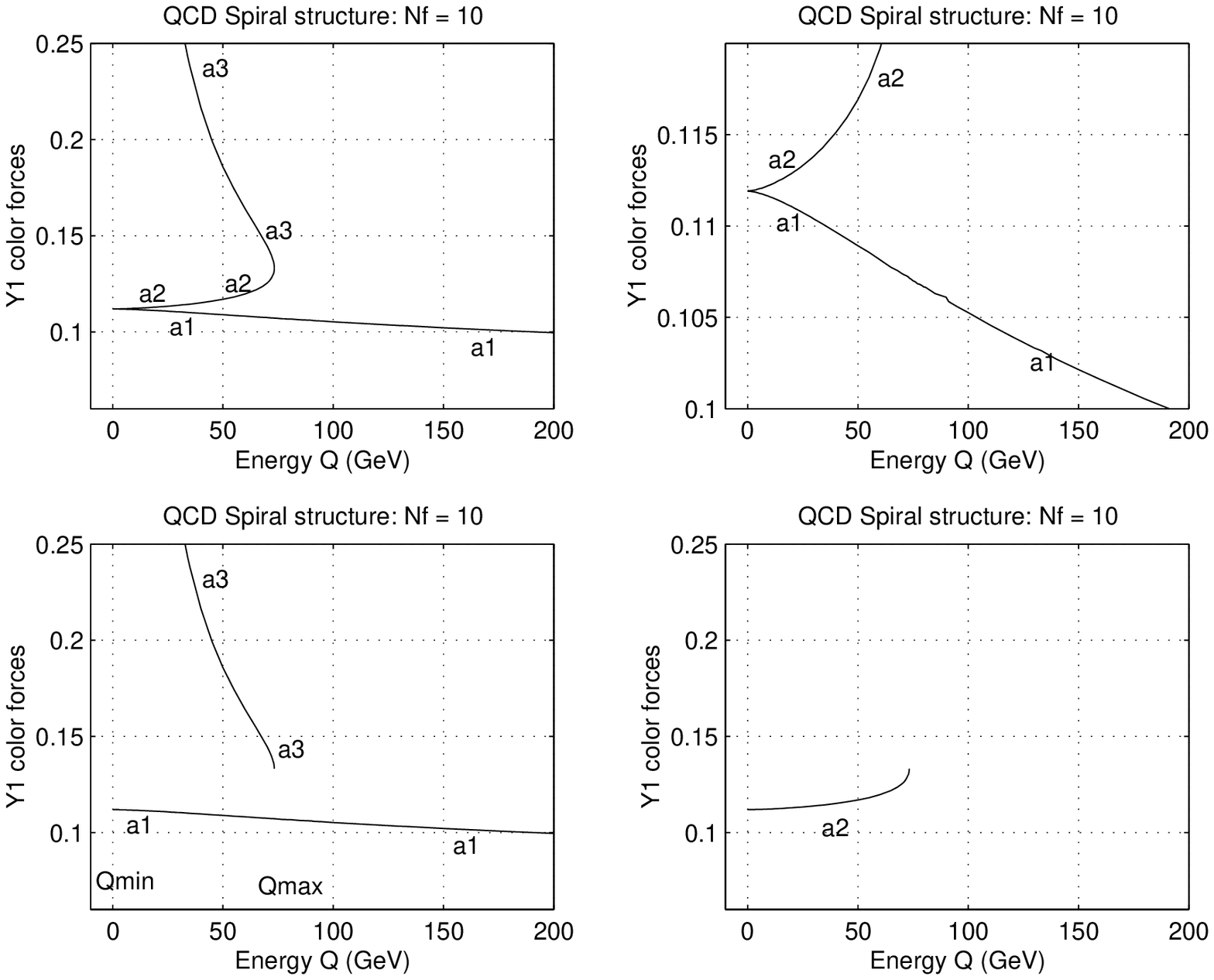}}
\caption{A plot showing a  multi-component couplant structure in Pad\'{e} QCD: Nf = 10 case}
\label{fig: ndili46}
\centering
\end{figure}

\begin{figure}
\scalebox{1.0}{\includegraphics{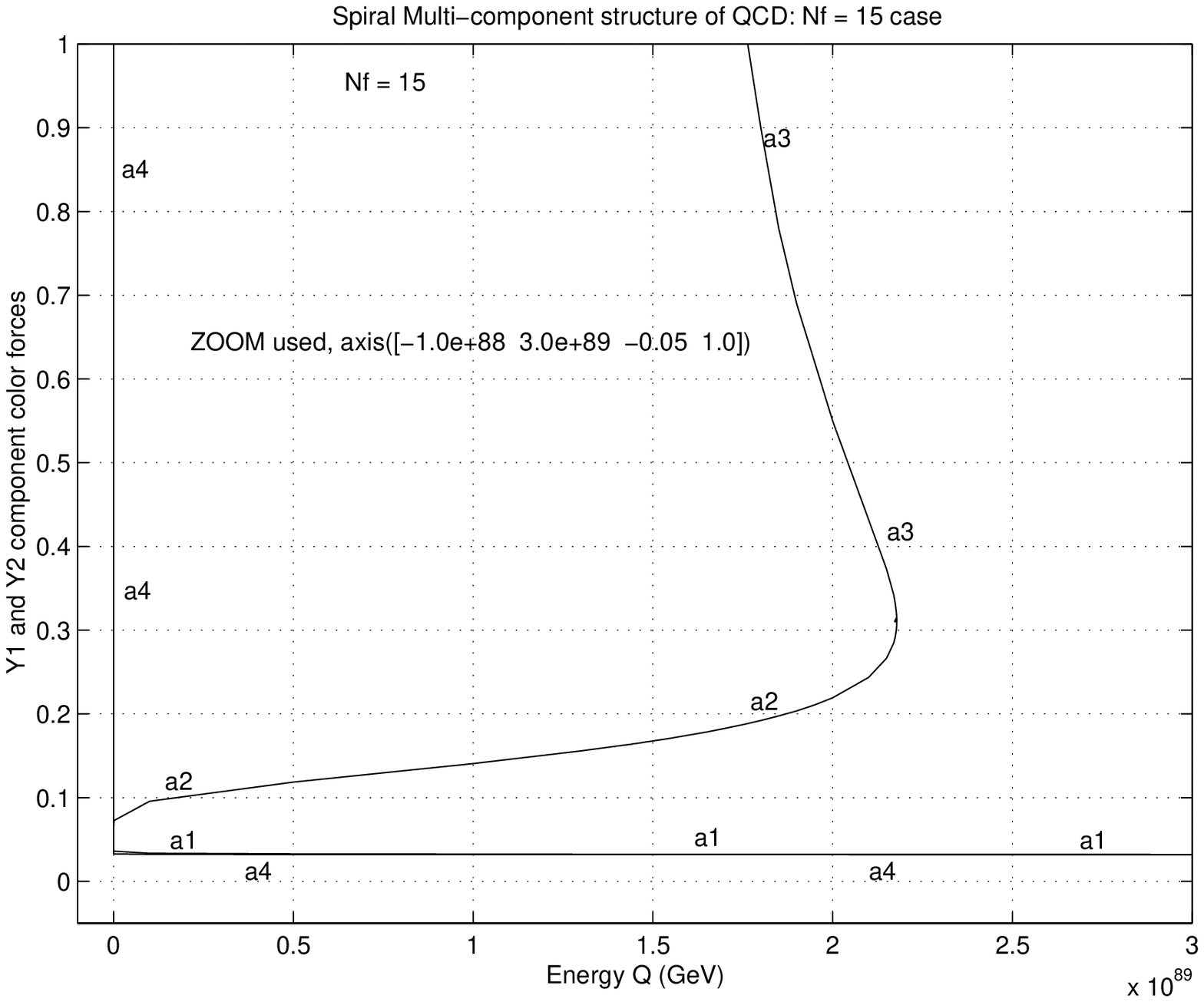}}
\caption{A plot showing a  spiral multi-component structure in Pad\'{e}  QCD: Nf = 15 case}
\label{fig: ndili47}
\centering
\end{figure}

\begin{figure}
\scalebox{1.0}{\includegraphics{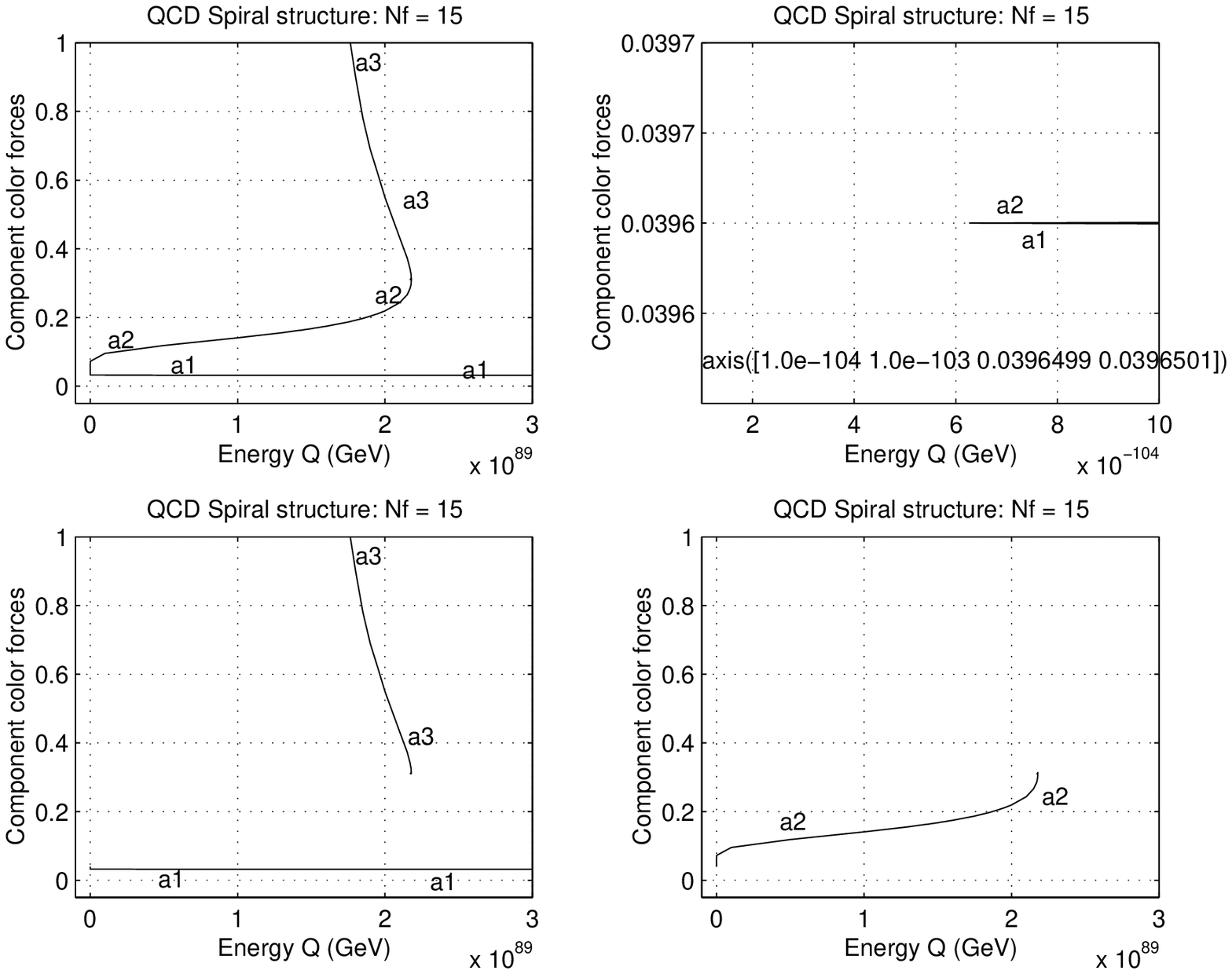}}
\caption{A plot showing a  multi-component couplant  structure of Pad\'{e}  QCD: Nf = 15 case}
\label{fig: ndili48}
\centering
\end{figure}

\begin{figure}
\scalebox{1.0}{\includegraphics{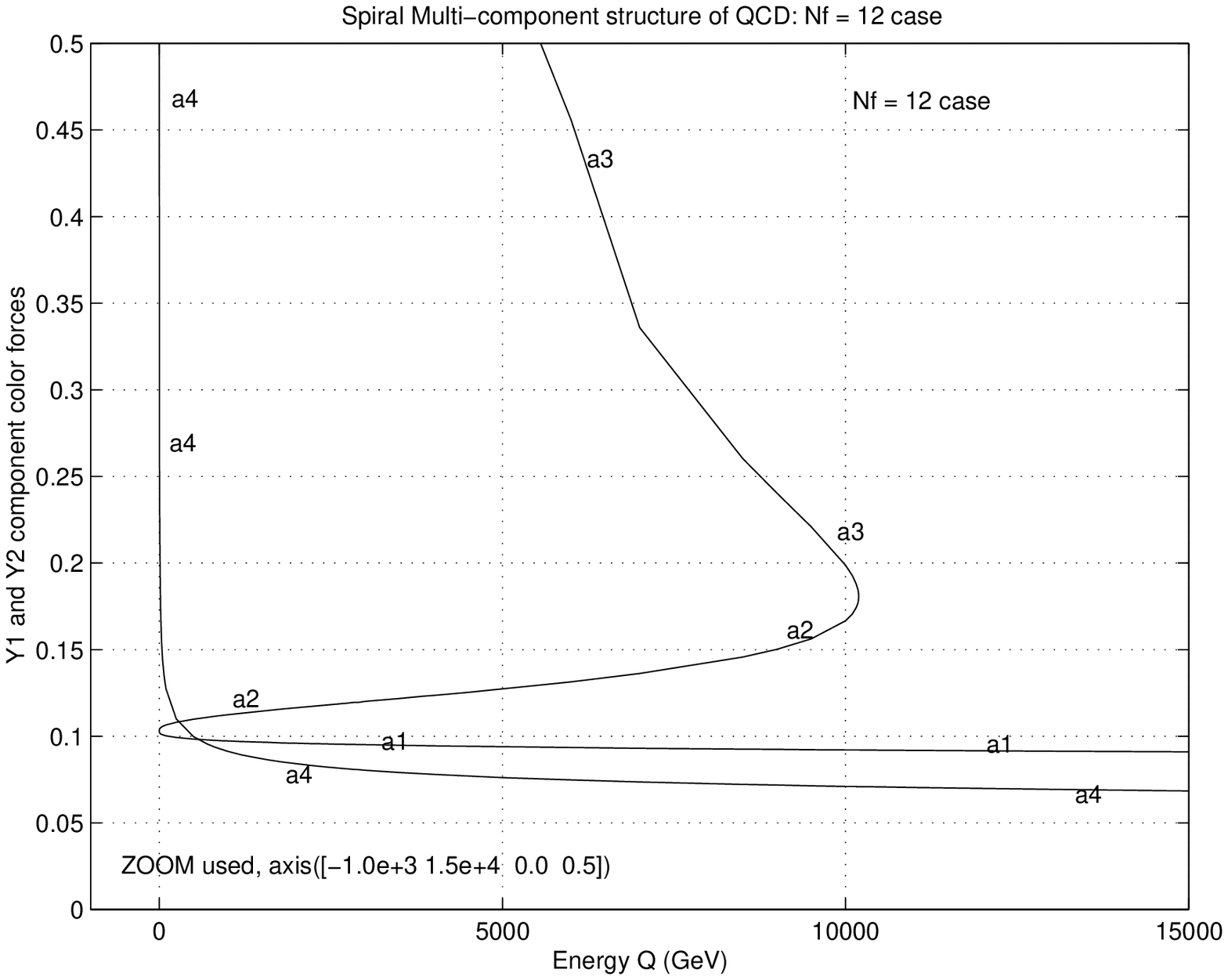}}
\caption{A plot showing a  spiral multi-component structure in Pad\'{e} QCD: Nf = 12 case}
\label{fig: ndili51}
\centering
\end{figure}

\begin{figure}
\scalebox{1.0}{\includegraphics{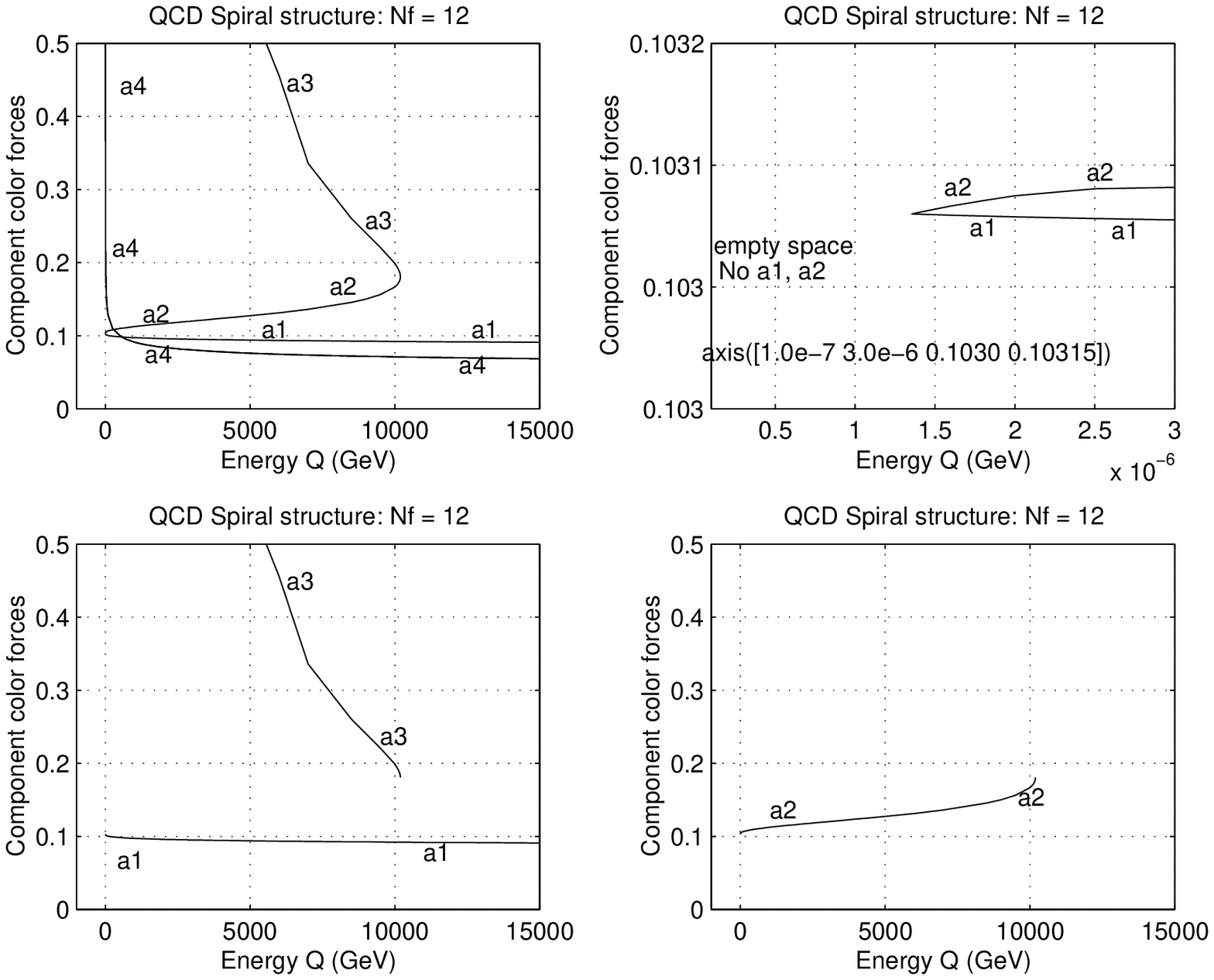}}
\caption{A plot showing a  multi-component couplant structure of Pad\'{e} QCD: Nf = 12 case}
\label{fig: ndili52}
\centering
\end{figure}

Arising from the fact stated above that while the $a_{1}, a_{2}, a_{3}$ are independent solutions of
Pad\'{e} eqn.~(\ref{eq: ndili60}), they are however joined together at the two critical 
(bifurcation) momentum points $Q_{\mathrm{min}}$ and $Q_{\mathrm{max}}$, in the manner given by: 
\[
a_{1} = a_{2} \ne  0, \mathrm{at} :  Q = Q_{\mathrm{min}},   
\]
and\\
\[
a_{2} = a_{3}  \ne 0,  \mathrm{at} :   Q = Q_{\mathrm{max}}. 
\]

we find that a combined plot of the individual variations
$(a_{1}, Q); (a_{2}, Q); (a_{3}, Q)$ gives one unbroken chain-like spiral structure,  
shown  in figs.~\ref{fig: ndili37} to~\ref{fig: ndili52} for various flavors.  The $a_{4}$
component has also been plotted in.

These chain-like spiral structures together with our computed values of the bifurcation
point critical momenta $Q_{\mathrm{min}}$ and $Q_{\mathrm{maxc}}$ shown in Table~\ref{tab: ndili2},
lead us to explicitly answer the question, for which flavor states the infra-red scenario I or 
scenario II holds.   Additionally, we have also shown in Table~\ref{tab: ndili2}, the values of the 
Pad\'{e} couplants $a_{1}(Q_{\mathrm{min}})$ and $a_{3}(Q_{\mathrm{max}})$ at these critical cut-off 
momentum points. They all enable us to answer the same question definitively.  Before discussing the 
question however, we first show that the above  features of the optimized $[1|1]$ Pad\'{e} QCD
are also shared to a large extent by the other Pad\'{e} QCDs, $[1|1]$, $[2|1]$ and $[1|2]$ that we 
also analyzed.

\section{ANALYSIS AND FEATURES OF THE NON-OPTIMIZED $[1|1], [2|1]$ AND $[1|2]$ PAD\'{E} QCDs}
\begin{figure}
\scalebox{1.0}{\includegraphics{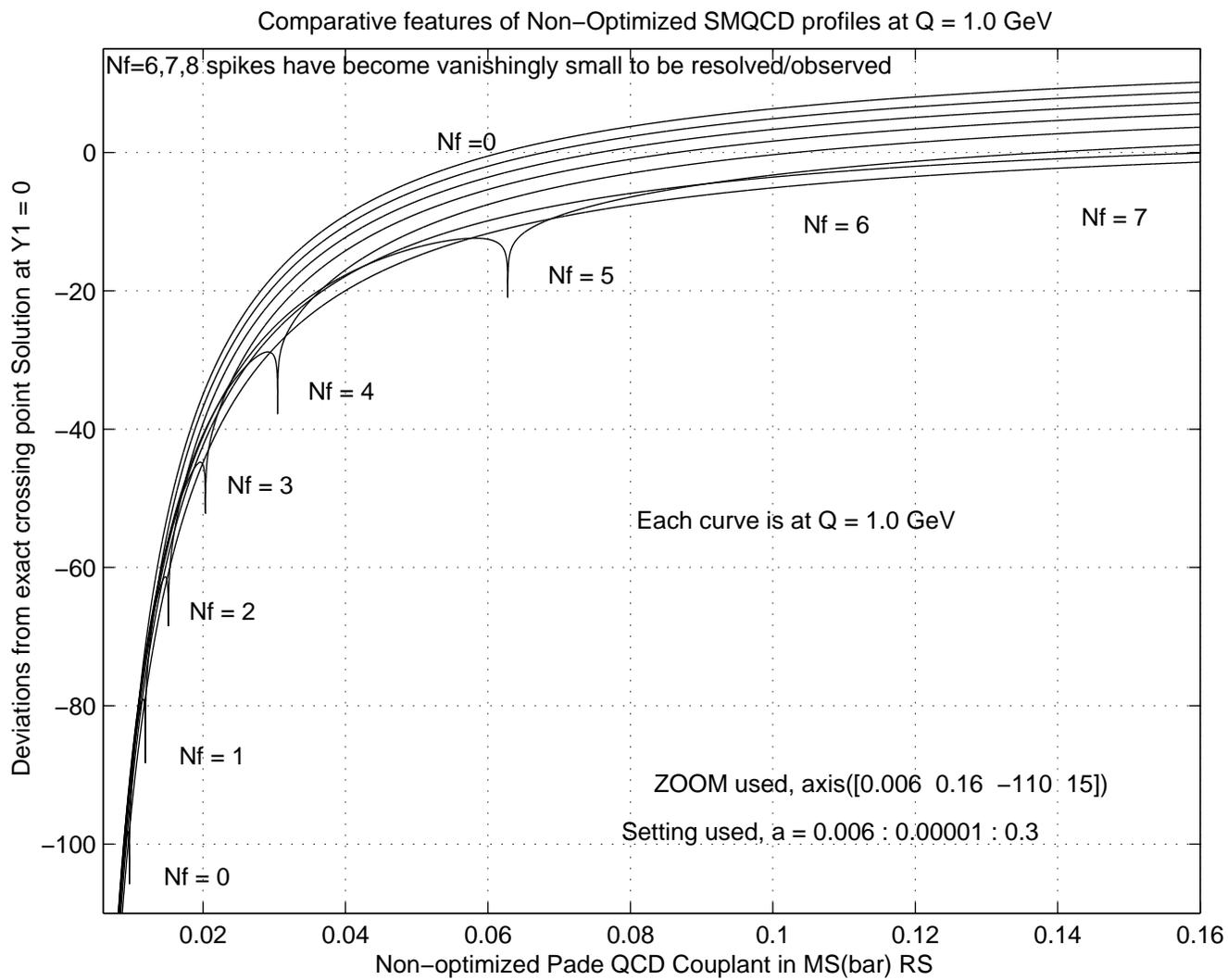}}
\caption{Combined Profile features of Non-optimized  $[1|1]$ Pad\'{e} QCD 
 at Q = 1.0 GeV, for $0 \le N_{f} \le 8$}.
\label{fig: ndili93}
\centering
\end{figure}

.
Using exactly the same computational and graphical procedures  described above for the optimized $[1|1]$
Pad\'{e} QCD,  we also analyzed  the non-optimized $[1|1], [2|1]$, and $[1|2]$ Pad\'{e} QCD couplant
equations~(\ref{eq: ndili37})~,~(\ref{eq: ndili59e}) and~(\ref{eq: ndili59f}).  We found essentially 
the same Pad\'{e} QCD features, with small differences which we  summarize as follows.
\begin{enumerate}
\item The $a_{4}$ component solution found in optimized Pad\'{e} QCD is not exhibited by any of the three
non-optimized Pad\'{e} QCDs.  As will be recalled, the $a_{4}$ solution arose from the quadratic
nature of the optimized $c_{2}$ coefficient given in eqns.~(\ref{eq: ndili54}) and~(\ref{eq: ndili55}).
This $c_{2}$ quadratic  feature is not present in the non-optimized $[1|1], [2|1], [1|2]$ Pad\'{e} QCDs,
and as such the $a_{4}$ solution does not arise in these cases. We can therefore for the moment play
down the existence and role of the Pad\'{e}  $a_{4}$ component solution.

\item  However, the triple crossing point feature and hence the $a_{1}, a_{2}, a_{3}$  component structure, 
are present in exactly the same form in all the four Pad\'{e} QCDs we analyzed, optimized or not.
Correspondingly, the spiral chain-like structure exists in all of them and  can be taken
as intrinsic  characteristic feature of a Pad\'{e} QCD. 

\item  We found that while each Pad\'{e} QCD  at a given flavor, is characterized by two critical or cut off momenta
$Q_{\mathrm{min}}$ and $Q_{\mathrm{max}}$, the actual values of these momentum pairs differed
substantially from one Pad\'{e} QCD to other. The values found for example, for the non-optimized $[1|1]$
Pad\'{e} QCD are shown in Table~\ref{tab: ndili6} to be compared with Table~\ref{tab: ndili2} for the
optimized $[1|1]$ Pad\'{e} QCD.  The $[2|1]$ and $[1|2]$ paired critical momenta differ just as much.
The trend of the variation is not yet  clear but is being separately studied.

\item  In contrast, the values of the critical (cut-off) couplants $a_{1}(Q_{\mathrm{min}})$ and
$a_{3}(Q_{\mathrm{max}})$ exhibit some consistency, especially in high flavor states seen in 
Tables~\ref{tab: ndili2} and~\ref{tab: ndili6}.

\item We found that the optimized $[1|1]$ Pad\'{e} QCD has the feature of resolving clearly, the spiral
nature of the $N_{f} = 6, 7, 8$ flavor states shown in figs.~\ref{fig: ndili43b} and~\ref{fig: ndili44a}, compared
to the non-optimized $[1|1]$  Pad\'{e} QCD where the couplant structure of the same 
$N_{f} = 6, 7, 8$ flavor states, is not so  resolved, as can be seen in fig.~\ref{fig: ndili93}.

\end{enumerate}

The above small differences apart, we affirm that our Pad\'{e} QCD of various orders,
optimized or not, has the characteristic feature of  triple multiplicity of solutions 
$(a_{1}, a_{2}. a_{3})$; two characteristic bifurcation points, $Q_{\mathrm{min}}$ and 
$Q_{\mathrm{max}}$; and a spiral chain-like structure continuously connecting the three component
solutions, $(a_{1}, a_{2}, a_{3})$.  Using figs.~\ref{fig: ndili37} to~\ref{fig: ndili52} 
of the $[1|1]$ NNLO optimized Pad\'{e} QCD as typical of these behaviors and properties of a
Pad\'{e} QCD, we now consider the question of which of infra-red scenario I or II holds at a 
given flavor.

\section{THE QUESTION OF  IR SCENARIO I  OR  SCENARIO II  IN PAD\'{E} QCD}
As already stated, in our approach, the question of whether we have infra-red  scenario I
(IR stable fixed point and frozen couplant) or IR scenario II (IR attractor point and bifurcated couplant)
is directly decided by whether the momentum gap $0 \le Q < Q_{\mathrm{min}}$ exists and is finite.
Where this is the case, the IR scenario II  necessarily holds, ruling out scenario I.
Conversely, if the momentum gap does not exist for a given flavor, the IR structure of that
flavor state is necessarily of scenario I type, and scenario II becomes ruled out.  We now
examine our spiral plots as well as our Table~\ref{tab: ndili2} to see which scenario holds. We find as
follows:
\begin{enumerate}
\item {\bf The case $0 \le N_{f} \le 5$}\\
From our spiral plots figs.~\ref{fig: ndili37} to~\ref{fig: ndili43a} it is clear that IR scenario II
is what operates and not scenario I.  The same conclusion is drawn from our Tables~\ref{tab: ndili2}
and~\ref{tab: ndili6}, where  it is seen that the momentum gap $0 \le Q \le Q_{\mathrm{min}}$ is far
from being zero.  These findings agree with those of  Elias et. al. who found that scenario II
holds for all $0 \le N_{f} \le 5$, regardless of which Pad\'{e} approximants, 
$[1|2], [2|1], [1|3], [3|1]$, or $[2|2]$ they used.

\item {\bf The case $N_{f} = 6, 7, 8$}\\
From our spiral plots shown in figs.~\ref{fig: ndili43b} and~\ref{fig: ndili44a} it is also clear that
IR scenario II holds for $N_{f} = 6, 7, 8$, and not scenario I.  The $Q_{\mathrm{min}}$ values shown 
in  Tables~\ref{tab: ndili2}  confirm this further.  The momentum gap $0 \le Q \le Q_{\mathrm{min}}$ is
far from being zero in all these cases.

\item{\bf The case $9 \le N_{f} \le 16$}\\
For these flavor states,   Elias et. al. by their method, found scenario I IR behavior. In contrast,
our method shows the opposite scenario II behavior  for all $9 \le N_{f} \le 16$ as for all
$0 \le N_{f} \le 8$.  Our finding of scenario II for all $9 \le N_{f} \le 16$ is seen clearly in
figs.~\ref{fig: ndili45} to~\ref{fig: ndili52} for the cases of $N_{f} = 10, 12, 15$, chosen for 
illustration.  All the other flavor states $N_{f} = 9, 11, 13, 14. 16$, have exactly the same
scenario II behavior.
 
We can explain this particular difference in the two results, and establish 
that our finding is the correct Pad\'{e} position. Looking at figs.~\ref{fig: ndili45} 
to~\ref{fig: ndili52} as well as Tables~\ref{tab: ndili2} and~\ref{tab: ndili6},  it is seen that the 
infra-red attractor points exist at progressively lower and lower momentum in these high flavor states,
and only a method designed to follow this trend can establish the continued existence of the 
scenario II behavior. The Elias et. al. method would appear not  equipped for this.

 Even with our method, looking only at our figs.~\ref{fig: ndili45},~\ref{fig: ndili47},
 and~\ref{fig: ndili51}, one would gain the impression that our $a_{1}$ and $a_{2}$ Pad\'{e} component 
 couplants  which are the exact analogue of Elias et. al. bifurcated couplants at $\mu = \mu_{c}$, 
 stretch on to zero momentum point, $Q = \mu = 0$ in which case scenario I could be concluded. 
 
 However, if one follows these
 two branch Pad\'{e} component couplants $a_{1}$ and $a_{2}$, to  sufficiently low momentum regions, 
 shown typically in our  figs.~\ref{fig: ndili48} and~\ref{fig: ndili52}, one sees unambiguously,  
 that these two component couplants still meet and turn back at some finite critical momentum
 $Q_{\mathrm{min}} = \mu_{c} \ne 0$, leaving a clearly visible gap, $0 \le Q < Q_{\mathrm{min}}$, 
 where  both the $a_{1}$ and $a_{2}$ are totally absent.  This gap is present and completely visible 
 in all cases $9 \le N_{f} \le 16$, leaving no doubt that scenario II is intrinsically  what holds
 in these Pad\'{e} flavor cases as for the $0 \le N_{f} \le 8$ cases.
 
 It just happens that for the higher flavor states, the spiral (bifurcation)  point where $a_{1}$ and 
 $a_{2}$ meet, shifts to lower and lower momentum point, and unless this shifting is followed from flavor
 to flavor, one can conclude erroneously that scenario I operates in these higher flavor states. 
\end{enumerate} 
 
 Against the above, we can affirm that our results  are actually consistent  with those of Elias et. al. 
 and  that the features  we found and analyzed  in our Pad\'{e} QCDs, are the intrinsic features of
 Pad\'{e} QCD of any order $[1|1], [2|1], [1|2], [3|1], [1|3]$ or $[2|2]$.  However our graphical
 method of analyzing these Pad\'{e} QCD features, has the advantage of showing clearly that the 
 infra-red scenario II behavior is the intrinsic behavior of Pad\'{e} QCD in all flavor states 
 $0 \le N_{f} \le 16$, and not IR scenario I.
 
 We are led also to firmly identify our critical momentum point $Q_{\mathrm{min}} \equiv \mu_{c}$ 
 as fundamentally an infra-red attractor point or a pole singularity of Pad\'{e} QCD beta function, for  
 all flavor states, $0 \le N_{f} \le 16$.  This means that the critical Pad\'{e} QCD couplants
 $a_{1}(Q_{\mathrm{min}})$ shown in Table~\ref{tab: ndili2} are  fundamentally  infra-red attractor
 points and not IR stable fixed points, although for sufficiently high flavor $N_{f}$, the
 attractor behavior (point) can be mistaken for a  IR stable  fixed point  or frozen couplant behavior.

 \section{COMPARISON WITH  TRUNCATED PQCD  INFRA-RED FIXED POINT RESULTS}
 Having settled the above question, we can  check for further consistency of our Pad\'{e} QCD 
 compuatations and findings by comparing them with the earlier work of Banks and Zaks~\cite{Banks82};
 and of Stevenson et. al.~\cite{Mattingly92,Mattingly94,Stevenson94,Kubo84}, who used purely
 truncated PQCD beta function to search for infra-red (IR) fixed points in PQCD.
 Banks and Zaks~\cite{Banks82} as well as Stevenson~\cite{Stevenson94} analyzed the NLO truncated 
 PQCD of  eqn.~(\ref{eq: ndili7}) and predicted PQCD IR fixed points shown against flavor numbers 
 in our Table~\ref{tab: ndili4}. Separatelyly, 
 Stevenson et. al.~\cite{Mattingly92,Mattingly94,Stevenson94,Kubo84}, using optimized NNLO truncated 
 PQCD of eqn.~(\ref{eq: ndili8}) investigated  the same  IR  fixed point structure of PQCD. The IR fixed 
 points they found are shown in the same Table~\ref{tab: ndili4}.  We can now compare these results
 with our $a_{1}$ values computed from our $[1|1]$ optimized (NNLO) Pad\'{e} QCD. These $a_{1}(Q_{\mathrm{min}})$ values are 
 also shown side by side in Table~\ref{tab: ndili4}. We now observe as follows.
 
 \begin{enumerate}
 
 \item  For low flavors $N_{f}$, there is no agreememt  between our $a_{1}$ values and the 
 NLO or NNLO  IR fixed point values of the above authors.  This is consistent with the earlier stated fact 
 that the point $(Q_{\mathrm{min}}, a_{1}(Q_{\mathrm{min}}))$ is not an IR stable fixed point of QCD 
 but a pole singularity (IR attractor point) of the beta function.

 \item For very high $N_{f} = 16, 15, 14, 13 ..$,  our values of $a_{1}$ agree almost exactly with the NLO
 and NNLO  IR stable  fixed point values of the above authors. This agreement is evidence that our Pad\'{e} QCD
 computations and graphical solution method are correct and reliable. However, the agreement comes not because  our $a_{1}(Q_{\mathrm{min}})$ point
 has suddenly become IR stable fixed point of scenario I, but from the spiral structure shifting to lower and lower
 momentum points discussed earlier.  As a result of this shifting, the IR attractor point 
 $a_{1}(Q_{\mathrm{min}})$ becomes  mimiced more and more effectively by an IR stable fixed point of a
 seemingly frozen PQCD (or $a_{1}$) couplant. But when one follows this  bifurcation point shifting to 
 sufficiently low energy as shown typically in our figs.~\ref{fig: ndili48} and~\ref{fig: ndili52}, one
 soon  sees that $a_{1}$ and $a_{2}$ are still separate or bifurcated component couplants  and that the
 PQCD couplant $a_{1}$ did not freeze at all. 
 
 \end{enumerate}
 
 We conclude finally that  our Pad\'{e} QCD graphical  computations  and findings, can be relied upon, 
 being consistent with the independent Pad\'{e} beta function numerator/denominator zero analysis of
 Elias et. al.;  and also consistent with the direct PQCD IR computations of Banks and Zaks, 
 and Stevenson et. al.

 \section{SUMMARY AND CONCLUSIONS}
 We have presented a reliable and consistent graphical computational method of analyzing the infra-red 
 structure of Pad\'{e} approximant QCDs.  We have shown with the method that the intrinsic behavior of
 Pad\'{e} QCD is one in the PQCD bifurcates at some critical Kogan-Shifman type momentum
 $\mu_{c} = Q_{\mathrm{min}}$, leaving the infra-red region $0 \le \mu < \mu_{c}$ totally decoupled from the
 PQCD dynamics.  We found however that besides this PQCD bifurcation, there exists a second bifurcation
 by the upper branch  $a_{2}$ couplant, into a third branch or component couplant $a_{3}$, at 
 another critical momentum $Q_{\mathrm{max}}$ where in all cases, $Q_{\mathrm{max}} > Q_{\mathrm{min}}$.
 The three components or branches $(a_{1}, a_{2}, a_{3})$ are arranged in a chain-like spiral structure.
 The $a_{3}$ in general runs into the original  infra-red region $\mu < \mu_{c}$ but soon rises sharply to very
 large values in a manner suggestive of a Landau pole behavior of the Pad\'{e} couplant at its second
 bifurcation point.  The extent to which the above Pad\'{e} QCD features and structure can be taken over as
 those of real physical QCD is a matter for further studies which we are now following up.

{\bf Acknowledgements}
The author is grateful to Dr. Billy Bonner, Director T.W. Bonner Nuclear Laboratory, Rice University, for
providing him office space in the laboratory and for general hospitality. He thanks Dr. P. M. Stevenson,
Professor of Physics at the same Laboratory for discussion concerning his optimization principle used in
this paper.  The author thanks Dr. V. Elias for reading through an earlier version of this paper,
and making useful comments. Some computing assistance received from  Niki Serakiotou and Pablo Yepes
is gratefully acknowledged.
The work would however never have been possible neither would the author's visiting stay at Rice University,
without the sustained financial  support made available to the author from outside  sources.  He thanks
all contributors. The author is particularly grateful to Dr. Awele Ndili of Stanford University, California,
who provided him a lot of technical advice with regard to computing and software, and specifically purchased
a modern PC for the author`s use,  on which all the computations and figures  reported in this
paper were done.

\appendix

\section{(Appendix A): The exact MATLAB  Program (QCD3.m) used by the author to solve numerically, the optimized SMQCD 
eqn.~(\ref{eq: ndili59}) of this paper}

\% The footnotes given below are to guide the actual execution of the program; they are not part of the program.\\
\% The program now begins.\\

clear all; \\
$N = 2$; \% see footnote No. 1 below.\\
Rh2 = -10.91120013471925;  \% see footnote No. 2 below.\\
$Q = 1.0;$  \%  see footnote No. 3 below.\\
$L = 0.230;$ \% L  = $\Lambda_{\mathrm{QCD}} = 0.230$ GeV, kept constant all through the SMQCD computations.\\
$r1 = 1.9857 - 0.1153 * N;$  \% $r1$ is a constant all through the (optimized) SMQCD computations.\\
$b = (33 -2*N)/6;$  \% $b$ is a constant all through the SMQCD computations.\\
$c = (153 - 19*N)/(66 - 4*N);$ \%  $c$  is a constant all through the SMQCD computations.\\
Rh1 = $b * log(Q/L) + c*log(2*c/b) - r1;$  \% see footnote No. 4 below.\\
$D = 8*c*c  - 4*Rh2;$  \% $D$ varies as we vary input Rh2.\\
$P1 = 3*c/2  + sqrt(D)/2;$ \% $P1$ varies as we vary input Rh2.\\
$P2 = 3*c/2  - sqrt(D)/2;$ \% $P2$ varies as we vary input Rh2.\\
X1 = Rh1 - P1;  \% X1 varies as we vary Q and Rh2.\\
X2 = Rh1 - P2;  \% X2 varies as we vary Q and Rh2.\\
$a = 0.02  :  0.00001  :  0.12;$  \%  see footnote No. 5 below.\\
$F1 = 1.0  + c*a  -  2*a*P1;$  \%  $F1$ is just a notation.\\
$F2 = 1.0  + c*a  -  2*a*P2;$  \%  $F2$ is just a notation.\\
$H1 = ((c*a) ./F1);$  \% note the space and the  dot before /F1 \\
$H2 = ((c*a) ./F2);$  \% note the space and the  dot before /F2 \\
$X3 = (1 ./a)  + c*log(abs(H1));$  \% note the space and the  dot before /a \\
$X4 = (1 ./a)  + c*log(abs(H2));$  \% note the space and the  dot before /a \\
Y1  = X1  -  X3;\\
Y2 = X2  -  X4;\\

plot(a, Y1); grid on  \% see footnote No. 6 below.\\
xlabel(`Optimized couplant`)\\
ylabel('Y1 Deviation from optimization solution (crossing point)')\\
title('Graphical search for crossing point solution of the Optimization Eqn. for N =2')\\
\% use next a suitable axis command axis($[x_{1} x_{2} y_{1} y_{2}]$), to zoom closely into a crossing region to be read off.
\% This zooming in with the axis command needs to be used all the time, and with great skill, to obtain accurate readings.
\% The MATLAB program ends here.\\

\% FOOTNOTES begin (not part of the program).\\

\% Footnote No. 1: After varying Q over all desired ranges at one fixed flavor number N, we change N  by hand,
by simply inserting any other value:  N = 0, 1, 2, 3, .......16.\\

\% Footnote No. 2:  The Rh2 values we used were those computed previously  by Mattingly and 
Stevenson~\cite{Mattingly92,Mattingly94,Stevenson99}. Explicitly, these $\rho_{2}$ values are:

\% $N_{f} = 0,  \rho_{2} =  - 8.410032589173554$;  $N_{f} = 1,  \rho_{2} =  -  9.996607149709793$;\\  

\% $N_{f} = 2,  \rho_{2} =  - 10.91120013471925$;  $N_{f} = 3,  \rho_{2} =  -  12.20710268197531$;\\  

\% $N_{f} = 4,  \rho_{2} =  - 13.90995802777778$;  $N_{f} = 3,  \rho_{2} =  -  15.49181836878120$;\\  

\% $N_{f} = 6,  \rho_{2} =  - 17.66469557734694$;  $N_{f} = 7,  \rho_{2} =  -  19.78668878025239$;\\  

\% $N_{f} = 8,  \rho_{2} =  - 22.74511792421761$;  $N_{f} = 9,  \rho_{2} =  -  25.96983971428571$;\\  

\% $N_{f} = 10,  \rho_{2} =  - 30.64825148592373$;  $N_{f} = 11,  \rho_{2} =  - 36.70527878387512$;\\  

\% $N_{f} = 12,  \rho_{2} =  - 46.58505774333333$;  $N_{f} = 13,  \rho_{2} =  - 63.56012999671129$;\\  

\% $N_{f} = 14,  \rho_{2} =  - 101.9145678055556$;  $N_{f} = 15,  \rho_{2} =  - 229.8874551066667$;\\  

\% $N_{f} = 16,  \rho_{2} =  - 1724.404563921111$.\\

\% Footnote No. 3: Start in general with Q = 1.0 GeV (for any flavor), and after observing the SMQCD
profile pattern, you vary Q upwards or downwards in whatever steps (large or small) you desire, but state the value of Q 
in GeV always, e.g. $Q = 1.0e-305$   for $Q = 10^{-305}$ GeV.\\

\%Footnote No. 4: For $9 \le N \le 16$, you must replace $c*log(2*c/b) - r1;$ by $c*log(abs(2*c/b) - r1;$  This is because
$c$ changes sign as we enter the $N \geq 9$ or $(c^-)$ phase of QCD, so we need to use the absolute value. This apart, there is no 
other difference with the $(c^+)$ or $0 \le N \le 8$ computations.  Notice however, that the SMQCD profile now turns upside down.\\

\% Footnote No. 5: The two end points:  $a  = 0.02$  and  $a = 0.12$ are in general not fixed for any one N value but can be 
chosen differently or varied in the course of any one computation. The optimal choice to suit an already chosen Q value will
be found by experience. In general, when Q is very low, we need to make the terminal point of the $a$ range very large, in
order to be able to read at all  the $a_{3}(Y_{1})$ crossing point. However, whatever the range chosen for $a$, the same 
incremental step $\Delta a = 0.00001$  of the floating couplant should be used throughout these computations to achieve the 
same degree of accuracy of crossing point readings (solutions).  Also for any one Q value the chosen range of $a$ should not
be too large, otherwise the computer goes into a time consuming loop and floods its memory unduly.\\

\% Footnote No. 6: After plotting the (a, Y1) curves for all desired Q values, we replace the command:  plot(a, Y1), with
the command: plot(a, Y2). Then we vary Q again over all desired values (with N still kept fixed), in order to plot out the
(a, Y2) profiles and read off the $a_{4}$ crossing points (solutions).\\

\% Footnote No. 7:  Corresponding changes are to be made by hand in the ylabel and title commands.

\newpage

\begin{table}
\caption{Values of the denominator zeros $(a_{d})$, and the numerator zeros $(a_{n})$ of the Pad\'{e}
beta function eqn.~(\ref{eq: ndili35}) computed as a function of flavor number $N_{f}$.}\label{tab: ndili1a}
\begin{center}

\begin{tabular}{|c|c|c|}\hline
Flavor Number $N_{f}$   &  Denominator zero $a_{d}$   &  Numerator zero $a_{n}$ \\
\hline

0  &  0.2856   &   0.8453\\

1  &  0.3094   &   0.9339\\

2  &  0.3433   &   1.0749\\

3  &  0.3976   &   1.3565\\

4  &  0.5053   &   2.2780\\

5  &  0.8549   &   -10.9638\\

6  &  - 3.2000   &   -0.8058\\

7  &  - 0.2280   &   -0.2036\\

8  &  -0.0063   &   - 0.0063\\

9  &   0.0.0799   &   0.0763\\

10  &  0.1288  &   0.1089\\

11  &  0.1625  &   0.1150\\

12  &  0.1886  &   0.1056\\

13  &  0.2108  &   0.0873\\

14  &  0.2307  &   0.0640\\

15  &  0.2494  &   0.0384\\

16  &  0.2676   &   0.0126\\

\hline
\end{tabular}
\end{center}
\end{table}

\begin{table}
\caption{Values of $Q_{\mathrm{min}}$ and $Q_{\mathrm{max}}$ in optimized $[1|1]$ Pad\'{e} QCD, 
tabulated here as a function of flavor number $N_{f}$ , together with the corresponding values
of $a_{1}(Q_{\mathrm{min}})$ and $a_{3}(Q_{\mathrm{max}})$ at these spiral (bifurcation) points.}\label{tab: ndili2}
\begin{center}

\begin{tabular}{|c|l|l|l|l|}\hline
Flavor Number $N_{f}$   &  $Q_{\mathrm{min}}$ (GeV)  &  $a_{1}(Q_{\mathrm{min}})$  &    $Q_{\mathrm{max}}$  (GeV)  &  $a_{3}(Q_{\mathrm{max}})$\\
\hline

 0   &  25.725856708   &  0.063660   &  1014.5456575   &  0.074680\\

 1   &  30.04865615  &   0.065450   &   722.2234986368   &  0.076230\\

 2   &  33.44905052   &  0.068430   &  907.79549390   &  0.079170\\

 3   &  37.90878530  &   0.071720   &   679.6977179750   &  0.082210\\

 4   &  43.8476891550  &  0.075430  &   1763.271161880  &  0.085340\\

 5   &  49.2803656516  &  0.080480  &  435.1917430243  &   0.089580\\

 6   &  56.6158149   &   0.086260   &  316.669113500   &  0.093780\\

 7   &  60.714770460   &  0.094350   &  184.198105110   &  0.099280\\

 8   &  56.96621400   &  0.103870   &  59.14194290  &   0.104190\\

 9   &  5.1612395010   &   0.10950   &  47.126555910   &  0.117200\\

 10   &  6.858398820$(10^{-2})$   &   0.111920  &  73.2831370350  &   0.133120\\

 11   &  3.742331065$(10^{-3})$   &   0.111070   &  276.8901326   &  0.154840\\

 12   &  1.3532233520$(10^{-6})$   &   0.103080  &   1.019000784$(10^{+4})$   &  0.180700\\

 13   &  8.5876890$(10^{-15})$   &   0.087960  &   1.771666446$(10^{+8})$    &  0.214860\\

 14   &  3.540839347$(10^{-49})$  &   0.065770  &   1.10405153$(10^{+22})$  &   0.256120\\

 15   &  6.26397660$(10^{-104})$  &   0.039650   &  2.17827906$(10^{+89})$   &  0.310460\\

 16   &  $\ll 10^{-307}$    &  0.01280   &  $\gg 10^{+308}$   &   0.380525\\
\hline
\end{tabular}
\end{center}
\end{table}

\begin{table}
\caption{Values of $Q_{\mathrm{min}}$ and  $Q_{\mathrm{max}}$, in the non-optimized $[1|1]$ Pad\'{e} QCD.
The gap  between $Q_{\mathrm{min}}$ and $Q_{\mathrm{max}}$ closed up so fast after $N_{f} = 5$, that we
could not resolve or read off the crossing point values for $N_{f}$ = 6, 7 and 8 in this non-optimized case.}\label{tab: ndili6}
\begin{center}
\begin{tabular}{|c|l|c|l|c|}\hline
Flavor Number   &  $Q_{\mathrm{min}}$   &  $a_{1}(Q_{\mathrm{min}})$  &    $Q_{\mathrm{max}}$   &  $a_{3}(Q_{\mathrm{max}})$\\
$N_{f}$         &     (GeV)             &                             &         (GeV)           &                   \\
\hline

0   &  $5.6468615736(10^{+7})$   &  0.009440  &  $2.2356088581(10^{+8})$   &  0.009650\\

1   &  $4.4044582296(10^{+6})$   &  0.011590   &  $2.6026228349(10^{+7})$   &  0.011890\\

2   &  $3.2521338660(10^{+5})$   &  0.014690   &  $1.40497854250(10^{+6})$   &  0.015140\\

3   &  $2.1002809749(10^{+4})$   &  0.019640   &  $1.0761648724(10^{+5})$  &   0.020350\\

4   &  $1.0134866543(10^{+3})$   &  0.02911   &  $8.5575884074(10^{+3})$   &  0.03047\\

5   &  $2.53433811590(10^{+1})$   &  0.058180   &  $2.351206362(10^{+2})$   &  0.062790\\

6   &                    &              &                    &    \\

7   &                    &              &                    &    \\

8   &                    &              &                    &    \\

9   &  $7.928886017(10^{+12})$  &   0.012690   &  $1.5839622507(10^{+13})$   &  0.012790\\

10   &  $2.083364761(10^{+5})$   &  0.026420   &  $3.5918756304(10^{+6})$   &  0.027450\\

11   &  $2.064339838(10^{-1})$   &  0.043050   &  $5.9926835830(10^{+3})$   &  0.048350\\

12   &  $5.940796088(10^{-8})$   &  0.062140   &  $6.6365417260(10^{+1})$   &  0.083840\\

13   &  $3.1465791130(10^{-22})$   &  0.075920   &  $8.845270086(10^{-1})$   &  0.154850\\

14   &  $2.617118166(10^{-56})$   &  0.069880   &  $3.096682427(10^{-3})$   &  0.3321469\\

15   &  $4.098229275(10^{-168})$   &  0.043470   &  $2.848034499(10^{-8})$   &  0.99750\\

16   &  $\ll 1.0(10^{-307})$   &  0.013230    &     $1.401656210675(10^{-36})$    &  9.634880\\
\hline
\end{tabular}
\end{center}
\end{table}

\begin{table}
\caption{Comparative values of the IR fixed points of PQCD from NLO  \&  NNLO
perturbative QCD calculations, and the values of $a_{1}(Q_{\mathrm{min}})$ and $a_{3}(Q_{\mathrm{max}})$
shown earlier in Table~\ref{tab: ndili2}, computed from the optimized $[1|1]$ Pad\'{e} .}\label{tab: ndili4}
\begin{center}
\begin{tabular}{|c|l|c|c|c|}\hline
Flavor Number  &  $Q_{\mathrm{min}}$   &  $a_{1}(Q_{\mathrm{min}})$  &  $a_{\mathrm{IR}}^*(PQCD)$   &   $a_{\mathrm{IR}}^*(PQCD)$\\
$N_{f}$        &    (GeV)              &   op. $[1|1]$                     &     at NLO             &   at NNLO (optimized)      \\
\hline
0  &  25.725856708  &  0.063660  &  0.4112  &  0.313284\\

1  &  30.04865615  &  0.065450  &  0.3863  &  0.280270\\

2  &  33.44905052  &  0.068430  &  0.3614  &  0.2634796\\

3  &  37.90878530  &  0.071720  &  0.3364 &  0.244217\\

4  &  43.8476891550  &  0.075430  &  0.3115  &  0.224065\\

5  &  49.2803656516  &  0.080480  &  0.2866  &  0.208085\\

6  &  56.6158149  &  0.086260  &  0.2617  &  0.190693\\

7  &  60.714770460  &  0.094350  &  0.2368  &  0.176191\\

8  &  56.96621400  &  0.103870   &  0.2118  &  0.160361\\

9  &  5.1612395010  &  0.10950  &  0.1869  &  0.146072\\

10  &  $6.858398820 \times 10^{-2}$  &  0.111920  &  0.1620  &  0.1304388\\

11  &  $3.742331065 \times 10^{-3}$  &   0.11107  &  0.1371  &  0.1150355\\

12  &  $1.3532233520 \times 10^{-6}$  &   0.103080  &  0.1121  &  0.0979828\\

13  &  $8.5876890 \times 10^{-15}$  &  0.087960  &  0.0872  &  0.0797984\\

14  &  $3.540839347 \times 10^{-49}$  &  0.065770  &  0.0623  &  0.05940013\\

15 &   $6.26397660 \times 10^{-104}$  &  0.039650  &  0.0374 &   0.03688832\\

16  &  $\ll 10^{-307}$  &  0.0128  &  0.0125  &  0.01248992\\
\hline
\end{tabular}
\end{center}
\end{table}

\newpage
\listoftables

\newpage
\listoffigures

\end{document}